\documentclass{aa}  
\usepackage{graphicx}	
\usepackage{amsmath}	
\usepackage{amssymb}	
\usepackage{bm}	        
\usepackage{csvsimple}
\usepackage{ragged2e}
\usepackage{multirow}
\usepackage{siunitx} 
\usepackage[utf8]{inputenc}
\usepackage{dirtytalk}
\usepackage{xparse}
\usepackage{enumitem} 
\usepackage{array}
\usepackage{CJKutf8}
\usepackage[flushleft]{threeparttable}
\usepackage{txfonts}
\usepackage[colorlinks, citecolor=blue, linkcolor=blue]{hyperref}
\usepackage{graphicx}
\usepackage{txfonts}
\usepackage{svg}

\begin{document}

   \title{Emergence of high-mass stars in complex fiber networks (EMERGE)}

   \subtitle{I. Early ALMA Survey: observations and massive data reduction}

   \titlerunning{EMERGE I. Early ALMA Survey: observations and massive data reduction}

   \author{A. Hacar
          \inst{1} 
          \and
          A. Socci
          \inst{1}
          \and
          F. Bonanomi
          \inst{1}
          \and
          D. Petry
          \inst{2}
          \and
          M. Tafalla
          \inst{3}
          \and
          D.~Harsono  
          \inst{4}
          \and
          J. Forbrich
          \inst{5}
          \and
          J. Alves
          \inst{1}
          \and
          J. Grossschedl
          \inst{1}
          \and
          J.~R. Goicoechea
          \inst{6}
          \and
          J. Pety
          \inst{7,8}
          \and
          A. Burkert
          \inst{9,10,11}
          \and
          G.X. Li
          \inst{12}
          }

   \institute{Department of Astrophysics, University of Vienna,
              T\"urkenschanzstrasse 17, A-1180 Vienna\\
              \email{alvaro.hacar@univie.ac.at}
         \and
             European Southern Observatory, Karl-Schwarzschild-Strasse 2, D-85748 Garching, Germany
        \and   
            Observatorio Astronómico Nacional (IGN), Alfonso XII 3, E-28014 Madrid, Spain
        \and
        Institute of Astronomy, Department of Physics, National Tsing Hua
University, Hsinchu, Taiwan
        \and
            University of Hertfordshire, Centre for Astrophysics Research, College Lane, Hatfield AL10 9AB, UK
        \and   
            Instituto de Física Fundamental (CSIC). c/ Serrano 121-123, 28006, Madrid, Spain
        \and
            IRAM, 300 rue de la Piscine, 38406 Saint Martin d'H\`eres, France
        \and
            LERMA, Observatoire de Paris, PSL Research University, CNRS, Sornonne Universit\'es, 75014 Paris, France
        \and
        Universit\"ats-Sternwarte, Ludwig-Maximilians-Universit\"at M\"unchen, Scheinerstrasse 1, D-81679 Munich, Germany 
        \and
        Excellence Cluster ORIGINS, Boltzmannstrasse 2, D-85748 Garching, Germany
        \and
        Max-Planck Institute for Extraterrestrial Physics, Giessenbachstrasse 1, D-85748 Garching, Germany
        \and 
            South-Western Institute for Astronomy Research, Yunnan University, Kunming, 650500 Yunnan, P.R. China
        }

   \date{Received XXX ; accepted 06/03/2024}

 
  \abstract
   {Recent molecular surveys have revealed a rich gas organization of sonic-like filaments at small-scales (aka fibers) in all kind of environments prior to the formation of low- and high-mass stars. These fibers form at the end of the turbulent cascade and are identified as the fine sub-structure within the hierarchical nature of the gas in the Interstellar Medium (ISM). 
   }
   {Isolated fibers provide the sub-sonic conditions for the formation of low-mass stars. This paper introduces the EMERGE project aiming to investigate whether complex fiber arrangements (networks) could also explain the origin of high-mass stars and clusters.}
   {We analyzed the EMERGE Early ALMA Survey including 7 star-forming regions in Orion (OMC-1/2/3/4 South, LDN~1641N, NGC~2023, and Flame Nebula) homogeneously surveyed in both molecular lines (N$_2$H$^+$ J=1-0, HNC J=1-0, plus HC$_3$N J=10-9) and 3mm-continuum using a combination of interferometric ALMA mosaics and IRAM-30m single-dish (SD) maps together with a series of {\it Herschel}, {\it Spitzer}, and {\it WISE} archival data. We also developed a systematic data reduction framework allowing the massive data processing of ALMA observations. 
   }
   {We obtained independent continuum maps and spectral cubes for all our targets and molecular lines at different (SD and interferometric) resolutions and exploring multiple data combination techniques. 
   Based on our low-resolution (SD) observations (30\arcsec or $\sim$~12\ 000~au), we describe the global properties of our sample covering a wide range of physical conditions including low- (OMC-4 South, NGC~2023), intermediate (OMC-2, OMC-3, LDN~1641N), and high-mass (OMC-1, Flame Nebula) star-forming regions in different evolutionary stages. 
   The comparison between our SD maps and ancillary YSO catalogs denotes N$_2$H$^+$ (1-0) as the best proxy for the dense, star-forming gas in our targets showing a constant star formation efficiency and a fast time evolution of $\lesssim$~1~Myr.
   While apparently clumpy and filamentary in our SD data, all targets show a much more complex fibrous sub-structure at the enhanced resolution of our combined ALMA+IRAM-30m maps (4\farcs5 or $\sim$~2\ 000~au). A large number of filamentary features at sub-parsec scales are clearly recognized in the high-density gas ($\gtrsim 10^5$~cm$^{-3}$) traced by N$_2$H$^+$  (1-0) directly connected to the formation of individual protostars. Surprisingly, this complex gas organization appears to extend further into the more diffuse gas ($\sim10^3-10^4$~cm$^{-3}$) traced by HNC (1-0). 
   }
   {
   This paper presents the EMERGE Early ALMA survey including a first data release of continuum maps and spectral products for this project to be analysed in future papers of this series.
   A first look at these results illustrates the need of advanced data combination techniques between high-resolution interferometric (ALMA) and high-sensitivity, low-resolution single-dish (IRAM-30m) datasets to investigate the intrinsic multi-scale, gas structure of the ISM. 
   }



   \maketitle
%

\section{Introduction}

A classical observational dichotomy separates the origin of solar-like stars from their high-mass counterparts. Low-mass stars (few M$_\odot$; e.g., HL~Tau) can be found in isolation or small groups in quiescent, low-mass environments such as Taurus. In contrast, high-mass stars ($>$10~M$_\odot$; e.g., 
$\theta^1$ Ori C in the \object{ONC}) are typically found within clusters (which also form low-mass stars) and under the influence of intense stellar feedback (i.e. UV radiation, winds, outflows) in giant molecular clouds such as Orion \citep[e.g.,][]{Pabst2019}. 
These findings led to the proposal of independent formation scenarios for low- \citep[][]{Shu1987} and high-mass \citep[][]{Bonnell2001,McKee2003} stars. Based on similar (spherical) initial conditions (either a low-mass, dense core or a massive clump, respectively) these models have dominated the star-formation field during the last decades.

A series of observational studies have highlighted a direct link between the filamentary structure of the ISM and the initial conditions for star-formation \citep[see][for recent reviews]{Andre2014,Hacar2022,Pineda2023_PP7}.
According to large-scale {\it Herschel} FIR dust continuum emission surveys, most dense cores and young stellar objects (YSO) are preferentially formed in association to dense filaments in both nearby \citep{Andre2010} as well as in more distant Galactic Plane \citep{Molinari2010} molecular clouds. Low-mass cores are found along parsec-scale filaments \citep{Hartmann2001} and likely form out of their internal gravitational fragmentation \citep{Schneider1979,Inutsuka1997}. On the other hand, most of the dense clumps harbouring high-mass stars and clusters are found at the junction of massive filaments, forming the so-called Hub-Filament Systems \citep[HFS; ][]{Myers2009a,Kumar2020}. Filaments provide a preferential orientation with additional mass reservoir and could funnel large amounts of material towards cores and clumps \citep[e.g.,][]{Peretto2013}. Yet, the connection between the previous star formation models and the new filamentary conditions of the ISM remains under debate, particularly in the case of high-mass stars \citep[e.g.,][]{Motte2018}.

The analysis of the gas kinematics of many the above filamentary clouds has revealed a high level of gas organization prior to the formation of stars. Single-dish (low spatial resolution) observations demonstrate how many of the parsec-size filaments detected by {\it Herschel} in nearby low-mass clouds \citep[e.g., B213;][]{Palmeirim2013} are actually collections of velocity-coherent filaments at smaller scales identified by their internal velocity dispersion close to the sound speed, usually referred to as fibers \citep{Hacar2013,Andre2014,Pineda2023_PP7}. Sharing similar kinematic properties although smaller in size, further ALMA (high spatial resolution) observations resolved a similar fiber structure at sub-parsec scales in massive star-forming regions \citep{Hacar2018}.
Since their discovery, filamentary structures of different lengths  (observed to have between $\sim$0.1 and $\sim$7~pc) but similar sonic internal velocity dispersion than those identified in fibers have been systematically identified in all types of low-mass clouds \citep{Arzoumanian2013,Feher2016,Kainulainen2016,Hacar2016b}, clusters \citep{Fernandez-Lopez2014,Hacar2017b}, high-mass star-forming regions \citep{Trevino-Morales2019,Cao2021}, and Infrared Dark Clouds \citep[IRDCs; ][]{Henshaw2014,Chen2019}, to cite a few examples. 
While first identified from the gas velocity of low-density tracers and potentially affected by line-of-sight confusion \citep[see][]{Zamora2017,Clarke18}, the subsequent detection of fibers in high density tracers such as N$_2$H$^+$ \citep[e.g.,][]{Chen2019,Shimajiri2019}, NH$_3$ \citep{Monsch2018,Sokolov2019}, H$^{13}$CO$^+$ and NH$_2$D$^+$ \citep[e.g.,][]{Cao2021} guarantees a good correspondence with the true dense gas structure within clouds in large statistical samples.
Systematically produced in hydro- and turbulent simulations \citep{Smith2014a,Moeckel2015,Clarke2017,LiKlein2019}, these ``filaments-within-filaments'' appear as the first sub-sonic structures created at the end of the turbulent cascade as part of intrinsic hierarchical structure of gas in the ISM \citep[see][for a discussion]{Hacar2022}.

Fibers could potentially connect the formation of both low- and high-mass stars \citep[see][for a discussion]{Hacar2018}. Individual fibers promote the formation of small chains of low-mass, Jeans-like cores via quasi-static fragmentation \citep{Schneider1979,Hacar2011,Tafalla2015} which associated in large groups could become self-gravitating and agglomerate a high number of cores and stars \citep{Hacar2017a,Hacar2017b}. In parallel, interactions between fibers could also form super-Jeans cores via merging or collision \citep{Clarke2017,Hoemann2021}   
while, at the same time, provide large mass reservoirs favouring the enhanced accretion rates needed to form high-mass stars \citep{Bernasconi1996,Behrend2001}.
Massive fiber arrangements could therefore simultaneously generate stellar clusters and high-mass stars \citep{Hacar2018}. A similar scenario has been proposed from the merging of parsec-scale filaments \citep{Kumar2020,Kumar2022}.
Given these promising results, this new fiber scenario for star formation deserves further exploration. 

This paper introduces the {\it Emergence of high-mass stars in complex fiber networks} (EMERGE) project\footnote{EMERGE Project website: \url{https://emerge.univie.ac.at/}}.
As working hypothesis, EMERGE aims to investigate whether both high-mass stars and clusters could be created as 
emergent phenomena in densely populated fiber systems. 
Rather than by distinct processes, EMERGE will explore whether the formation of low- and high-mass stars together with clusters arises as a natural consequence of the global properties of these fiber systems showing a continuum spectrum of masses as function of the network density.

EMERGE will systematically analyze the substructure, internal interactions, and dynamical evolution of these filamentary systems across the Milky Way.
Typically located at kpc distances, resolution and sensitivity constraints have so far limited the analysis of these massive regions to individual targets or small samples. EMERGE will survey a large, homogeneous ALMA sample of massive fiber networks extracted from its public archive using a novel combination of data mining techniques \citep[][]{ALminer} and massive data processing \citep[][]{vanTerwisga2022}.
In this first work (hereafter Paper I) we present the EMERGE Early ALMA Survey (see Sect.\ref{sec:EMERGE_sample}) that includes 7 star-forming regions in Orion systematically analyzed using a combination of high spatial resolution resolution ALMA (interferometric) mosaics plus large-scale IRAM-30m (single-dish) observations at 3~mm (N$_2$H$^+$, HNC, HC$_3$N, HCN, and 3mm-continuum). 
Designed as a proof of concept of this project, Paper I introduces a standard set of procedures for data combination and massive data reduction.
In a series of accompanying papers we will investigate the effects of interferometric filtering on ISM observations carried out with ALMA \citep[][Paper II]{BonanomiPaperII} as well as present the analysis of the physical properties of dense fibers identified in this first EMERGE sample \citep[][Paper III]{SocciPaperIII}. Future EMERGE papers will extend this analysis to other regions.


\begin{figure*}[t]
\centering
\includegraphics[width=\textwidth]{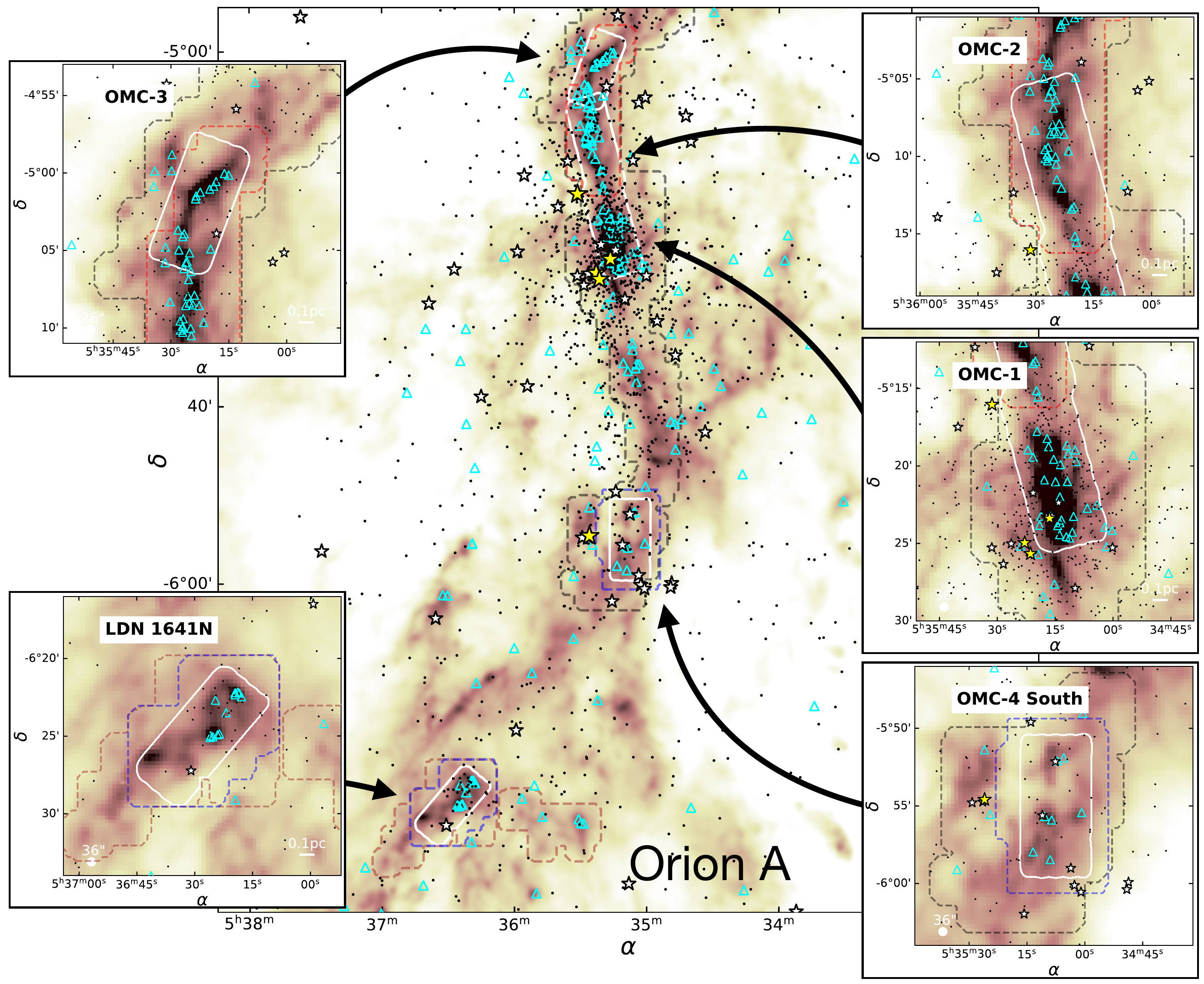}
      \caption{EMERGE Early Science survey in Orion A (central panel), including zoom-ins of the OMC-1, OMC-2, OMC-3, OMC-4 South, and LDN~1641N regions (see subpanels), shown over the total gas column density derived from Herschel observations \citep[background image at 36~arcsec resolution;][]{Lombardi2014}. The footprints of our ALMA mosaics (white lines) as well as our IRAM-30m observations (colored lines; see color-code in Table~\ref{table:molecules}) are indicated in all panels together with the positions of O (yellow stars) and B stars (white stars) at the distance of Orion. The positions of young protostars (cyan triangles) and disk stars (black dots) \citep[][]{Megeath2012} and the corresponding Herschel beamsize and 0.1~pc scale bar are marked in the individual zoom-ins.
              }
\label{fig:OriA_footprint}
\end{figure*}

\begin{figure*}[t]
\sidecaption
\includegraphics[width=0.7\textwidth]{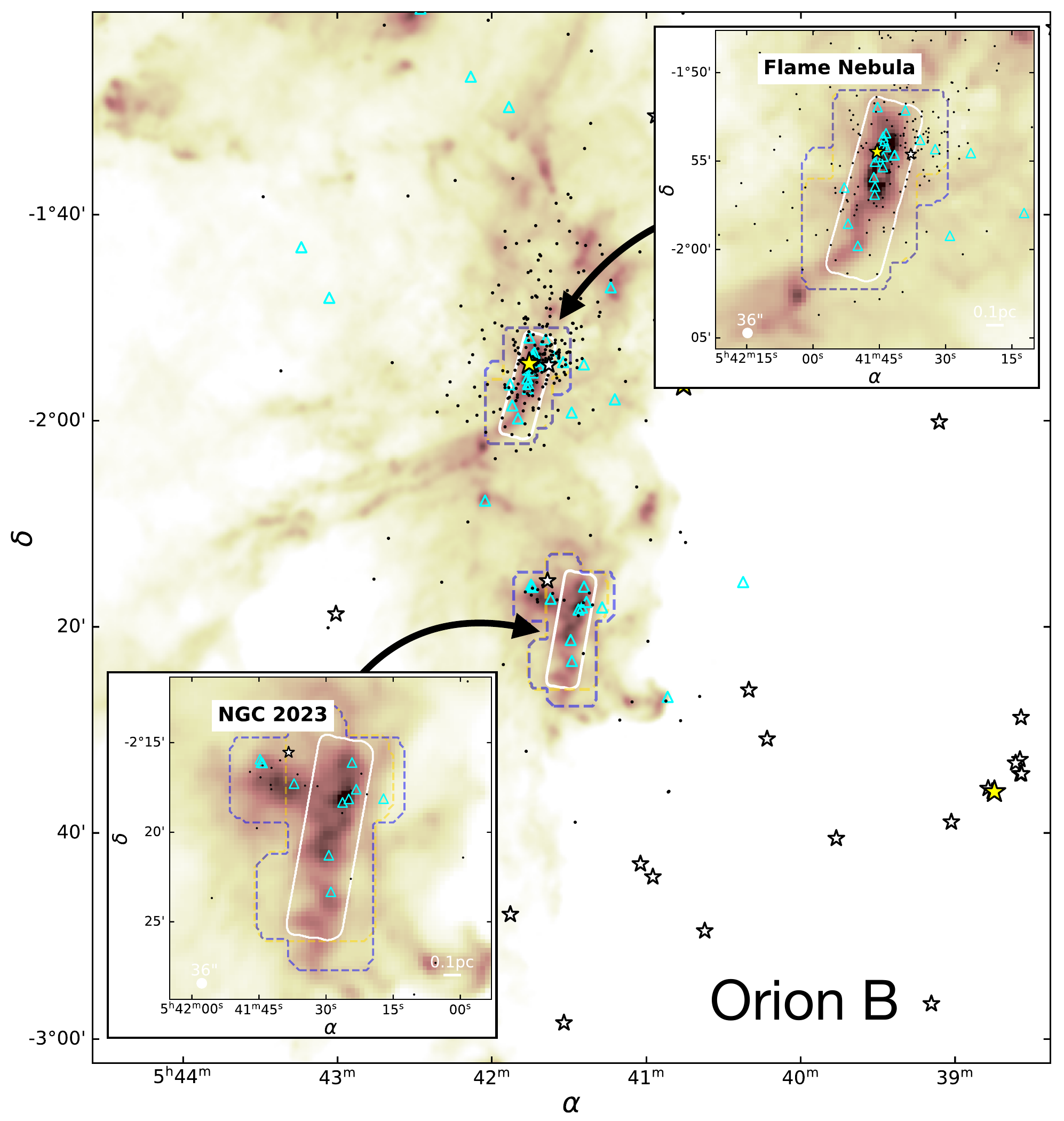}
      \caption{Similar to Fig.\ref{fig:OriA_footprint} but for the case of Orion B with zoom-ins of NGC~2023 and Flame Nebula (NGC~2024).
              }
\label{fig:OriB_footprint}
\end{figure*}

\begin{table*}
\caption{EMERGE Early ALMA Survey: Targets}             
\label{table:Targets}      
\centering          
\begin{tabular}{l c c c c c c c c c c}     
\hline\hline       
Target   & Region    & D$^{(1)}$     &   M$_{tot}$$^{(2)}$  & $\mathrm{M}_{dense}$$^{(3)}$    & $f_{dense}^{(4)}$ & $\left<\mathrm{T}_\mathrm{K}\right>$$^{(5)}$ & O-star? &  P$^{(6)}$ & P+D$^{(6)}$  & P/D \\ 
        &           & (pc)  & (M$_\odot$) & (M$_\odot$) &    & (K)        &    &      &     \\ 
\hline                    
OMC-1   & Orion A   & 400   & 1032$^{\star}$       & 582.2    & 0.56       & 33.7 & Yes           & 36   & 423  & 0.09 \\
OMC-2   & Orion A   & 400   & 581.0       & 372.8   & 0.64       & 26.3 & No         & 33   & 142  & 0.30 \\
OMC-3   & Orion A   & 400   & 643.2       & 351.1   & 0.55       & 25.4 & No         & 26   & 102  & 0.34 \\
OMC-4 South & Orion A   & 400   & 469.4      & 73.6 & 0.16       & 20.3 & ?         & 11    & 59  & 0.23  \\
LDN~1641N & Orion A & 400   & 478.6       & 174.3    & 0.36       & 25.4 & No          & 13    & 51  & 0.34  \\
\hline
NGC~2023 & Orion B & 423   &  330.6       & 70.7    & 0.21       & 21.9 & No          & 6     & 8  & 0.75 \\
Flame~Nebula  & Orion B & 423   & 287.8$^{\star}$     &  94.1 & 0.33       & 28.8 & Yes          & 21   & 141  & 0.18  \\
\hline                  
\end{tabular}
\tablefoot{Physical parameters derived from low-resolution (Herschel and/or IRAM-30m) maps within an area of 1.5$\times$1.5 pc$^2$ ($\sim$~700$\times$700 arcsec$^2$) around the phase center of our ALMA maps (see coordinates in Table~\ref{table:Fields}): 
(1) Adopted cloud distances; 
(2) Total mass, M(total), derived from Herschel column density maps \citep{Lombardi2014};
(3) Total dense gas mass, M(dense), derived from N$_2$H$^+$(1-0);
(4) Fraction of dense gas (i.e. $f=\mathrm{M}_{tot}/\mathrm{M}_{dense}$;
(5) Mean gas kinetic temperature, $\left<\mathrm{T}_\mathrm{K}\right>$;
(6) Number of Protostars (P) and Disk stars (D) identified by {\it Spitzer} \citep{Megeath2012} (see also Fig.~\ref{fig:ALL_YSOs}).
($\star$) Note how the mass contribution in stars could significantly increase the mass load of high-mass, clustered regions such as OMC-1 and Flame Nebula.
}
\end{table*}

\section{The EMERGE Early ALMA Survey}\label{sec:EMERGE_sample}

With our EMERGE Early ALMA Survey we aim to investigate the internal gas structure and physical conditions for star-formation in different environments of Orion \citep[D$\sim$400~pc; [; see also Sect.~\ref{sec:sample_prop}]{Menten2007,Grossschedl2018}.  
This first survey sampled a total of 7 prototypical star-forming regions: OMC-1, OMC-2, OMC-3, OMC-4 South, and LDN~1641N in Orion A, and NGC~2023 and the Flame Nebula (or NGC 2024) in Orion B (see Table~\ref{table:Targets}). We designed our ALMA survey to cover both low- and high-mass star-forming regions showing different levels of stellar activity, mass, and evolution. We show the location, column density maps, and stellar populations in these regions in Figures \ref{fig:OriA_footprint} and \ref{fig:OriB_footprint}.


Previous stellar surveys have classified these regions as function of their clustering properties and star-formation activity \citep[see][]{PetersonMegeath2008,Bally2008}. OMC-1 and Flame Nebula are identified as the two most massive embedded clusters in Orion showing largest peak stellar densities in this complex ($>$~1500~stars~pc$^{-2}$) including several high-mass stars. Although less concentrated, OMC-2, OMC-3, and LDN~1641~N contain large embedded populations of YSOs forming smaller groups. In contrast, OMC-4 South and NGC~2023 only form a hand full of low-mass sources mainly in isolation \citep[see][for a full comparison]{Megeath2015}.

Our source selection also covers a wide range of environments. OMC-1 and Flame Nebula are exposed exposed to intense  EUV ionizing radiation (producing HII regions) and to strong FUV dissociating radiation (creating neutral Photo Dissociated Regions, PDRs) produced by the O-type $\theta^1$~Ori C (O6Vp) and NGC~2024 IRS-2{\bf b} (O8V) stars, respectively \citep[see][]{Bally2008}. Although less severe, the stellar winds and UV radiation of nearby clusters and high-mass stars seem to influence the OMC-3 \citep[NGC~1977 and 42~Orionis; ][]{PetersonMegeath2008}, OMC-4 South \citep[NGC~1980 and $\iota$~Ori; ][]{Alves2012}, and NGC~2023 \citep[$\sigma$~Ori; ][]{Brown1994} regions as denoted in previous velocity-resolved observations of the FIR [CII]~158$\mu$m fine-structure line \citep[see][]{Pabst2017,Pabst2020}. Despite active in star formation as denoted by a large population of young outflows \citep[e.g.,][]{Stanke2007,Sato2023}, OMC-2 and LDN~1641N appear as more pristine regions and therefore become good candidates to also investigate unperturbed gas.

Most of our targets have been systematically surveyed at long wavelengths, particularly at low spatial resolution ($\theta \gtrsim$15"). 
The Orion complex has been mapped using large-scale FIR \citep[36";][]{Lombardi2014} plus (sub-) millimeter (sub-mm) \citep[14";][]{Stanke2022} continuum surveys as well as in some CO transitions \citep[11"-14";][]{Berne2010,Goicoechea2020,Stanke2022}.
In the Orion A cloud, the OMC clouds (OMC-1/2/3/4 South; see Fig.\ref{fig:OriA_footprint}) are part of the famous Integral Shaped Filament \citep[ISF; ][]{Bally1987}, the most massive filamentary cloud in the solar neighbourhood, and the one containing the most massive cluster within the local kpc, the Orion Nebula Cluster \citep[see][for a review]{PetersonMegeath2008}.
The ISF has been mapped at sub-mm wavelengths in continuum \citep[14";][]{Johnstone1999}, by molecular surveys \citep[$\sim$30";][]{Kauffmann2017,Brinkmann2020} and dedicated observations of both diffuse \citep{Shimajiri2014,Stanke2022} and dense molecular tracers \citep[$\sim$30";][]{Tatematsu2008,Hacar2017a}, to cite few examples \citep[for more information see][and references therein]{PetersonMegeath2008,Hacar2018}. Away from this main star-forming site, LDN~1641N has received less attention, although it has been covered in some large-scale surveys in Orion A \citep[$\gtrsim$15"; e.g.,][]{Nishimura2015,Mairs2016}. NGC~2023 and Flame Nebula have been observed as part of cloud-size studies in Orion B (see Fig.\ref{fig:OriB_footprint}) both in single-line observations \citep[$\geq$19";][]{Lada1991,Stanke2022} and unbiased molecular surveys \citep[$\gtrsim$22";][]{Pety2017,Orkisz2019,Santamaria2023}. 

The clouds in our sample have been classified as filamentary in previous low spatial resolution studies \citep[e.g.,][]{Johnstone1999,Orkisz2019,Konyves2020,Gaudel2023}.
Nonetheless, a close inspection of Figs.\ref{fig:OriA_footprint} and \ref{fig:OriB_footprint} reveals a variety of cloud morphology. Regions such as OMC-2, OMC-3, and Flame Nebula appear elongated, following the large-scale filamentary structure of their parental cloud \citep{Johnstone1999}. The OMC-1 region is identified as a Hub-Filament System (HFS) structure characteristic of massive clumps \citep{Rodriguez-Franco1992}. On the other hand, OMC-4 South and NGC~2023 show more irregular and diffuse organization, while LDN~1641N exhibits a cometary shape \citep{Mairs2016,Kirk2016}. 

Compared to the large number of low spatial resolution observations, the study of these targets at higher spatial resolution ($\theta\leq$10") is more limited. Only those targets in Orion A (OMC-1/2/3/4 South, and LDN~1641N) have been systematically observed using large-scale CO maps \citep[$\sim$7";][]{Kong2018,Suri2019} with no similar counterpart study of Orion B. Local surveys mainly focused the northern part of the ISF region (OMC-1/2/3) mapped in FIR \citep[$\sim$8"; ][]{Schuller2021} or millimeter \citep[$\sim$4\farcs5; ][]{Takahashi2013} continuum. On the other hand interferometric observations of high density tracers ($>10^4$~cm$^{-3}$) have been restricted to regions of specific interest \citep[$\sim$4"; e.g., OMC-1 by ][]{Wiseman1998,Hacar2018}, sections of these regions ($\sim$3-5"; e.g., parts of OMC-2 by \citet{Kainulainen2017} or Flame Nebula by \citet{Shimajiri2023}), and/or peculiar targets within these fields ($\sim$1-2"; e.g., the head of LDN~1641N by \citet{Stanke2007}, OMC-2 FIR-4 by \citet{Chahine2022}).

We surveyed our EMERGE Early ALMA sample with a homogeneous set of high spatial resolution, high sensitivity ALMA observations (4\farcs5) in combination with additional large-scale, low spatial resolution, single-dish (30") maps (Sect.\ref{sec:Observations}) for which we also created a new data reduction framework (Sect.\ref{sec:obs_datareduction}). 
We investigated the gas properties in these regions (star-formation, column density, temperature, etc) using a dedicated suite of molecular tracers (N$_2$H$^+$, HNC, and HC$_3$N) and continuum maps (Sect.~\ref{sec:sample_prop}). By selecting targets within the same star-forming complex we carry our a direct comparison of these regions at similar resolution while avoiding additional variations depending on their local galactic environment. As primary goal of our study, we aim to systematically investigate the physical conditions in both low- and high-mass star-forming regions from pc-scales down to 2000~au resolutions (see also Papers II and III).

\section{Observations}\label{sec:Observations}

\begin{table*}
\caption{EMERGE Early ALMA Survey: ALMA Fields}             
\label{table:Fields}      
\centering          
\begin{tabular}{l c c c c l l l}     
\hline\hline       
Field & \multicolumn{2}{c}{Phase center} & Map size & No. of & Spectral Windows & Proj. ID\\ 
  & $\alpha$(J2000) & $\delta$(J2000) &  (arcsec$^2$) & pointings & & \\ 
\hline                    
   OMC-1$^{\#}$        & 05$^h$35$^m$27$^s$.2 & -05$^\circ$09$'$42$"$    & $220"~\times~600"$ & 148 & N$_2$H$^+$, 3mm-Cont. & 2015.1.00669.S \\
   OMC-2$^{\#}$        & 05$^h$35$^m$14$^s$.2 & -05$^\circ$22$'$21$"$    & $220"~\times~600"$ & 148 & N$_2$H$^+$, 3mm-Cont. & 2015.1.00669.S\\
\hline
   OMC-3        & 05$^h$35$^m$22$^s$.5 & -05$^\circ$02$'$00$"$    & $210"~\times~500"$ & 124 & N$_2$H$^+$, HNC, HC$_3$N, 3mm-Cont. & 2019.1.00641.S\\
   OMC-4 South  & 05$^h$35$^m$07$^s$.5 & -05$^\circ$55$'$00$"$    & $230"~\times~540"$ & 149 & N$_2$H$^+$, HNC, HC$_3$N, 3mm-Cont. & 2019.1.00641.S\\
   LDN~1641N     & 05$^h$36$^m$27$^s$.0 & -06$^\circ$25$'$00$"$    & $210"~\times~540"$ & 132 & N$_2$H$^+$, HNC, HC$_3$N, 3mm-Cont. & 2019.1.00641.S\\
   NGC~2023     & 05$^h$41$^m$29$^s$.0 & -02$^\circ$20$'$20$"$    & $150"~\times~650"$ & 123 & N$_2$H$^+$, HNC, HC$_3$N, 3mm-Cont. & 2019.1.00641.S\\
   Flame~Nebula & 05$^h$41$^m$46$^s$.0 & -01$^\circ$56$'$37$"$    & $150"~\times~600"$ & 111 & N$_2$H$^+$, HNC, HC$_3$N, 3mm-Cont. & 2019.1.00641.S\\

\hline                  
\end{tabular}
\tablefoot{(\#) Observations presented in \citet{Hacar2018}.}
\end{table*}

\begin{table*}
\caption{EMERGE Early ALMA Survey: observational setup}             
\label{table:molecules}      
\centering          
\begin{tabular}{l c c c c c c c c c}     
\hline\hline       
Species    & Frequency     & E$_u^{(1)}$ &  n$_{eff}^{(2)}$ & \multicolumn{2}{c}{ALMA}  & \multicolumn{2}{c}{IRAM-30m}$^{(3)}$ & \multicolumn{2}{c}{ALMA+IRAM}\\ 
                       &               &       &               & Proj.ID         & $\delta v$      & Proj.ID          & $\delta v$ & ID & Resol. \\ 
                       & (GHz)         &  (K)  &    (cm$^{-3}$) &               &   (km~s$^{-1}$)       &  &  (km~s$^{-1}$) &  & (km~s$^{-1}$) \\ 
\hline                 
   N$_2$H$^+$ (J=1-0)      & 93.173764 & 4.47 & 1.0$\times 10^{4}$  & 2015.1.00669.S & 0.098 & 032-13 &     0.06 & narrow &  0.10\\
                            &           &      &                       & 2019.1.00641.S &    0.114 & 034-16  & 0.06 & narrow &0.12 \\
                                          &   &   &    &   &   & 120-22   & 0.06 & narrow & 0.12\\
   HNC (J=1-0)      & 90.663564 & 4.53 & 3.7$\times 10^{3}$  & 2019.1.00641.S &   0.233 & 032-13  & 0.66 & broad &0.66\\
                          &   &   &    &   &   & 034-16   & 0.66 & broad & 0.66\\
              &   &   &    &   &   & 060-22   & 0.06 & narrow & 0.25\\
               &   &   &    &   &   & 133-22   & 0.06 & narrow & 0.25\\
   HC$_3$N (J=10-9)     & 90.979023 & 24.1 & 4.3$\times 10^{5}$  & 2019.1.00641.S & 0.233 & 032-13    & 0.66 & broad & 0.66 \\
                             &   &   &    &   &   & 034-16   & 0.66 & broad & 0.66\\
                            &   &   &    &   &   & 060-22   & 0.16 & narrow & 0.25\\
                            &   &   &    &   &   & 133-22   & 0.16 & narrow & 0.25 \\
   HCN (J=1-0)      & 88.631602 & 4.25 & 8.4$\times 10^{3}$  & ---$^{(\star)}$ &  ---$^{(\star)}$ & 032-13   & 0.66 & broad & ---\\
                             &   &   &    &   &   & 034-16   & 0.66 & broad & ---\\
              &   &   &    &   &   & 133-22   & 0.66 & broad & --- \\
   3mm-Cont.        & 93, 104 & ---  & ---    & 2015.1.00669.S & 6.0 \& 5.4 & ---$^{(\star)}$  & ---$^{(\star)}$ & --- & Cont.$^{(\star\star)}$ \\
   3mm-Cont.        & 93.2, 91.2 & ---  & ---    & 2019.1.00641.S & 3.71 & ---$^{(\star)}$ & ---$^{(\star)}$ & --- & Cont.$^{(\star\star)}$ \\
\hline                  
\end{tabular}
\tablefoot{(1) Values for the upper energy levels (E$_u$), without hyperfine structure, are taken from the Leiden Atomic and Molecular Database \citep[LAMDA;][]{LeidenLAMDA_2005,LeidenLAMDA_2020}. 
(2) Effective excitation density at 10~K \citep{Shirley2015}.
(3) The footprints of the different IRAM-30m data are color-coded in Figs.\ref{fig:OriA_footprint} and \ref{fig:OriB_footprint}: 032-12 (black), 034-16 (brown), 120-20 (gold), 133-22 (blue), and 060-22 (red). 
($\star$) No data available. ($\star\star$) ALMA-alone maps. 

}
\end{table*}

\subsection{Interferometric ALMA observations}\label{sec:obs_ALMA}

We surveyed the molecular and continuum emission of the OMC-3, OMC-4 South, LDN~1641, NGC~2023 and Flame Nebula regions using the Atacama Large Millimeter Array (ALMA) (proj. ID: 2019.1.00641.S; PI: Hacar) in Chajnantor (Chile) during ALMA Cycle-7. We summarize the main parameters of our ALMA observations in Table~\ref{table:Fields}. We used independent large-scale ALMA mosaics, containing between 111 and 149 pointings each, with a typical map sizes of $\sim 200\times 600$~arcsec$^2$ (or $\sim 0.4 \times 1.2$~pc$^2$ at the distance of Orion). The footprints of the individual ALMA fields are shown in Fig.\ref{fig:OriA_footprint} and \ref{fig:OriB_footprint} (white contours) superposed to the Herschel total gas column density maps of these regions \citep[background image;][]{Lombardi2014}. Our choice for the phase center, size, and orientation of our maps follows the distribution of the N$_2$H$^+$~(1-0) integrated emission reported in previous single-dish surveys of the Orion A \citep{Hacar2017a} and Orion B \citep{Pety2017} clouds.

We mapped the molecular and continuum emission of all of our targets with the ALMA-12m array (only) in its most compact configuration (C43-1), and with baselines between 15 and $\sim$~310 meters, achieving a native resolution of $\theta\sim$~3.5~arcsec. All fields, each observed as individual Scheduling Blocks (SB), were covered at least twice and observed for a total of 2-3 hours per field with a minimum of 45 antennas under average weather conditions (Precipitable Water Vapor PWV=2.5-5~mm). Following standard procedures pointing, bandpass, and flux calibrations were carried out at the beginning of each observing run which phase calibration were obtained every 10-15 minutes using different sources. 

Our observations were carried out in Band 3 using a single spectral setup combining three narrow line plus two broad continuum spectral windows (SPW) simultaneously. We include a detailed description of the different observational setup in Table~\ref{table:molecules}. 
In our new ALMA observations we used a narrow SPW to map the N$_2$H$^+$ (1-0) (93.17 GHz) emission at high spectral resolution ($\delta v=$~0.114 km~s$^{-1}$). We set independent SPWs for both HNC (1-0) (90.66 GHz) and HC$_3$N (10-9) (90.97 GHz) lines at intermediate spectral resolution ($\delta v=$~0.233 km~s$^{-1}$). Moreover, we set two additional broad band, 1.875 GHz wide SPWs to map the continuum emission at 93.2 and 91.2 GHz.

To complete our survey we added similar ALMA Cycle-3 mosaics of the OMC-1 and OMC-2 regions in Band 3 (proj. ID: 2015.1.00669.S; PI: Hacar). The description of these previous datasets are presented in \citet{Hacar2018} (see also Tables~\ref{table:Fields} and \ref{table:molecules}). Independent N$_2$H$^+$ (1-0) maps are carried out along OMC-1 and OMC-2 at high spectral resolutions ($\delta v=$~0.098 km~s$^{-1}$) together with two 1.875~GHz wide, broad continuum bands at 93 and 104~GHz, respectively \citep[see also][]{vanTerwisga2019}. Unfortunately no HNC (1-0) nor HC$_3$N (10-9) observations were possible at the time due to the limitations on the data rates during Cycle-3. No ALMA map is therefore available for these transitions in OMC-1 or OMC-2. 

When needed (e.g., during data combination), we assumed standard flux conversions, i.e., $\left(\frac{\mathrm{T}_\mathrm{mb}}{1~\mathrm{K}}\right) = \left(\frac{S_\nu}{1~\mathrm{Jy}}\right) \left[13.6 \left(\frac{300~\mathrm{GHz}}{\nu}\right)^2 \left(\frac{1~\mathrm{arcsec}}{\theta}\right)^2 \right]$ \citep[e.g., 6.96 and 0.16 K~Jy$^{-1}$ for beams of 4\farcs5 and 30\arcsec, respectively; see ALMA technical handbook,][]{ALMAHandbook}. The reduction of our ALMA observations, including data combination with additional single-dish observations (Sect.~\ref{sec:obs_30m}), is discussed in Sect.~\ref{sec:obs_datareduction}.

\subsection{Single-dish IRAM-30m observations}\label{sec:obs_30m}

We complement our ALMA survey with additional single-dish observations at 3~mm carried out at the 30-meter Intitute Radioastronomie Millimetric telescope (IRAM-30m) in Granada (Spain) in multiple observing campaigns between 2013 and 2023. We mapped the molecular emission of all our ALMA targets using independent, large-scale IRAM-30m mosaics all observed with the Eight Mixer Receiver (EMIR) receiver connected to both the VErsatile SPectrometer Array (VESPA) and Fast Fourier Transform Spectrometer (FTS) backends. We summarize these observations in Table~\ref{table:molecules}.
We observed all our fields with two independent spectral setups. In a first set of observations (proj. IDs: 032-13, 034-16, 120-20) we obtained large-scale maps of N$_2$H$^+$ (1-0) (93.17 GHz) using VESPA at high-spectral resolution ($\delta v=$~0.06 km~s$^{-1}$). Simultaneously, we mapped the emission of HNC (1-0) (90.66 GHz), HCN (1-0) (88.63 GHz) and HC$_3$N (10-9) (90.97 GHz), among other lines, using multiple FTS units in broad configuration at low spectral resolution ($\delta v=$~0.66 km~s$^{-1}$). In a second set of maps (proj. IDs: 060-22, 133-22) we then focused on the local HNC (1-0) (90.66 GHz) and HC$_3$N (10-9) (90.97 GHz) emission and reobserved our ALMA fields using smaller mosaics at higher spectral resolution with VESPA ($\delta v=$~0.06 km~s$^{-1}$) and FTS in narrow configuration ($\delta v=$~0.16 km~s$^{-1}$). 

Each large single-dish mosaic is constructed by several individual tiles, typically of sizes of $200\times 200$~arcsec$^2$, observed multiple times in orthogonal directions using On-The-Fly (OTF), Position Switching (PSw) observations. The footprints of our low- and high-resolution IRAM-30m maps is again shown in Figs. \ref{fig:OriA_footprint} and \ref{fig:OriB_footprint}. We carried out our data reduction using the GILDAS-CLASS software \citep{GILDASoriginal,GILDAS} using standard procedures for OTF observations. We reduced each tile independently and applied facility provided efficiencies to calibrate our spectra into main beam temperature scale (T$_\mathrm{mb}$).  For each of our target lines we combine our calibrated spectra, observed at a native resolution $\sim$~27~arcsec, into a regular grid and convolved them into a final resolution of $\theta=$~30~arcsec in order to obtain individual Nyquist-sampled cubes. 
As final data products of our single-dish observations we obtained individual N$_2$H$^+$ (1-0), HNC (1-0), HCN (1-0), and HC$_3$N (10-9) datacubes (including their entire hyperfine structure) for each of the sources in our survey (OMC-1/2/3/4 South, LDN~1641N, NGC~2023, and Flame~Nebula). In the case of HNC and HC$_3$N, we reduced these cubes at low and high spectral resolution. 

Following \citet{Hacar2020}, we combined our low spectral resolution (broad) HCN (1-0) and HNC (1-0) observations to investigate the gas temperature of the gas in our survey. We obtained individual integrated intensity maps of each of these tracers (I(HCN) and I(HNC)) adjusted to the velocity range of each target see in HCN.
Both HCN and HNC present bright emission usually following the column density distribution of our clouds and extend beyond the coverage of our maps \citep[see][]{Pety2017,Hacar2020,Tafalla2023}.
Yet, we consider only those positions with I(HNC)~$\geq$~2~K~km~s$^{-1}$ and then convert the resulting I(HCN)/I(HNC) intensity ratios into the corresponding gas kinetic temperature $T_\mathrm{K}$(HCN/HNC) per pixel using the empirical correlations defined by \citet{Hacar2020}. The I(HCN)/I(HNC) intensity ratios provides a good proxy of the gas kinetic temperature at column densities N(H$_2$)~$>10^{22}$~cm$^{-2}$ and within an optimal temperature range between $\sim$15 and 40~K, with only minor contamination in local regions affected by outflows (e.g., Ori BN/KL in OMC-1). Using this method we then obtain large-scale, kinetic temperature $T_\mathrm{K}$(HCN/HNC) maps (hereafter $T_\mathrm{K}$ maps) for all the sources in our sample (see Sect.\ref{sec:EMERGE_globalprop}).

\subsection{Complementary IR data}

We aim to investigate the connection between stars and gas in our different ALMA fields.
In addition to our millimeter observations, our project benefits from a wealth of ancillary, public surveys in Orion at multiple wavelengths. In particular, we used the total H$_2$ column density maps obtained by \citet{Lombardi2014} from the FIR {\it Herschel} observations of the entire Orion A and B clouds. We converted these maps, originally measuring the dust opacity at 350~$\mu$m ($\tau_{350\mu\mathrm{m}}$), first into equivalent K-band extinction values (A$_K$) using the conversion values provided by Lombardi et al, and later into total H$_2$ column density maps assuming a standard reddening law \citep[A$_K$/A$_V$=0.112;][]{Rieke1985} and dust-to-gas conversion factor \citep[N(H$_2$)/A$_V = 0.93\times 10^{21}$~cm$^{-2}$~mag$^{-1}$;][]{Bohlin1978}. We therefore obtained total H$_2$ column density maps (N(H$_2$)) of all our sources at 36" resolution (Sect.\ref{sec:EMERGE_globalprop}). 

For illustrative purposes we also make use of archival 3.4~$\mu m$ emission images obtained by the Wide-field Infrared Survey Explorer ({\it WISE}) \citep{WISEtelescope2010}. Together with the detection of most of the young stars in our targets, the {\it WISE} 3.4~$\mu m$ band includes prominent Polycyclic Aromatic Hydrocarbons (PAH) features, such as the C-H stretching band of FUV-pumped PAH \citep[e.g.,][]{Chown2023}, highlighting the effects of external FUV radiation onto the dust grains. Compared to the equivalent 12~$\mu m$ images, the  {\it WISE} 3.4~$\mu m$ band provides a wide dynamic range in intensity without undesired saturation effects in bright regions such as OMC-1. These {\it WISE} images offer a complementary view compared to the molecular gas traced by our (sub-)millimeter maps. 

The YSOs in both Orion A and B clouds have been systematically surveyed by \citet{Megeath2012} using {\it Spitzer} observations. From the analysis of their mid-IR colours, Megeath et al classified these sources as protostars (P) or pre-main sequence stars with disks (D), typically corresponding to Class 0/I and Class II objects \citep{Greene1994}, respectively. We adopt the \citet{Megeath2012} catalog as reference for our study. We refined this catalog by including additional young objects (Protostars and Flat spectrum, all included as P objects in our catalogue) reported in follow-up {\it Herschel} observations \citep{Furlan2016,Stutz2013}. Overall, we expect a high homogeneity and completeness of this catalogue in describing the low-mass, YSO population within our ALMA fields with perhaps the exception of the OMC-1 and Flame Nebula due to their bright nebulae and crowded populations \citep[see also][for a detailed discussion]{Megeath2015,Grossschedl2019}. The lack of large-scale, systematic surveys of Class III objects in Orion B (e.g. in X-rays) limits the extension of this analysis to older populations. In a separate catalog for massive stars, we collected sources in Orion classified as spectral types O and B using Simbad \citep{SIMBAD}.

\section{Massive data reduction of ALMA observations}\label{sec:obs_datareduction}

\begin{figure*}[t]
\centering
\includegraphics[width=1.\textwidth]{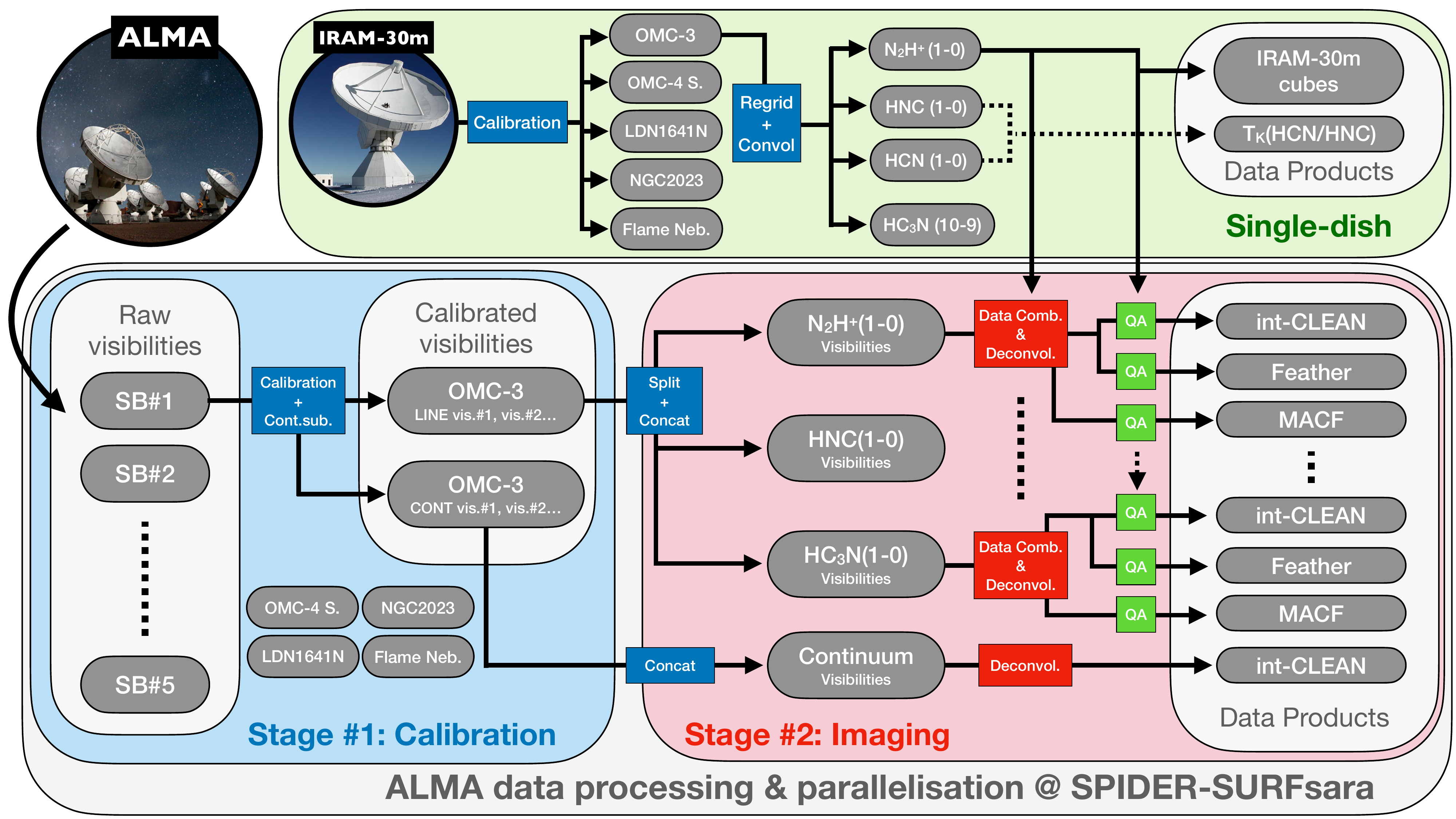}
      \caption{EMERGE Early Science data reduction workflow for our single-dish, IRAM-30m  {\bf (Upper panel)} and interferometric ALMA {\bf (Lower panel)} observations. In the case of our ALMA reductions, both calibration (stage \#1; light blue) and imaging (stage \#2; light red) processes are carried out at the SPIDER-SURFsara supercomputer using parallel processing.
              }
\label{fig:workflow}
\end{figure*}

The nature and volume of our EMERGE Early ALMA Survey, including mosaics of a large number of target areas (5+2) combining different molecular (3) and continuum (1) maps, sometimes in multiple spectral configurations (2), requires the development of automatic data reduction techniques. The large number of potential data products plus the use of our single-dish data using different data combination techniques (see below) make the manual and customized reduction of each target not only highly inefficient but also costly in time. The optimization of this process and the production of homogeneous datasets is of paramount interest for our study.

In this section we introduce a data reduction framework where the standardization of the calibration and imaging steps allows the full automation of the ALMA data reduction process shortening the effective processing time by more than an order of of magnitude.
First,  we detail the description of the calibration and imaging steps for individual observing setups in Sect.~\ref{sec:ALMA_calibration}. Second, we describe how these standardized processes can be easily parallelized using a supercomputer in Sect.~\ref{sec:ALMA_supercomputer}. Third, we quantify the quality of our reductions using advance quality assessments in Sect.~\ref{sec:ALMA_dataquality}.
Finally, we present the resulting molecular line cubes and continuum maps, including when possible combined ALMA+IRAM-30m observations, as final data products of our survey in Sect.~\ref{sec:ALMA_dataproducts}. 

\subsection{Calibration, imaging, and data combination}\label{sec:ALMA_calibration}

We present the workflow of our data reduction scheme in Figure~\ref{fig:workflow} (lower panel). We exemplify this reduction process for the data obtained in the OMC-3 region, from the original ALMA Scheduling Block (SB) containing raw visibilities including continuum and lines to the final data products in FITS format in both interferometric-alone and combined maps and cubes. 
Overall, the full data reduction process is divided in two stages: (1) the calibration of the raw visibilities, and (2) the imaging (or deconvolution) process. 
We carried out both calibration and imaging steps using the Common Astronomy Software Applications (CASA) package \citep{casa2022,CASAsoftware2022_2}. We discuss each of these steps independently below.

As part of stage \#1, we applied standard calibration to the raw visibilities in order to prepare our data for imaging (Fig.~\ref{fig:workflow}, bottom panel, left part). Different executions of the same SB lead into different Measurement Sets (MS) including multiple SPWs (lines + continuum). We recreated the ALMA pipeline calibration of each MS file applying facility provided scripts ({\it scriptforPI}) using CASA versions 4.7.2 or 5.4.0 depending on the target. We then applied a continuum subtraction for all science SPWs in each calibrated MS file independently and created individual files for both continuum and continuum-subtracted line visibilities. 

In stage \#2, we imaged each line and continuum mosaic independently (Fig.~\ref{fig:workflow}, bottom panel, right part). We first split (only for line SPWs) and then concatenated the previous calibrated visibilities producing single visibility files for each molecular species and continuum dataset per field. Each ALMA mosaic was then imaged using the CLEAN deconvolution algorithm \citep{CLEAN_hogbom1974,CLEAN_clark1980} using the task \texttt{clean} in CASA version 5.7\footnote{We note that the ALMA task \texttt{tclean} showed unstable results when using a \texttt{startmodel} option during data combination in CASA version 5.7. These issues are not present in \texttt{clean} which motivated our preference for this latter task. Newer versions of CASA seem to have fixed this problem and \texttt{tclean} will be used in future EMERGE data reductions. }. 
When possible, we applied data combination to include our IRAM-30m data (Sect.~\ref{sec:obs_ALMA}) as short-spacing information (see below). To facilitate its implementation, we homogenized most of the \texttt{clean} parameters in our reductions. Our line and continuum mosaics are deconvolved adopting the phase centers indicated in Table~\ref{table:Fields} with a \texttt{mapsize} of 1600"$\times$1600" with a  \texttt{cellsize} of 0\farcs45, and reduced using standard \texttt{clark} plus \texttt{briggs} weighting schemes with \texttt{robust} parameter equal to 0.5 using 5$\cdot 10^4$ \texttt{iterations}. 
As main differences, we set thresholds of 1~mJy for maps in continuum reduced in \texttt{msf} mode and of 10~mJy per channel for our line cubes obtained in \texttt{velocity} mode. In the case of the line cubes, we also optimized the spectral resolution (see last column in Table~\ref{table:molecules}) and number of channels for deconvolution depending on the species and resolution (narrow and broad) of our interferometric and single-dish data (see Table~\ref{table:molecules}).

\citet{Plunkett2023} recently highlighted the role of data combination in the analysis of extended sources such as our ALMA targets. Due to the so-called {\it short-spacing problem} \citep[][]{Wilner1994}, the lack of sensitivity at short-baselines leads to scale-dependent filtering effects in both continuum and line observations \citep[e.g.,][]{Pety2013}. As shown in previous works \citep[e.g.,][]{Leroy2021_PHANGS}, these issues can critically affect the fidelity of interferometric observations.
We explored the effects of data combination using three independent reductions for each of our molecular line mosaics: int-CLEAN, Feather, and MACF. A detailed description of these data combination methods and their implementation can be found in \citet{Plunkett2023}.

In a first and basic reduction, we obtained an interferometric-only map by deconvolving the ALMA-alone visibilities (int-CLEAN). Second, we use the task \texttt{feather} \citep{cotton2017} as standard data combination CASA method to combine our previous int-CLEAN maps with our IRAM-30m observations (Feather). Third, and finally, we applied a Model Assisted Cleaning and Feather combination (MACF)  which introduces the single-dish information both as initial cleaning model and feathering image. 
Contrary to the line observations, no single-dish information is available in continuum. As a result, we were able to produce only interferometric-alone (int-CLEAN) continuum maps for our sources. 
Additional data combination techniques \citep[see][for a review]{Plunkett2023} will be explored in future papers of this series.

The deconvolution of our ALMA observations produces images with typical beamsizes around $\theta_0\sim$~3\farcs5. We applied a primary beam (PB) correction to all our cleaned images and masked all emission channels (voxels) with PB values lower than 0.5.
To increase the stability of our maps, we also applied a small convolution after deconvolution (and before feathering) and resampled the resulting cubes every half beam, leading into Nyquist-sampled maps (i.e. $\theta/2$) and cubes with a final resolution of $\theta=$~4\farcs5.

Our specific choices for the above stage \#2 parameters aim to define single parameter values that can be homogeneously applied to our entire sample.
Spatial paramaters such as the \texttt{mapsize} are automatically defined to cover the largest dimensions of our dedicated IRAM-30m data around the corresponding ALMA fields. By design, we also adopt a single \texttt{cellsize} value of $\theta/10$ in order to correctly oversample the native beamsize $\theta_0$ (i.e. \texttt{cellsize}$< \theta_0 /5$) while at the same time facilitate the Nyquist resampling of our data products at their final resolution (final pixel size = 5$\times$ original \texttt{cellsize}).
Other cleaning parameters are chosen as a compromise between the maximization of the quality of our maps (see Quality Assessments in Sect.~\ref{sec:ALMA_dataquality}) and the simultaneously minimization of the total computing time per field.
Additional tests using larger number of iterations or lower thresholds produce no significant differences on the final data products while have a direct impact on the deconvolution timescales when applied to the large number of targets and lines included in our sample (76 individual reductions; see below). Dedicated reductions could potentially produce better results for individual observations at the expense of larger computing times and human intervention per field. In contrast, our EMERGE Early ALMA Survey leverages on a large statistical sample of targets and lines homogeneously reduced in short timescales (see Sect.\ref{sec:ALMA_supercomputer}). Similar optimization strategies appear of critical importance for future ALMA archival studies.

\subsection{Parallelization on a high-performance computer}\label{sec:ALMA_supercomputer}

The standardization of the calibration and imaging steps (Sect.~\ref{sec:ALMA_calibration}) makes our data reduction process easy to parallelize into separate CASA jobs. As shown in Fig.~\ref{fig:workflow}, we also take advantage of the original ALMA data structure. Each of our ALMA fields is stored in a single SB that can be recreated separately. After an early split, the imaging process of the resulting continuum and line calibrated visibilities can be then run independently in an automatic way. Also, each of these modular steps can be carried out using dedicated CASA versions. 

We executed parallel runs of both calibration and imaging steps for our entire ALMA sample in the SPIDER Data Processing Platform\footnote{\url{https://spiderdocs.readthedocs.io/en/latest/}} part of the SURF\footnote{\url{https://www.surf.nl/}} Data Processing facilities in Amsterdam\footnote{The EMERGE Early Science Survey was executed as part of the projects “Advance ALMA data reduction using SPIDER” (PI: A. Hacar) and "Advanced ALMA data reduction" (PI: A. Ahmadi) carried out at SURF-SPIDER in collaboration with the Allegro Dutch ARC node in Leiden.}.
The high performance and flexibility of SURF-SPIDER permits to run a large number of CASA instances in parallel. First, we simultaneously recreate all our calibrated visibilities at once (stage \#1). Second, we carry out all individual line and continuum deconvolutions (including data combination) simultaneously (stage \#2). Each CASA run is assigned to 5 CPUs providing a total memory of 40~GB per job to optimize the CPU and queuing times. Additional tests indicate no performance improvement when scaling our reductions neither with a larger number of cores\footnote{We remark here that the use of a higher number of cores (CPUs) indeed reduces the execution time per reduction although this does not significantly improve the overall performance of the sample processing. Our tests indicate that the reduction process inversely scales with the number of cores. However, this performance scaling becomes sub-linear with more than 5 cores and usually saturates at $\sim$10 cores. The request of $>5$ cores makes the reduction process not only more expensive in computational resources but also more difficult to schedule by the queuing system in large numbers. Our choice is therefore justified in order to optimize the use of computational resources while reducing the execution time (queuing + reduction) for the entire sample.}
nor with  CASA executions in parallel (\texttt{mpicasa}) mode in the SURF-SPIDER architecture. This approach is similar to the parallel reduction method introduced by \citet{vanTerwisga2022} for the massive data processing of ALMA single-pointing, continuum maps, this time extended into mosaics and spectral cubes. 

The efficiency of our data reduction process strongly benefits from the use of high-performance computers such as SURF-SPIDER. Our full ALMA survey consists of a total of (5 targets $\times$ [(Cont. + N$_2$H$^+$ + HNC + HC$_3$N) $\times$ 1~int-CLEAN +  [(1$\times$ N$_2$H$^+$ + 2$\times$ HNC + 2$\times$ HC$_3$N cubes)$\times$ 2~datacomb.] + 2 targets $\times$ [(Cont. + N$_2$H$^+$) $\times$ (1~int-CLEAN + 2~datacombs)] ) = 76 individual data products. The processing time per individual target line (including the three data combination methods) is about 30-60 hours of computing time depending on the spectral resolution and queuing time. On the other hand, with the implementation of our parallel data processing the entire EMERGE Early ALMA Sample can be reduced in $~\lesssim$~72~hours (including overheads), that is, at least an order of magnitude less time than the standard (linear) reduction. The application of this parallel methodology in massive ALMA archive research will be presented in a future work.

\subsection{Quality assessments}\label{sec:ALMA_dataquality}

The use of automatic reductions (see Sect.~\ref{sec:ALMA_supercomputer}) requires a careful evaluation of the quality of our data products. To homogeneously quantify the goodness of our reduction and combination processes, we have developed several of the Quality Assessment (QA) metrics described by \citet{Plunkett2023}. We focus on the comparison of the flux recovered in our spectral cubes (N$_2$H$^+$, HNC, HC$_3$N) using different methods (int-CLEAN, Feather, and MACF) with respect the total flux present in our single-dish, IRAM-30m observations (see Fig.~\ref{fig:workflow}).  A detailed discussion of our QA results can be found in Appendix~\ref{appendix:QA}.

Our analysis demonstrates how the addition of the the short-spacing information significantly improves the image fidelity of our ALMA observations (Appendix~\ref{appendix:QA_maps}; see also Paper II). Similar to the results discussed in \citet{Plunkett2023}, our int-CLEAN reductions exhibit flux losses of more than 70\% of the flux detected in the single-dish data due to interferometric filtering. In contrast, the implementation of data combination techniques such as Feather and MACF improve the flux recovery to $\gtrsim$~90\%. The lack of short-spacing information has a dramatic impact also on the gas kinematics at all scales (Appendix~\ref{appendix:QA_spectra}). Different velocity components are selectively (and largely unpredictably) filtered in velocity altering derived parameters such as line intensities and ratios, centroid velocities, and linewidths. Our QAs highlights the need of data combination for the analysis of regions with complex and extended emission similar to those included in our sample.

While both Feather and MACF methods produce satisfactory reductions, MACF at high spectral resolution (narrow) shows the most stable results in our assessments (total flux recovery, flux per intensity bin, and flux per channel) in most of our fields and spectral setups (see Appendix~\ref{appendix:QA_results} for a full discussion). 
For a homogeneous analysis, and to facilitate the comparison with the OMC-1 and OMC-2 results \citep{Hacar2018}, we adopted our MACF reductions as final science products in all our fields (see Sect.~\ref{sec:ALMA_dataproducts}) and will use them during the analysis of our EMERGE sample (hereafter referred to as ALMA+IRAM-30m data). On the other hand, no single-dish data were available to complement our continuum observations. Our ALMA Band-3 continuum int-CLEAN maps should be then treated with appropriate caution. 

\subsection{Final data products \& Data Release 1}\label{sec:ALMA_dataproducts}

Our final EMERGE Early ALMA Survey consists of {\bf 76} ALMA datasets, including spectral and 3~mm continuum maps as well as multiple data combinations, in the OMC-3, OMC-4 South, LDN 1641N, NGC~2023, and Flame Nebula star-forming regions, all with a final resolution of 4\farcs5 or $\sim$~2000~au at the distance of Orion. Our sample also includes a total of 30 IRAM-30m-alone spectral maps of the same line tracers (see Sect.~\ref{sec:obs_30m}) at 30\arcsec resolution. Moreover, we obtained the $T_\mathrm{K}$ (30\arcsec; Sect.~\ref{sec:obs_30m}) and derived a set of total column density maps at high spatial resolution (4\farcs5) using a new technique to be presented in Hacar et al (in prep) in all these targets (5$\times$ each) as high-level data products. Our new observations are also complemented by similar ALMA (only Continuum and N$_2$H$^+$) plus IRAM-30m observations in OMC-1 and OMC-2 \citep{Hacar2018,vanTerwisga2019}.

All our fully-reduced, ALMA plus IRAM-30m data products will be included in different Data Releases (DR) accessible in a dedicated page on our website\footnote{\url{https://emerge.univie.ac.at/results/data/}}. Accompanying this paper, we have released all integrated intensity maps of our ALMA and IRAM-30m observations in FITS format as part of the EMERGE Data Release 1 (DR1). Future DRs will include all spectral cubes and column density maps of our sample.

\section{Sample properties}\label{sec:sample_prop}

In this section we present the main physical properties of the target sample observed in our EMERGE Early ALMA Survey based on the analysis of their integrated intensity and continuum maps at both (low-) single-dish (Sect.~\ref{sec:EMERGE_globalprop}) and (high-) interferometric (Sect.~\ref{sec:EMERGE_ALMAprop}) resolutions. To illustrate our results (e.g. emission distribution, gas structure, etc) we use OMC-3 as representative showcase of our dataset (e.g., Fig.~\ref{fig:OMC3_IRAM30m}). Similar plots for other regions can be found in Appendix \ref{appendix:DP}. To facilitate the comparison between targets we display all regions within the same plotting ranges in each of the figure panels. Our ALMA targets cover a wide range of physical conditions in terms of mass, structure, stellar content, and evolution (see below). We quantify and summarize some of these properties in Table~\ref{table:Targets}. Additional analysis using the full spectral information will be presented in future papers of this series (e.g., gas kinematics; see Paper III).

In our calculations we adopted standard distances to our targets (see Table~\ref{table:Targets}). For sources in Orion A (OMC-1/4 and LDN~1641N) we assume a typical distance of 400~pc as a compromise between the VLBI measurements of the ONC \citep[414$\pm7$;][]{Menten2007} and the Gaia estimates for the head of the ISF \citep[393$\pm$13;][]{Grossschedl2018}. On the other hand, we use a distance of 423~pc for Orion B (NGC~2023 and Flame Nebula) in agreement with the Gaia \citep[423$\pm$21;][]{Zucker2019} and VLBI \citep[$\sim$~420~pc;][]{Kounkel2017} results. Uncertainties of $\pm$~10-20~pc (or $\sim$~5\% of the cloud distance; see references above) are expected from these values producing only minor effects on our mass and size calculations. For a typical distance of 400~pc, the corresponding physical resolution of our data would then be 0.058~pc (or 12000~au) for our IRAM-30m observations (30\arcsec) and 0.009~pc (or 1800~au) for our ALMA maps (4\farcs5).

\subsection{Low-resolution observations}\label{sec:EMERGE_globalprop}

\subsubsection{Gas column density and temperature}\label{sec:EMERGE_evolution}

\begin{figure*}[t]
\centering
\includegraphics[width=1.\textwidth]{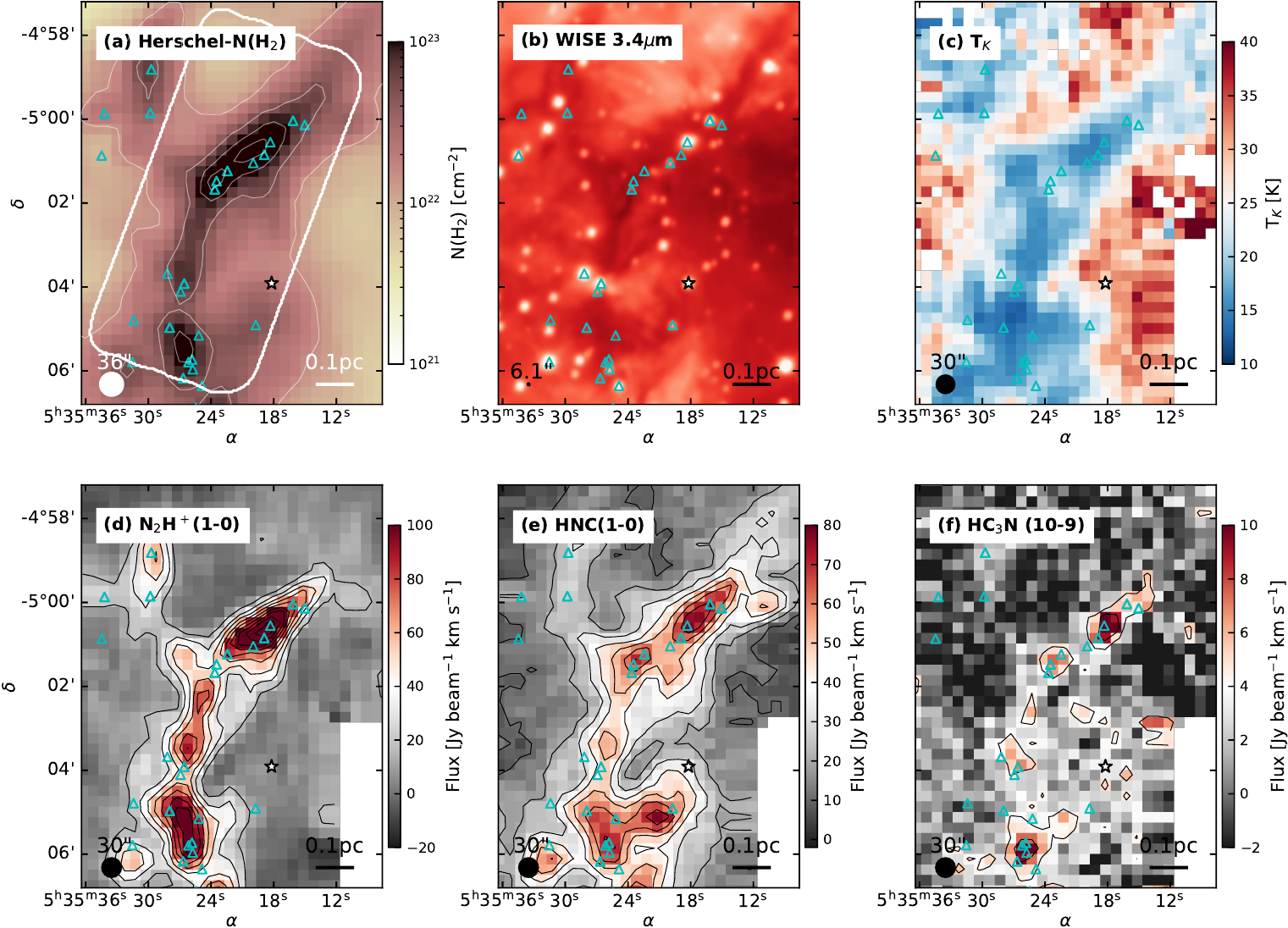}
      \caption{Low-resolution observations in OMC-3. From left to right and from top to bottom: 
      {\bf (a)} Herschel, total gas column density, N(H$_2$) (36"); 
      {\bf (b)} {\it WISE} 3.4~$\mu$m emission (6.1");
      {\bf (c)} gas kinetic temperature, $T_\mathrm{K}$,
      {\bf (d)} N$_2$H$^+$ (1-0), {\bf (e)} HNC (1-0), and {\bf (d)} HC$_3$N (10-9) integrated intensity maps obtained in our IRAM-30m (broad) observations (30").
      Symbols are similar to those in Fig.~\ref{fig:OriA_footprint}. The corresponding beam size and scale bar are indicated in the lower par of each panel. For reference, panel a includes contours at N(H$_2$) = [15, 50, 100, 150, 200] $\times 10^{21}$~cm$^{-2}$ as well as the footprint of our ALMA observations shown in Fig.~\ref{fig:OMC3_ALMA+IRAM30m}. Similar plots for all the other regions can be found in Appendix~\ref{appendix:DP}.
              }
\label{fig:OMC3_IRAM30m}
\end{figure*}

While limited to only 7 targets, our EMERGE Early ALMA Sample covers an extensive range of physical conditions and environments.
In addition to the overall description of our survey (Sect.~\ref{sec:EMERGE_sample}), we quantified additional evolutionary properties of our sources considering all positions covered by our IRAM-30m (broad) observations within representative area of 1.5$\times$1.5 pc$^2$ from the center of our maps ($\sim$~700$\times$700 arcsec$^2$; roughly similar to the zoom-in maps in Figs.~\ref{fig:OriA_footprint} and \ref{fig:OriB_footprint}). 

The wide range of stellar activity in our sample is reflected in the populations of protostars (P) plus disk (D) YSOs (Table~\ref{table:Targets}). Our survey covers almost two orders of magnitude in average stellar density (i.e. (P+D)/Area or simply P+D) from the highly clustered OMC-1 region to the more diffuse NGC~2023 forming stars almost in isolation.
We note here that the stellar density is largely underestimated in the case of OMC-1 and Flame Nebula since our stellar counting does not include any of the optical stars identified in these clouds as well as the bright emission of their nebulae rapidly increases the incompleteness of IR surveys \citep{Megeath2015}.

\begin{figure*}[ht!]
\centering
\includegraphics[width=0.8\textwidth]{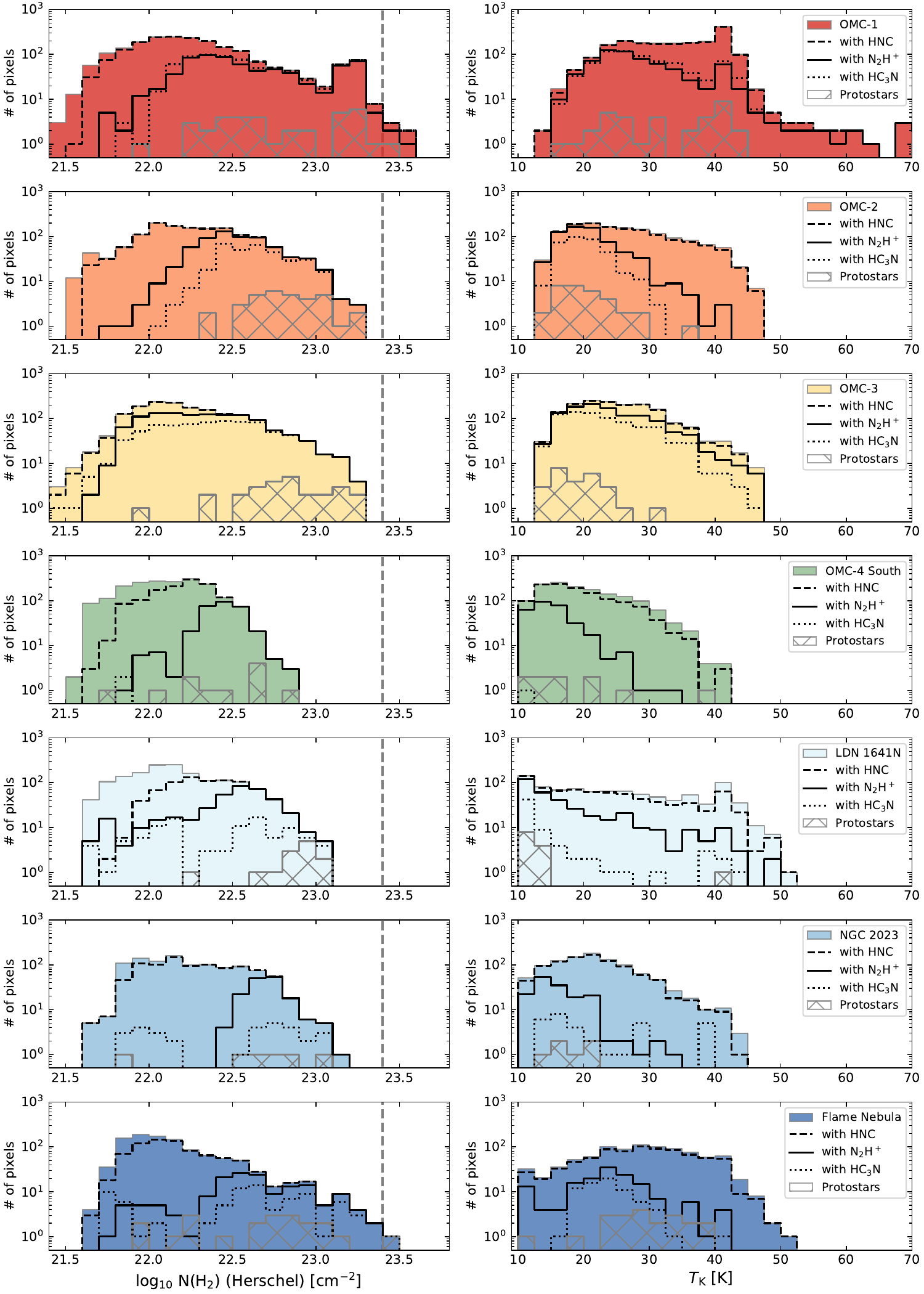}
      \caption{Histograms of the {\bf(Left)} gas column density N(H$_2$) and {\bf(Right)} gas kinetic temperature $T_\mathrm{K}$ distributions detected at low spatial resolution (Herschel or IRAM-30m) within the central 1.5$\times$1.5 pc$^2$ ($\sim$~700$\times$700 arcsec$^2$) of all regions part of the EMERGE Early ALMA Survey (see also Tables \ref{table:Targets} and \ref{table:Fields}). From top to bottom: OMC-1, OMC-2, OMC-3, OMC-4 South, LDN~1641N, NGC~2023, and Flame~Nebula.
      Positions showing detected emission above the first contour in Fig.~\ref{fig:OMC3_IRAM30m} (and Figs.\ref{fig:DP_OMC1_IRAM}-\ref{fig:DP_FlameNebula_IRAM}) of HNC (black dashed lines), N$_2$H$^+$ (black solid lines), and HC$_3$N (black dotted lines), as well as including protostars (grey hatched areas) are highlighted in all histograms. A vertical grey dashed line in the N(H$_2$) histograms indicates the minimum column density of 10$^{23.4}$~cm$^{-2}$ (or 1~g~cm$^{-2}$) needed to form a high-mass star \citep{Krumholz2008}.
              }
\label{fig:ALL_NH2+TK}
\end{figure*}

The diversity of our sample becomes apparent in the distribution of the Column Density Probability Distribution Functions (N-PDF) displayed in Fig.~\ref{fig:ALL_NH2+TK} (left panel)\footnote{We note that the Herschel maps used in \citet{Lombardi2014} show several saturated pixels at the brightest positions in both OMC-1 and Flame Nebula fields. The derived N(H$_2$) values at these positions should be then taken with caution (see secondary peak in the N-PDF of OMC-1 in Fig.~\ref{fig:ALL_NH2+TK}, top left panel).}.  
Large amounts of gas at N(H$_2$)~$>$~10$^{23}$~cm$^{-2}$ are seen in regions forming O-type stars (OMC-1 and Flame Nebula), close to the theoretical predictions for the formation of high-mass stars \citep[i.e. N(H$_2$)~$=$~10$^{23.4}$~cm$^{-2}$;][]{Krumholz2008}. In contrast, low-mass and diffuse regions (e.g., OMC-4 South or NGC~2023) reach maximum column densities of N(H$_2$)~$\sim$~10$^{23}$~cm$^{-2}$ in only few locations. Likewise, the high-end of the N-PDF distributions in active regions with higher fractions of dense gas (OMC-1/2/3, LDN~1641N, and Flame Nebula) shows shallower slopes than those found more diffuse clouds (OMC-4 South and NGC~2023) in agreement with previous works in some of these clouds \citep{Stutz2015}. 

Our gas kinetic temperature $T_\mathrm{K}$ maps (see Fig.~\ref{fig:OMC3_IRAM30m}c) illustrate the thermal evolution of gas during the star formation process. Well shielded, high column density regions with N(H$_2$)~$>2\times 10^{22}$~cm$^{-2}$ usually show cold temperatures of $\lesssim 25$~K. In contrast, lower column density material is usually found at higher temperatures up to $\sim$~40~K heated up by external radiation resulting in a positive thermal gradient towards the edge of the cloud. Local heating effects are seen associated to some locations with strong star formation activity (e.g., compare the positions of the {\it WISE} sources with our $T_\mathrm{K}$ maps around OMC-2 FIR-2 in Figs. \ref{fig:DP_OMC2_IRAM}b and \ref{fig:DP_OMC2_IRAM}c). More prominently, regions under the influence of high-mass stars (OMC-1 and Flame Nebula) show large areas of both low- and high column density gas at temperatures above 40~K coincident with the bright emission in the {\it WISE} 3.4$\mu$m maps (see Figs. \ref{fig:DP_OMC1_IRAM} and \ref{fig:DP_FlameNebula_IRAM}, respectively). These results demonstrate the sensitivity of low-resolution HCN/HNC measurements to describe the thermal properties across a wide range of column densities and environments \citep[see ][for a discussion]{Hacar2020}.

Feedback also shapes the overall temperature distribution within these clouds as seen in the $T_\mathrm{K}$ histograms (or Temperature PDF, $T_\mathrm{K}$-PDF) shown in Fig.~\ref{fig:ALL_NH2+TK} (right panels) as well as changes the average gas temperature in these regions (see Table~\ref{table:Targets}). Most of the gas in low-mass star-forming regions (e.g., OMC-3 or OMC-4 South) is found at temperatures between 10-30~K with mean values around $\sim$25~K. In contrast, regions with embedded massive stars show larger average values of $\gtrsim$~30~K with significant amounts of gas at T$_K>45$~K (see the tails of the $T_\mathrm{K}$-PDF in OMC-1 and Flame Nebula). 
Interestingly, LDN~1641N shows a significant fraction of low column density gas above 40~K surrounding the cloud (see external heating in Fig.\ref{fig:DP_LDN1641N_IRAM}c). The origin of this warm component is unclear 
although the cometary shape of this cloud pointing towards the NGC~1980 cluster might suggest a potential interaction with $\iota$~Ori and NGC~1980.

\subsubsection{Molecular emission and density selective tracers}\label{sec:EMERGE_emissionprop}

We describe the global emission properties of our sample using our low spatial resolution observations shown in Figure~\ref{fig:OMC3_IRAM30m} (lower panels). For that, we take advantage of the extended spatial coverage of our (broad spectral resolution) IRAM-30m maps at low spectral resolution (see Sect.~\ref{sec:obs_30m}). Together with the N(H$_2$), {\it WISE}, and $T_\mathrm{K}$ maps (panels a-c) already presented in Sect.~\ref{sec:EMERGE_evolution}, we display the N$_2$H$^+$ (1-0), HNC (1-0), and HC$_3$N (10-9) integrated intensity maps in Fig.~\ref{fig:OMC3_IRAM30m} (panels d-f) in OMC-3. We obtained each of these molecular maps adapting the range of integrated velocities to the corresponding cloud and transition. Similar plots for other regions can be found in Figs.~\ref{fig:DP_OMC1_IRAM}-\ref{fig:DP_FlameNebula_IRAM}.

The emission of all molecular tracers is clearly detected in all sources (panels d-f) with the only exception of the  HC$_3$N emission in OMC-4 South (see Fig.~\ref{fig:DP_OMC4_IRAM}). The overall emission properties in our maps are similar to the results reported in previous large-scale surveys in Orion \citep[][]{Pety2017,Kauffmann2017}. N$_2$H$^+$ exhibits the brightest integrated intensity values typically associated to regions of high-column density above N(H$_2$)~$>2\times 10^{22}$~cm$^{-2}$ (or A$_V\gtrsim$~20~mag). HNC (as well as HCN, not shown) displays the most extended emission in our sample down to column densities of N(H$_2$)~$\sim 10^{21}$~cm$^{-2}$ (or few A$_V$), in many cases extending over the limits of our IRAM-30m maps. On the other hand, the HC$_3$N emission is usually clumpier, located in smaller areas, and comparatively weaker, although showing bright emission features closely related to the position of some (but not all) young protostars. The relative contributions of the different tracers become apparent in the N(H$_2$) and $T_\mathrm{K}$ histograms in Fig.~\ref{fig:ALL_NH2+TK} where we highlight those pixels with N$_2$H$^+$ (black solid line), HNC (black dashed line), and HC$_3$N (black dotted line) emission within the first contours our of map (e.g., see Fig.~\ref{fig:OMC3_IRAM30m}), and well as those pixels including protostars (grey hatched area).

In terms of line brightness, our IRAM-30m maps display a large dynamic range over up to two orders of magnitude in integrated intensity and with regions such as OMC-1 showing peak values $>$200~Jy~beam$^{-1}$~km~s$^{-1}$ (or $>$30~K~km~s$^{-1}$; e.g., Fig.\ref{fig:OMC3_IRAM30m}). Moreover, the total luminosity of our maps systematically changes between the regions in our sample (see comparison between the total number of pixels includes in the histograms in Fig.\ref{fig:ALL_NH2+TK}) and is closely related to their physical characteristics (see Sect.\ref{sec:EMERGE_globalprop}).
Regions reaching high column densities (OMC-1/2/3 and LDN~1641N) and/or high temperatures (OMC-1 and Flame Nebula) show the brightest lines and total intensity maps in all transitions. In contrast, the emission of the same lines in more diffuse and colder regions (NGC~2023 and OMC-4 South) is always weaker.

The results of our molecular maps match the expected behavior for the suite of selective tracers included in our survey (see line upper energies and effective densities in Table~\ref{table:molecules}).
Formed after the depletion of CO \citep[see][for a review]{BerginTafalla2007}, N$_2$H$^+$ is a density selective species tracing the cold ($\lesssim$~20~K) and dense ($> 10^4$~cm$^{-3}$) gas \citep{Tafalla2002} while its ground transition is associated to dense cores \citep{Caselli2002} and filaments \citep{Hacar2017b,Hacar2018}. Complementary to N$_2$H$^+$, the HNC (1-0) line (as well as the HCN (1-0)) is effectively excited at densities of a feq $10^3$~cm$^{-3}$ \citep{Shirley2015} and its emission traces the cloud gas from low- to intermediate column densities before also depletes at high densities \citep{Tafalla2021,Tafalla2023}.
On the other hand, the excitation of the HC$_3$N (10-9) transition requires higher temperatures (E$_u=24.1$~K) and it is only effectively excited in lukewarm regions exposed to feedback (Hacar et al in prep.).

\subsubsection{Evolutionary state and dense gas Star Formation Efficiency}\label{sec:EMERGE_evolstate}

Additional insight on the evolutionary state of our targets can be inferred from the comparisons of their populations of P and D stars, respectively. From the comparison of the number of Class II (similar to D) and Class I (similar to P) stars in nearby clouds, and assuming a typical age of 2~Myr for a typical Class II objects, IR surveys of nearby stars have inferred a median evolutionary timescale for a Class I object of $\sim$~0.5~Myr \citep{Evans2009_C2D}. Given these typical ages, the observed P/D ratio can be used as proxy of the relative ages between targets (see Table~\ref{table:Targets}). 
Young regions in which most of the stars were formed within the last Myr are expected to show high P/D$\gtrsim$1.
Clouds continuously forming stars for long periods of time should approach a steady-state ratio in which P/D~$=\frac{0.5\ \mathrm{Myr}}{2.0\ \mathrm{Myr}} \sim 0.25$.
Finally, older regions with an already declining star-formation rate should exhibit ratios such as P/D$<$0.25, and approach P/D$\rightarrow$0 once they exhaust their gas reservoir for forming stars.
Within our sample, OMC-1, Flame Nebula, and OMC-4 South appear to be older and more evolved (P/D$\lesssim$0.2) compared to younger regions such as OMC-2/3 and LDN~1641N (P/D$>$0.3) and particularly NGC~2023 (P/D=0.75). 

We quantified the potential of these regions to form new stars from the amount of gas detected in our Herschel maps (see Fig.~\ref{fig:OMC3_IRAM30m}a). We obtained two measurements adding all Herschel column density measurements: (a) the total mass of gas in our maps (M$_{tot}$), (b) mass of dense gas M$_{dense}$ showing significant N$_2$H$^+$ emission ($\geq 15$~Jy~beam$^{-1}$~km~s$^{-1}$ or $\gtrsim$~2.5~K~km~s$^{-1}$ in T$_{mb}$ units; see also Sect.\ref{sec:EMERGE_timescales}) and (c) fraction of dense gas ($f_{dense}$) obtained from the ratio between M$_{tot}$ and M$_{dense}$. Active regions such as OMC-1, OMC-2, OMC-3, and LDN~1641N exhibit large mass reservoirs (M$_{tot}>$~500~M$_\odot$) with significant amount of dense gas (M$_{dense}>$~300~M$_\odot$ and $f_{dense}\gtrsim 0.4$) within their central $\sim$~1~pc$^2$. Although comparable in total mass, OMC-4 South and NGC~2023 are more diffuse (M$_{dense}<$~90~M$_\odot$ and $f_{dense}\lesssim 0.2$). Flame Nebula is a particularly interesting case showing the lowest mass available (M$_{tot}<$~300~M$_\odot$) although with a significant fraction of dense gas ($f_{dense}= 0.33$). In its latest stages of evolution (see above) the dense gas in the Flame Nebula (as well as in OMC-1) appear to be more resistant to the disruptive effects of feedback in agreement to simulations \citep{Dale2014}.

\begin{figure}[h!]
\centering
\includegraphics[width=0.85\linewidth]{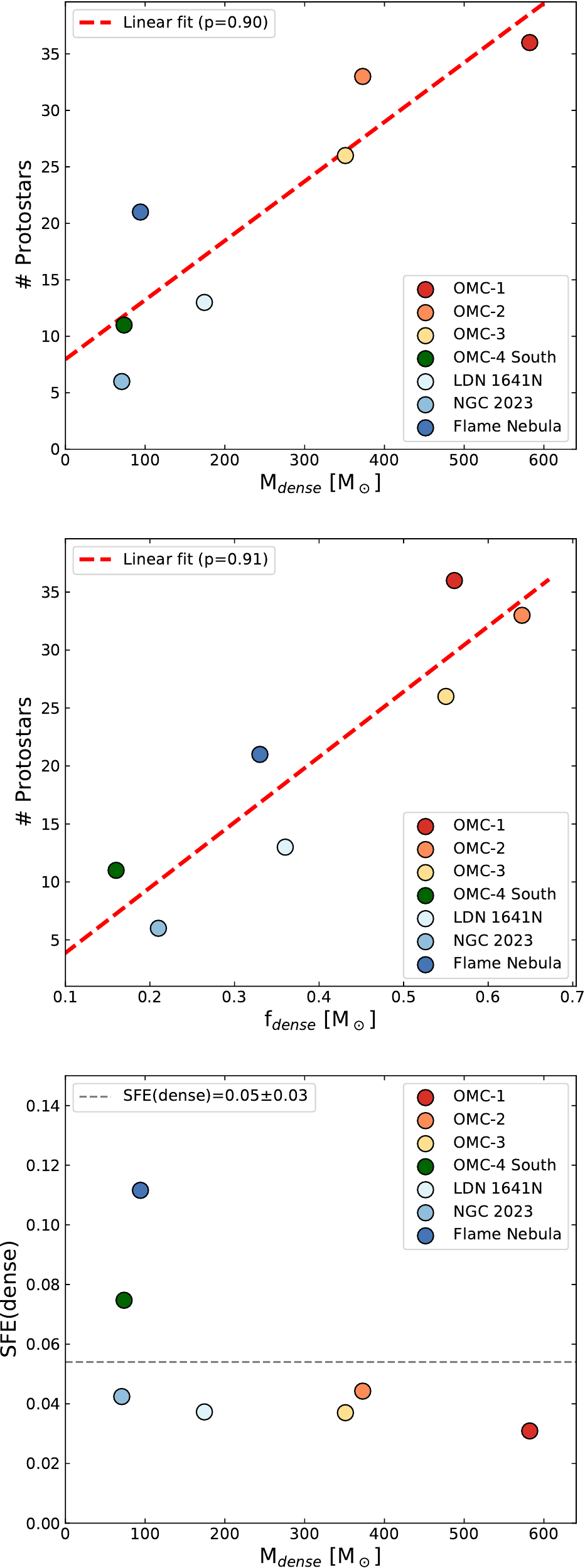}
      \caption{From top to bottom: {\bf (Upper panel)} number of protostars (P) with respect to the total amount of dense gas M$_{dense}$, {\bf (Middle panel)} 
      fraction of dense gas $f_{dense}$, {\bf (Lower panel)} and star formation efficiency in the dense gas across the EMERGE Early ALMA survey sample (see values in Table~\ref{table:Targets}). We display the linear fit of  all our targets (dashed red line) and their corresponding Pearson p-coefficients (see values in legend) in both upper and middle panels. 
              }
\label{fig:proto_vs_Mass}
\end{figure}

Despite the above exceptions, the values of M$_{dense}$ and $f_{dense}$, both correlated with the amount of N$_2$H$^+$ emission, appear as better proxies of the current star-formation than the total gas mass M$_{tot}$ of the region. 
We illustrate these correlations in Fig.~\ref{fig:proto_vs_Mass} by comparing the number of Protostars (P) with the amount of dense gas M$_{dense}$ (upper panel) and the fraction of dense gas $f_{dense}$ (middle panel) across our sample (see values in Table~\ref{table:Targets}). As shown in Fig.~\ref{fig:proto_vs_Mass} (upper panel), the number of protostars in all our targets shows a tight linear dependence (dashed red line) with M$_{dense}$ (with a Pearson coefficient p=0.90) stronger than the one obtained with M$_{tot}$ (p=0.77; not shown). The correlation with M$_{dense}$ further improves after removing the Flame Nebula (p=0.96; not shown).
A similar linear dependence is seen in Fig.~\ref{fig:proto_vs_Mass} (middle panel) when comparing the number of protostars with $f_{dense}$ (p=0.91). Figure~\ref{fig:proto_vs_Mass} reinforces the direct correspondence between the amount of dense gas (traced by N$_2$H$^+$) and the ongoing star formation in our targets. 

The linear correlation between total number of YSOs (P+D) and the amount of gas at column densities above A$_V\gtrsim 8$~mag, used as proxy of M$_{dense}$ for the dense gas above $10^4$~cm$^{-3}$, has been well documented in the past in large-scale surveys of nearby molecular clouds \citep[e.g.,][]{Lada2010}. 
A similar dependence is found when comparing the surface density of (only) protostars (P) and high column density gas \citep{Heiderman2010}. This positive correlation indicates how  high density material is needed to efficiently form stars within clouds \citep[e.g.,][]{Lada1992}. Figure~\ref{fig:proto_vs_Mass} indicates that this correlation gets tighter when considering the youngest protostars (P) and the highest density material (traced by N$_2$H$^+$) in our targets.

The linear correlation found in Fig.~\ref{fig:proto_vs_Mass} (upper panel) suggests a roughly constant star formation efficiency per unit of dense gas SFE(dense) of a few percent. In Fig.~6 (lower panel) we estimate this efficiency as SFE(dense)~$ = \frac{\mathrm{M}_{proto}}{\mathrm{M}_{dense}+\mathrm{M}_{proto}}$ \citep[e.g.,][]{Megeath2022}, where $\mathrm{M}_{proto}$ ($=0.5\times \# Protostars$) corresponds to the total mass in protostars each assumed with a typical mass of 0.5~M$_\odot$. Our targets exhibit a typical SFE(dense) around $0.05\pm 0.03$ (or $\sim$5\%; grey dashed line) for M$_{dense}$ values between $\sim 100$ and $600$ M$_\odot$.
Compared to the $\sim$~0.1~Myr free-fall time $\tau_{ff}=\sqrt{3\pi/32G\rho}$ for a typical dense gas density of $n\sim 10^5$~cm$^{-3}$ traced by N$_2$H$^+$, the above values for SFE(dense) translate into an approximate efficiency per free-fall time of $\epsilon_{ff}\sim$~1\% in agreement with previous predictions \citep[][]{krumholz2007} and surveys \citep[][]{pokhrel2021}.
The roughly constant SFE(dense) derived from our data indicates the total mass of dense gas as the primary factor determining the current star formation activity of our regions leading into a number of young protostars directly proportional to the amount of M$_{dense}$ available in each case \citep[see also][]{Lada2010}. 

This is expected as the cold ($T_\mathrm{K}\lesssim 20 $~K) and dense ($n(\mathrm{H}_2)>10^5$~cm$^{-3}$) material traced by N$_2$H$^+$ (Sect.~\ref{sec:EMERGE_evolution}) promotes the conditions for gravitational collapse and it is therefore prone to form stars. Active star-forming regions showing higher number of protostars (e.g., OMC-1) are simply the result of their (currently) higher content of this dense and cold material,  showing higher star formation rates (SFR(dense)=$\frac{\epsilon_{ff}}{\tau_{ff}}\times \mathrm{M}_{dense}$. Once this dense gas is generated, it forms stars at a constant SFE(dense) irrespective of the star formation regime (low- vs high-mass) and evolutionary state (young vs old, or P/D value; see above) of the cloud.

\subsubsection{Timescales for the evolution of dense gas}\label{sec:EMERGE_timescales}

\begin{figure*}[t]
\centering
\includegraphics[width=\textwidth]{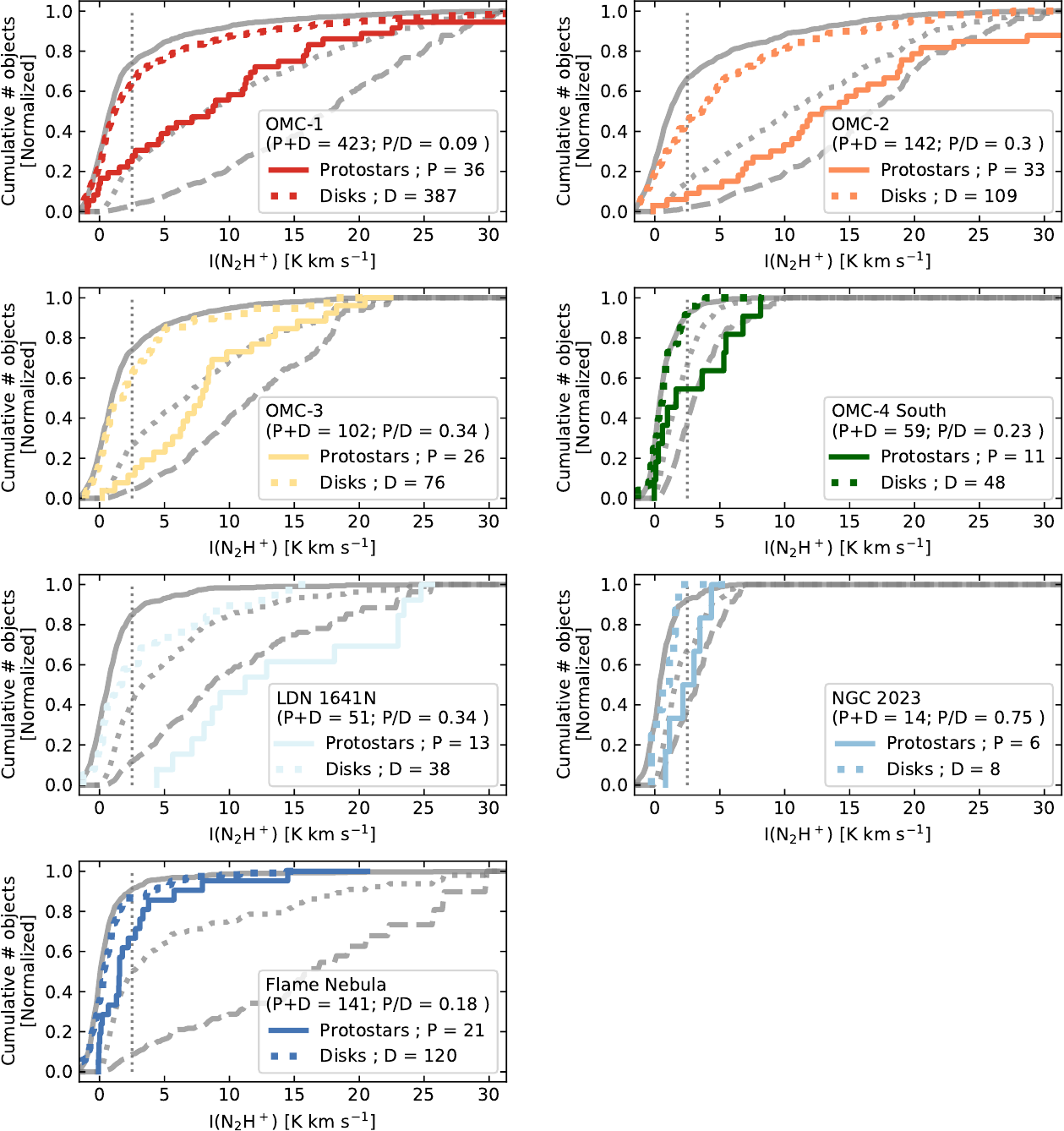}
      \caption{Cumulative distributions of the N$_2$H$^+$ (1-0) integrated  emission I(N$_2$H$^+$) at the positions of both protostars (solid colored line) and disk stars (dashed colored line) within the central 1.5$\times$1.5 pc$^2$ ($\sim$~700$\times$700 arcsec$^2$) in each of the regions part of the EMERGE Early Science sample. All panels are displayed within the same ranges of N$_2$H$^+$ intensity to facilitate their comparison.
      In each sub-panel, the local number of protostars (P) and disk stars (D), as well as their ratio (P/D), are indicated in the legend (see also Table~\ref{table:Targets}). In each of these regions, the average result of simulating the expected cumulative distributions of a P+D number of objects following (a) a random distribution (grey solid line), (b) a linear dependence (grey dotted line), and (c) a quadratic dependence (grey dashed line) are shown in the corresponding panels. 
              },
\label{fig:ALL_YSOs}
\end{figure*}

As demonstrated in previous sections, N$_2$H$^+$ appears as the most direct proxy of the young, star-forming gas in all targets of our survey. 
Following the methodology introduced by \citet{Hacar2017b}, in Figure~\ref{fig:ALL_YSOs} we display the cumulative distribution of the N$_2$H$^+$ integrated intensity (in K~km~s$^{-1}$) at the position of protostars (P; colored solid line) and disk stars (D; colored dotted line) found in the 7 regions of our study (see panels). In all cases, the distribution for P stars always runs towards the right side of the one for D stars, indicating that protostars are typically associated to brighter N$_2$H$^+$ emission in our maps, as expected for this dense gas tracer. 

We have compared the above observational results with the expected cumulative plots for a similar total number of objects per cloud (see legend) located following a series of simulated distributions: (a) a random distribution (grey solid line) as well as independent distributions in which the position of these objects is favoured at the position of the N$_2$H$^+$ emission in our maps using (b) linear (grey dotted line) and (c) quadratic (grey dashed line) weights, respectively. 
In all our targets, the distribution of D stars closely resembles a (quasi-)random distribution with respect to the dense gas traced in N$_2$H$^+$ indicating that the D stars and N$_2$H$^+$ emission are uncorrelated. In contrast, P stars have a much stronger connection with these emission features typically showing a linear or even quadratic correlation with the N$_2$H$^+$ intensities in regions such as OMC-2/3  and LDN~1641N. 

Deviations from the above general behaviour seen in OMC-1 (linear) and, more prominently, in the Flame Nebula (sublinear) can be explained by the evolutionary state of these targets (see Sect.~\ref{sec:EMERGE_evolstate}) and the direct impact of radiative and mechanical feedback in regions hosting high-mass stars.
Only small amounts of dense gas is detected in the immediate vicinity of the Trapezium stars (including $\theta^1$~Ori~C) and NGC~2024 IRS-2b, which demonstrates the rapid dispersion and photo-dissociation of the molecular material after the formation of these high-mass objects. Yet, the still visible correlation between the N$_2$H$^+$ emission and the location of the young protostars in these targets indicate that the dense gas is not heavily affected once the cloud manages to shield this intense radiation field \citep[e.g.,][]{odell2001}. In fact, large fractions of dense, undisturbed gas are still detected in regions such as OMC-1 Ridge and Orion South at a distance of ($\sim$~0.1-0.2~pc in projection) \citep[e.g.,][]{Wiseman1998}.

The results in Figure~\ref{fig:ALL_YSOs} suggest a fast evolution of the dense gas in the targers included in our survey. The strong correlation between P stars and N$_2$H$^+$ indicates that the dense gas survives at least the typical duration of the protostellar (class 0/I) phase with $\sim$0.5~Myr. On the other hand, the random distribution of D stars indicates that this correlation is lost at the typical timescales of $\sim$2~Myr for this older (Class II) YSO population. Analogous trends are reported in other filamentary clouds such as Taurus \citep[][comparing CO maps with the position of YSOs]{Hartmann2001} and Perseus  \citep[][using a similar analysis based on N$_2$H$^+$ observations]{Hacar2017b}. The combination of the above results suggests that the typical evolutionary timescales of the dense gas currently observed in N$_2$H$^+$ is therefore fast and within $\sim$1~Myr. 
A similar fast evolution for the dense, star-forming gas traced by N$_2$H$^+$ has been recently proposed from simulations \citep{Priestley2023_N2Hp}.
Within these timescales, a small fraction of the dense gas would be either accreted onto stars following its small SFE(dense) (Sect.~\ref{sec:EMERGE_evolstate}) while most of the remaining dense gas will be recycled by the turbulent motions inside these regions \citep[e.g.,][]{Padoan2016}.
The diversity of environments and regions explored here rules out exotic ejection mechanisms \citep[e.g.,][]{Stutz2018}. Instead, this fast evolution appears to be driven by the continuous assembly of dense gas within these regions and its subsequent local destruction once star-formation is ignited in them.
In contrast to more traditional quiescent scenarios (e.g., quasi-static fragmentation), these results depict a dynamical evolution for typical star-forming regions such as Orion in agreement with recent simulations \citep[][]{Padoan2016,Smith2020,Seifried2020,IbaMejia2022}.

\subsection{High-resolution observations: cloud sub-structure at 2000~au resolution}\label{sec:EMERGE_ALMAprop}

\begin{figure*}[t]
\centering
\includegraphics[width=1.\textwidth]{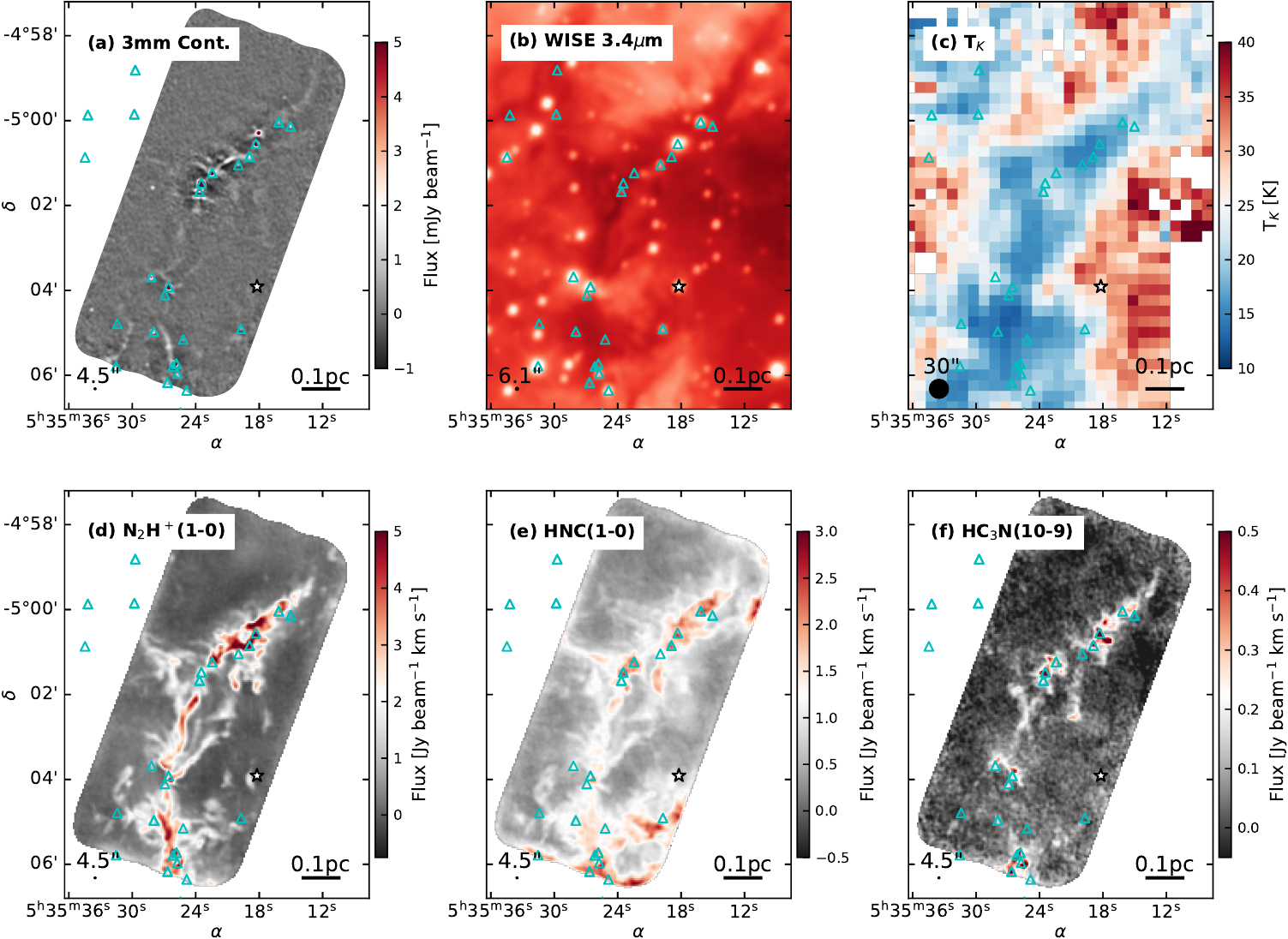}
      \caption{High-resolution observations in OMC-3. From left to right and from top to bottom: 
      {\bf (a)} ALMA interferometric-alone (int-CLEAN) continuum map (4\farcs5); 
      {\bf (b)} {\it WISE} 3.4~$\mu$m emission (6.1");
      {\bf (c)} gas kinetic temperature, $T_\mathrm{K}$ (30"),
      {\bf (d)} N$_2$H$^+$ (1-0), {\bf (e)} HNC (1-0), and {\bf (d)} HC$_3$N (10-9) integrated intensity maps obtained in our ALMA+IRAM-30m MACF observations (4\farcs5).
      Symbols are similar to those in Fig.~\ref{fig:OriA_footprint}. Similar plots for all the other regions can be found in Appendix~\ref{appendix:DP}.
      The enhanced resolution of our ALMA+IRAM-30m maps (panels a, d-f) can be seen from their comparison with the gas $T_\mathrm{K}$ map (panel c) displayed at the single-dish resolution.
      }
\label{fig:OMC3_ALMA+IRAM30m}
\end{figure*}

Our ALMA observations provide a new perspective of the gas structure down to $\sim$~2000~au resolution (or 4\farcs5 at the distance of Orion). In analogy to our low-resolution data (Sect.~\ref{sec:EMERGE_globalprop}), in Fig.\ref{fig:OMC3_ALMA+IRAM30m} we illustrate the high-resolution 3-mm continuum (panel a), as well as the integrated emission maps of N$_2$H$^+$ (1-0) (panel d), HNC (1-0) (panel e), and HC$_3$N (10-9) (panel f) lines, respectively, observed by ALMA in the OMC-3 region. Similar plots for all the other targets in our sample can be found in Figs.~\ref{fig:DP_OMC1_ALMA+IRAM}-\ref{fig:DP_FlameNebula_ALMA+IRAM} in Appendix~\ref{appendix:DP}.

The interferometric-alone (int-CLEAN) 3mm-continuum maps such in Fig.~\ref{fig:OMC3_ALMA+IRAM30m}a show a combination of two different emission mechanisms in our targets. We find large emission areas with fluxes of $>$10~mJy~beam$^{-1}$ are detected in both OMC-1 (Fig.\ref{fig:DP_OMC1_ALMA+IRAM}a) and Flame Nebula (Fig.\ref{fig:DP_FlameNebula_ALMA+IRAM}a) coincident with their bright emission nebulae and contaminated by the free-free (bremssthrahlung) emission in these HII regions up to 100~GHz \citep[see][for a discussion]{Mason2020}. Outside these areas, the 3mm emission is dominated by the thermal dust continuum reaching values of $\lesssim$~5~mJy~beam$^{-1}$. In fields such as OMC-3, we identify multiple bright, compact sources coincident with the position of previously identified millimeter compact objects \citep{Takahashi2013}, YSOs \citep{Megeath2012}, and protoplanetary disks \citep{vanTerwisga2019}, all of them unresolved in our maps. 
We detect a more extended cloud emission, with fluxes of $\lesssim$~2~mJy~beam$^{-1}$, towards column densities of N(H$_2$)~$\gtrsim 50 \times 10^{21}$~cm$^{-2}$ showing a large number of filamentary structures with sizes of few 10$^4$~au. Because of this limited sensitivity, however, regions such as OMC-4 (Fig.\ref{fig:DP_OMC4_ALMA+IRAM}a) or NGC~2023 (Fig.\ref{fig:DP_NGC2023_ALMA+IRAM}a) become mostly undetected in our survey. On the other hand, we identify clear negative side-lobes in regions with extended and bright emission such as OMC-1/2/3. Additional short-spacing information in continuum (via SD data) is thus needed to carry out a full analysis of these data products\footnote{We remark that, similar to our spectral observations (see Appendix~\ref{appendix:QA} for a discussion), the interferometric filtering of large-scale emission can artificially reduce the continuum emission at small scales. The lack of short-spacing information therefore reduces the effective dynamic range and sensitivity of our int-CLEAN maps to recover the true dust continuum emission.}. 

The enhanced dynamic range of our molecular ALMA+IRAM-30m (MACF) maps provides a more complete picture of the gas distribution in our targets.
The N$_2$H$^+$ (1-0) emission above $\gtrsim$~1~Jy~beam$^{-1}$~km~s$^{-1}$ ($>$7~K~km~s$^{-1}$) shows in a plethora of small-scale, elongated and narrow structures in all our targets (Fig.\ref{fig:OMC3_ALMA+IRAM30m}d).
The previous characterization of the N$_2$H$^+$ emission in OMC-1 and OMC-2 (see Figs.\ref{fig:DP_OMC1_ALMA+IRAM}d and \ref{fig:DP_OMC2_ALMA+IRAM}d) identified these filamentary features as velocity-coherent, sonic fibers \citep{Hacar2018}. The analogous emission features seen in all our maps suggests a similar fiber organization of the dense gas in all targets in our sample  \citep[see][for a full characterization]{SocciPaperIII}.
More striking, this filamentary gas organization continues into the more diffuse material traced by HNC (1-0) at $\gtrsim$~1~Jy~beam$^{-1}$~km~s$^{-1}$ ($\gtrsim$~7~K~km~s$^{-1}$; Fig.\ref{fig:OMC3_ALMA+IRAM30m}e) sometimes running perpendicular to the above dense fibers. Yet, the complexity and emission properties of the gas change across our sample: dense regions such as OMC-3 (Fig.\ref{fig:OMC3_ALMA+IRAM30m}e-d) and LDN~1641N (Fig.\ref{fig:DP_LDN1641N_ALMA+IRAM}e-d) show bright and highly contrasted emission in both N$_2$H$^+$ and HNC, while more diffuse clouds such as OMC-4 (Fig.\ref{fig:DP_OMC4_ALMA+IRAM}e-d) and NGC~2023 (Fig.\ref{fig:DP_NGC2023_ALMA+IRAM}e-d) present systematically weaker and shallower emission in these tracers. 
In contrast, we find the HC$_3$N (10-9) emission $\gtrsim$~0.15~Jy~beam$^{-1}$~km~s$^{-1}$ ($>$1~K~km~s$^{-1}$) to be typically more clumpy and concentrated towards local regions with dense gas (showing N$_2$H$^+$ emission) directly exposed to some stellar feedback (either by nebulae or YSOs) as seen in OMC-3 (Fig.\ref{fig:OMC3_ALMA+IRAM30m}f) or Flame Nebula (Fig.\ref{fig:DP_FlameNebula_ALMA+IRAM}f) although the lack of ALMA observations of this tracer (as well as HNC) in OMC-1 and OMC-2, both with bright emission in our SD maps (Figs.\ref{fig:DP_OMC1_IRAM} and \ref{fig:DP_OMC2_IRAM}), prevent drawing further conclusions on the origin of its emission.  

Figure~\ref{fig:OMC3_ALMA+IRAM30m} also highlights the ability of our molecular maps to reproduce the internal gas distribution at high spatial resolutions. 
OMC-3 appears as highly extincted, MIR-dark region showing multiple narrow structures in the {\it WISE} 3.4~$\mu$m image in Fig.\ref{fig:OMC3_ALMA+IRAM30m}b \citep[see][for a detailed discussion on the MIR absorption in OMC-3]{Juvela2023}. We find a direct spatial correlation (both in location and size) between these MIR absorption features and the emission distribution of our N$_2$H$^+$ maps (Fig.\ref{fig:OMC3_ALMA+IRAM30m}d).  These similarities extend to the more diffuse gas where we identify several ``whisker-like'' features in the {\it WISE} image (see northern part of OMC-3) also detected in HNC (Fig.\ref{fig:OMC3_ALMA+IRAM30m}e). 
Although less clear, we also observe these similarities in the case of NGC~2023 (Fig.\ref{fig:DP_NGC2023_ALMA+IRAM}). 
The nature of these diffuse gas features remains unclear at this point. Yet, 
this unique correspondence with the MIR absorption reinforces the potential of our N$_2$H$^+$ and HNC (ALMA+IRAM-30m) maps to describe the fine gas substructure within our cloud sample with high fidelity.

The inspection of our molecular ALMA+IRAM-30m (MACF) maps reveals a new and complex cloud substructure at high spatial resolution.
This is particularly remarkable in comparison with previous single-dish observations. Regions such as OMC-3 and Flame Nebula have been identified as single, elongated (and broad) filaments using low spatial resolution, continuum \citep[][]{Schuller2021,Konyves2020} and line \citep[][]{Orkisz2019,Gaudel2023} observations. Under the sharper eyes of ALMA, however, the gas organization in all our fields is clearly more complex, nebulous, and wispy, and sometimes still unresolved by our $\sim$~2000 au beamsize. 

Both low- and high-mass star-forming regions exhibit a strong fibrous and gaseous sub-structure. Far from idealize cylinders, the size, width, radial distributions, and morphology of these slender fibers vary both locally and between regions.
Yet, these fiber properties appear to vary smoothly across our sample showing no discontinuity between low- and high-mass star-forming regions.
Obtaining high-resolution, high dynamic range ALMA observations is therefore not only one of the main goals of the EMERGE project but also crucial to correctly interpret the ISM structure leading to the formation of all types of stars. We will characterize the physical properties of these gas substructure in following papers of this series \citep[e.g.,][]{SocciPaperIII} as well as compare these results with similar ongoing ALMA surveys \citep[e.g.,][]{sanhueza2019_ASHES, Anderson2021,barnes_2021_ALMAIRDC,motte2022_ALMAIMF, Liu2023_QUARKS}.

\section{Conclusions}

The EMERGE project aims to describe how high-mass stars and clusters originate as part of the complex interactions in dense filamentary (fiber-like) networks. This work (Paper I) introduces the EMERGE Early ALMA Survey and a series of novel methodologies for data reduction and exploration of ALMA observations. Accompanying papers will investigate the effects of filtering and data combination on similar ALMA observations \citep[][Paper II]{BonanomiPaperII} as well as characterize the physical properties of the dense fibers found in these regions \citep[][Paper III]{SocciPaperIII}. The main results of this paper can be summarized as follow:

\begin{enumerate}
    \item As part of our EMERGE Early ALMA Survey we systematically investigated 7 star-forming regions part of the Orion A and B clouds, namely OMC-1, OMC-2, OMC-3, OMC-4 South, LDN~1641N, NGC~2023, and Flame Nebula (Sect.\ref{sec:EMERGE_sample}). We homogeneously surveyed this sample combining large-scale single-dish (IRAM-30m; 30") observations together with dedicated interferometric mosaics (ALMA-12m array, Cycles 3+7; 4\farcs5) maps in both (3mm-)continuum and density selective molecular lines (N$_2$H$^+$ (1-0), HNC (1-0), and HC$_3$N (10-9)) in Band~3. We complemented these data with additional archival FIR observations and IR surveys (Sect.\ref{sec:Observations}). 

    \item We developed an optimized framework for the massive and automatic data reduction of our ALMA observations (Sect.\ref{sec:obs_datareduction}). This includes the paralellization of the calibration, imaging, and combination process of single-dish and interferometric (ALMA+IRAM-30m) observations carried out in high-performance computers.

    \item The analysis of the large-scale properties of our targets demonstrates the large variety of physical conditions sampled by our survey including low- (OMC-4 South, NGC~2023), intermediate- (OMC-2, OMC-3, LDN~1641N) and high-mass (OMC-1, Flame Nebula) star-forming regions covering a wide range of surface density of star formation, gas column densities, fractions of dense gas, temperatures, and evolutionary stages (Sect.\ref{sec:EMERGE_globalprop}). 

    \item Our suite of selective molecular tracers sample distinct gas regimes in our sample. N$_2$H$^+$ (1-0) highlights the densest and coldest gas in our targets at column densities $\gtrsim 20\times 10^{21}$~cm$^{-2}$. HNC (1-0) efficiently traces the cloud material at low- and intermediate densities down to $\sim 5\times 10^{21}$~cm$^{-2}$. On the other hand, HC$_3$N  (10-9) is connected to lukewarm temperatures in regions exposed to feedback (Sect.\ref{sec:EMERGE_emissionprop}). Among the three tracers, N$_2$H$^+$ {\bf (1-0)} appears as the best descriptor of the star-forming gas leading to the formation of protostars at constant efficiency  (Sect.\ref{sec:EMERGE_evolstate})
    and within timescales of $\sim$~1~Myr (Sect.\ref{sec:EMERGE_timescales}). 

    \item When observed in the (low-resolution) {\it Herschel} column density maps (36") our targets appear clumpy and filamentary. Similar structures are recognized in the N$_2$H$^+$ and HNC (1-0), IRAM-30m maps ($\sim$30" or $\sim$~12\ 000 au) showing total intensities that positively correlate with the total gas column density and kinetic temperatures of these regions (Sect.\ref{sec:EMERGE_emissionprop}).
    
    \item In contrast, our (high-resolution) ALMA+IRAM-30m observations (4\farcs5 or $\sim$~2000 au) provide a new perspective of the gas structure within these star-forming regions (Sect.\ref{sec:EMERGE_ALMAprop}). At the enhanced interferometric resolution the gas organization in all our targets exhibits an increasing level of complexity down to the beamsize of ALMA. Dozens of small-scale, dense fibers can be recognized in the dense gas traced by their N$_2$H$^+$ (1-0) emission similar to those previously identified in OMC-1 and OMC-2. Additional filamentary features are seen in more diffuse material traced by the HNC emission.
    
    \item The gas organization observed in our high-resolution ALMA maps suggest the common presence of dense fibers in all targets in our sample of our sample irrespective of their gas content (mass of diffuse vs dense gas), stellar density (isolated vs clustered), star-formation activity (low- vs high-mass), and evolutionary state (young vs evolved). The properties of these fibers will be fully characterized in other papers of this series (e.g., see Paper III). These results highlight the hierarchical nature of the ISM showing filaments down to $\sim$2000~au resolution.
    
\end{enumerate}

\begin{acknowledgements}
      This project has received funding from the European Research Council (ERC) under the European Union’s Horizon 2020 research and innovation programme (Grant agreement No. 851435).
      M.T. acknowledges partial support from project PID2019-108765GB-I00 funded by MCIN/AEI/10.13039/501100011033.
      D.H. is supported by Center for Informatics and Computation in Astronomy (CICA) grant and grant number 110J0353I9 from the Ministry of Education of Taiwan. D.H. acknowledges support from the National Technology and Science Council of Taiwan through grant number 111B3005191.
      J.R.G. thanks the Spanish MCIU for funding support under grant PID2019-106110GB-I00.
      This paper makes use of the following ALMA data: ADS/JAO.ALMA\#2019.1.00641.S., ADS/JAO.ALMA\#2013.1.00662.S. ALMA is a partnership of ESO (representing its member states), NSF (USA) and NINS (Japan), together with NRC (Canada), MOST and ASIAA (Taiwan), and KASI (Republic of Korea), in cooperation with the Republic of Chile. The Joint ALMA Observatory is operated by ESO, AUI/NRAO and NAOJ.
      This work is based on IRAM-30m telescope observations carried out under project numbers 032-13, 120-20, 060-22, and 133-22. IRAM is supported by INSU/CNRS (France), MPG (Germany), and IGN (Spain). 
      This research has made use of the SIMBAD database, operated at CDS, Strasbourg, France.
      This research has made use of NASA’s Astrophysics Data System.
      This work was carried out on the Dutch national e-infrastructure with the support of SURF Cooperative.
      The authors acknowledge assistance from Allegro, the European ALMA Regional Center node in the Netherlands.
\end{acknowledgements}
\bibliographystyle{aa}
\bibliography{EMREGE_PaperI.bib}

\begin{appendix}
\section{Assessing data quality in interferometric observations}\label{appendix:QA}

\subsection{On the effects of the short-spacing information}\label{appendix:QA_maps}

\begin{figure*}
\centering
\includegraphics[width=0.9\textwidth]{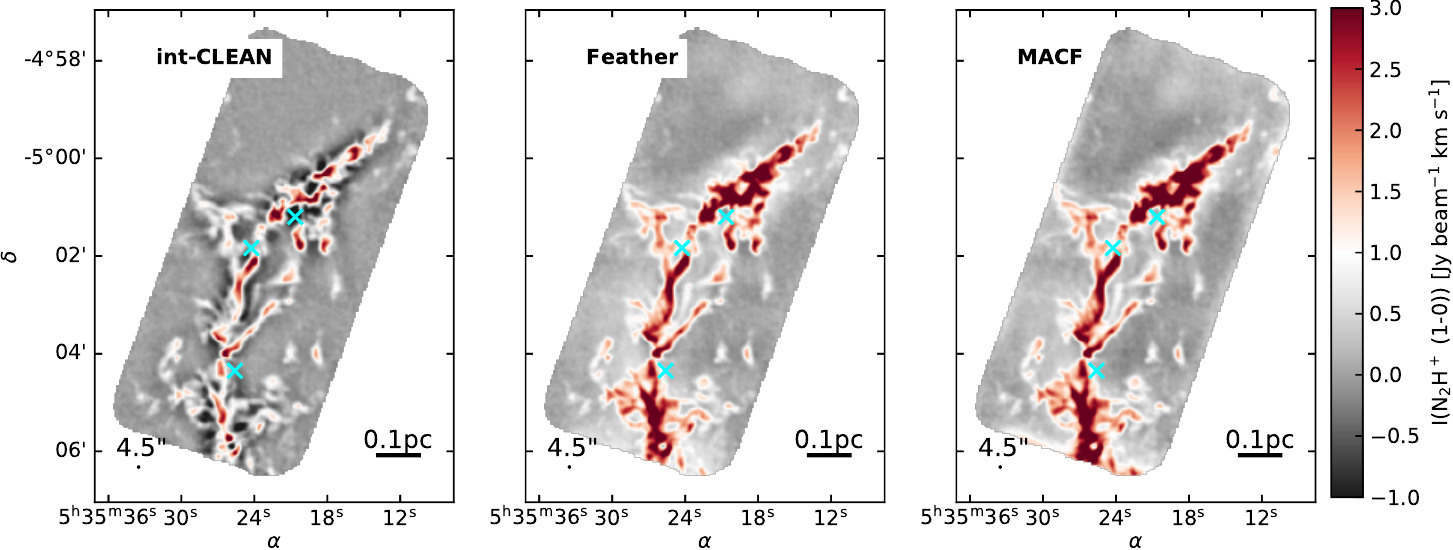}\\
\includegraphics[width=0.9\textwidth]{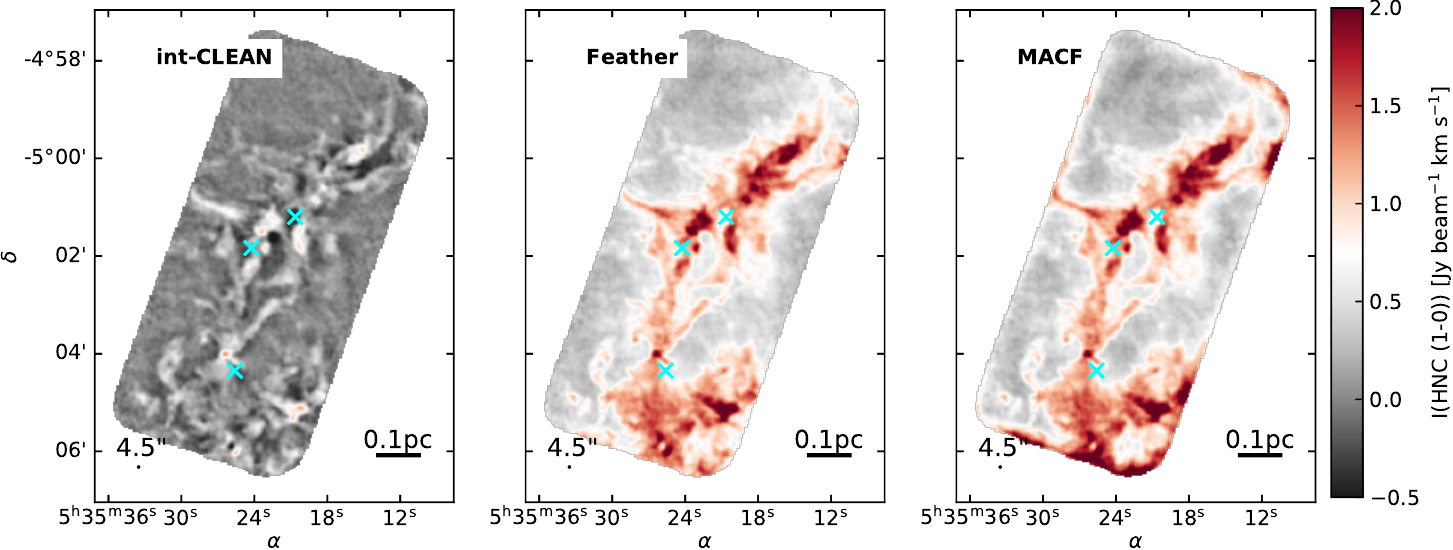}\\
\includegraphics[width=0.9\textwidth]{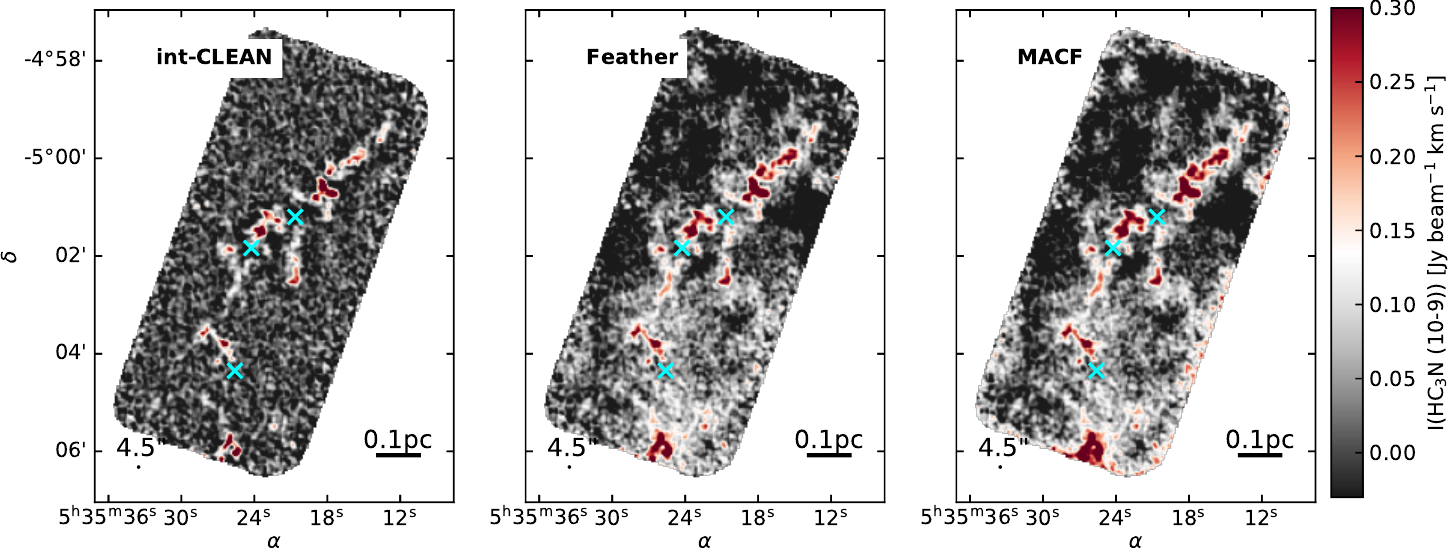}
      \caption{Effects of the different data combination methods in OMC-3: {\bf(Top)} N$_2$H$^+$~(1-0), {\bf(Middle)} HNC (1-0), and {\bf(Bottom)} HC$_3$N (10-9) integrated intensity maps. Different columns show the results for the
      {\bf(Left)} interferometric-alone (int-CLEAN),
      {\bf(Centre)} feather (Feather),
      and {\bf(Right)} Model Assisted Cleaning plus Feather (MACF) data combination methods, respectively.
      To facilitate their comparison all integrated intensity maps of the same molecule are displayed with the same intensity scale (see colorbars). The location of the spectra shown in Fig.~\ref{fig:OMC3_spectra} are indicated by cyan crosses in all maps.
      Compared to our standard PB cut above 0.5 (Sect.\ref{sec:ALMA_calibration}), these maps are shown down to PB values of 0.2 in order to identify potential issues at their edges (e.g., HNC MACF map). We remark here that these edges issues do not affect our scientific results as they are removed when our stricter PB cut is applied. Similar features can be identified in all the ALMA fields observed in these three tracers shown in Figs.~\ref{fig:OMC4_DCs}-\ref{fig:Flame_Nebula_DCs}.
              }
\label{fig:OMC3_DCs}
\end{figure*}

As recognized in multiple studies in the literature \citep[see][and references therein]{Leroy2021_PHANGS,Plunkett2023}, the lack of short-spacing information fundamentally impacts the recovery of extended emission in interferometric observations.
We illustrate the severity of these issues in real observations in Fig.~\ref{fig:OMC3_DCs} by comparing different high spectral resolution N$_2$H$^+$~(1-0), HNC (1-0), and HC$_3$N (10-9) maps in OMC-3 obtained using the int-CLEAN, Feather, and MACF data combination methods (see descriptions in Sect.\ref{sec:ALMA_calibration}). 
Similar comparisons for all the other targets observed in these tracers can be found in Figs.~\ref{fig:OMC4_DCs}-\ref{fig:Flame_Nebula_DCs}.
An eye inspection of these integrated intensity maps reveal how interferometric-alone (int-CLEAN; left panels) reductions misses most of the line emission of all our tracers. These filtering effects are particularly prominent in the case of molecules with extended and bright emission, such as N$_2$H$^+$ and HNC (top and bottom panels), easily recognized by the presence of prominent negative sidelobes surrounding large emission features. Less obvious, filtering also affects more compact emission features such as those seen in HC$_3$N (bottom panels) down to the ALMA resolution. Similar issues are seen in all targets of our sample.

The addition of the short-spacing information significantly changes the amount of recovered emission in all our spectral maps. Both Feather (central planels) and MACF (right panels) reductions reveal significant levels of extended emission detected in bright tracers such as N$_2$H$^+$ (top panels). Yet, data combination improves the flux recovery of both compact and extended emission features seen. The most dramatic improvements are seen however in our HNC maps (middle panels), the most diffuse tracer in our sample (see Sect.\ref{sec:EMERGE_emissionprop}). There, Feather and MACF manage to recover large amounts of extended emission systematically filtered our by the interferometer. Little or no sign of remaining negative sidelobes are observed in these combined maps.

\begin{figure}
\centering
\includegraphics[width=1\linewidth]{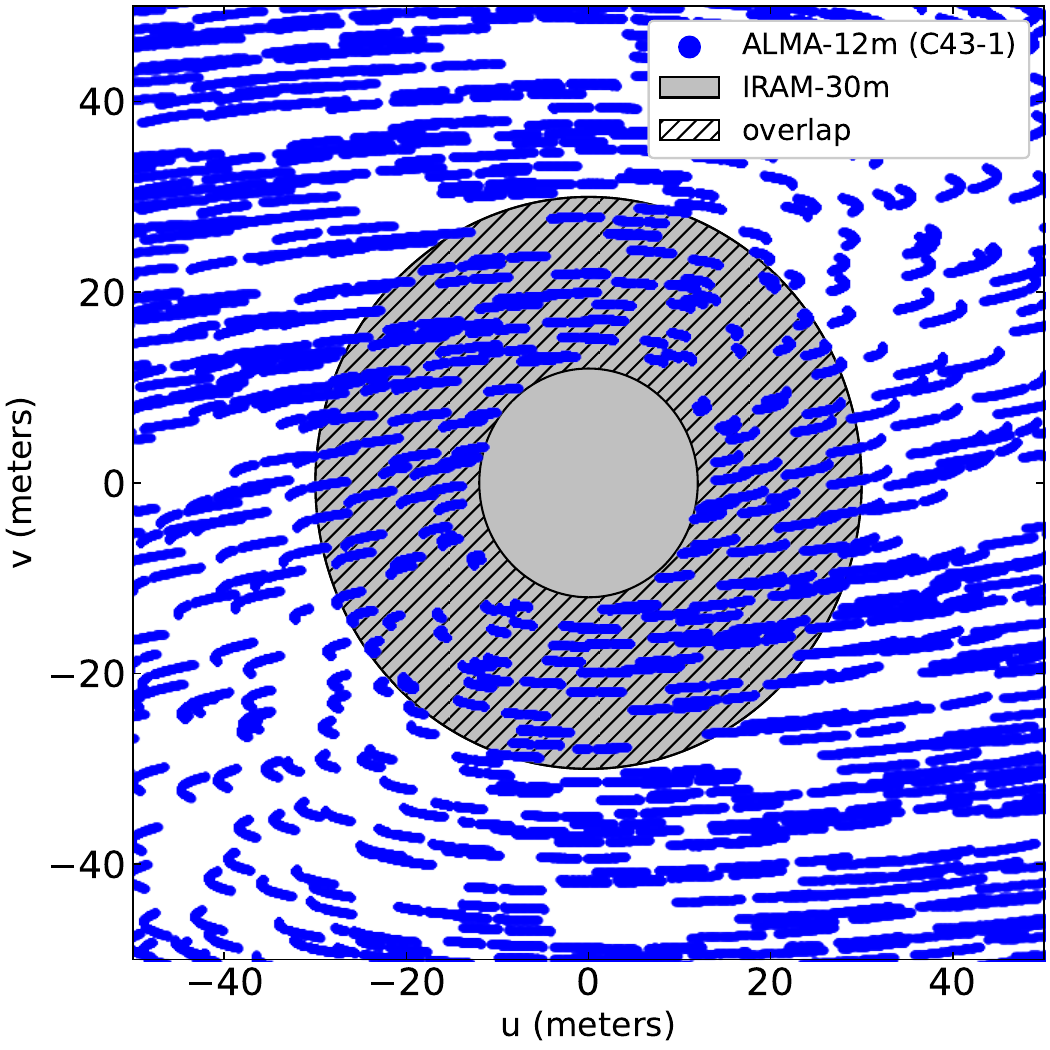}
      \caption{ALMA-12m array uv-coverage (blue), with effective baselines between $\sim$~12 and 300 meters (C43-1), obtained in our OMC-3 observations. We highlight the uv-area sampled by the IRAM-30m data ($\le$~30~m; grey) as well as its overlap with the ALMA baselines (hatched area). Note that we display only those baselines at u,v-distances $\le$~50~meters. 
              }
\label{fig:OMC3_uvplot}
\end{figure}

The above improvement on the recovery of extended emission after data combination can be understood from the uv-coverage in our ALMA+IRAM-30m observations. In Figure~\ref{fig:OMC3_uvplot} we show the uv-sampling of our ALMA-12m data (blue) in OMC-3 within the inner 50~meters of the uv-plane. The large number of baselines in the ALMA-12m array produces a densely sampled coverage of uv-plane above (projected) distances of $\ge$~12~meters. On the other hand, the inner gap in the uv-plane is uniformly sampled by our IRAM-30m data (grey area) up to 30~meters and with a sufficient overlap with the ALMA-12m array in C43-1 configuration (hatched area). While ALMA provides unprecedented sensitivity at large baselines ($\sim$12-300~meters in our case), the inclusion of the IRAM-30m data as short-spacing information becomes essential for recovering the extended emission in these targets.
As demonstrated in Paper II \citep[][]{BonanomiPaperII}, the homogeneous uv-sampling provided by a large SD such as IRAM-30m produces a significant improvement on the image quality during the study of the filamentary ISM 
compared to the more sparse coverage obtained using equivalent ALMA Compact Array (ACA; sampling intermediate uv-distances) plus total power (TP; sampling the innermost baselines) observations.

\subsection{Statistical results}\label{appendix:QA_results}

\begin{table*}
\caption{Quality assessment results}             
\label{table:QA_results}      
\centering          
\begin{tabular}{l | c | c c | c c | c c}     
\hline\hline       

              &         &  \multicolumn{2}{c}{OMC-3} &  \multicolumn{2}{c}{Flame Nebula} &  \multicolumn{2}{c}{NGC~2023} \\
Transition    & Method     &  A-par$^{(\star)}$ & FR$^{(\star\star)}$  &  A-par & FR   &  A-par & FR \\ 

\hline                 
N$_2$H$^+$ (1-0)    & int-CLEAN    & -0.79$\pm$0.04    & 18.8\% & -0.72$\pm$0.06    & 35.3\% & -0.90$\pm$0.02    & 4.7\% \\
 (narrow)    & Feather      & -0.02$\pm$0.05    & 90.6\% & -0.13$\pm$0.05    & 84.9\% & -0.07$\pm$0.01    & 91.4\% \\
             & MACF         & -0.00$\pm$0.03    & 99.5\% & -0.08$\pm$0.03    & 91.7\% & -0.02$\pm$0.01    & 97.0\% \\

\hline
HNC (1-0)    & int-CLEAN    & -0.94$\pm$0.01    & 6.3\% & -0.94$\pm$0.01    & 8.6\% & -0.97$\pm$0.02    & 4.5\% \\
(narrow)     & Feather      & -0.01$\pm$0.03    & 94.6\% & -0.03$\pm$0.03    & 95.7\% & -0.05$\pm$0.04    & 93.6\% \\
             & MACF         & -0.03$\pm$0.01    & 99.3\% & 0.00$\pm$0.02    & 99.1\% & -0.01$\pm$0.03    & 98.1\% \\
\hline
HC$_3$N (10-9) & int-CLEAN    & -0.69$\pm$0.03    & 32.9\% & -0.59$\pm$0.01    & 43.7\% & -0.72$\pm$0.04    & 32.7\% \\
 (narrow)      & Feather      & -0.16$\pm$0.01    & 83.5\% & -0.04$\pm$0.03   & 91.6\% & -0.16$\pm$0.03    & 80.6\% \\
             & MACF           & -0.07$\pm$0.01    & 92.9\% & 0.00$\pm$0.02    & 98.1\% & -0.09$\pm$0.02    & 87.1\% \\
\hline 
\hline
HNC (1-0)    & int-CLEAN    & -0.94$\pm$0.01    & 6.3\% & -0.94$\pm$0.01    & 8.6\% & -0.97$\pm$0.01    & 4.4\% \\
(broad)      & Feather      & -0.02$\pm$0.07    & 93.0\% & -0.02$\pm$0.03    & 94.3\% & -0.03$\pm$0.02    & 93.6\% \\
             & MACF         & -0.07$\pm$0.06    & 84.9\% & 0.02$\pm$0.02    & 89.9\% & 0.01$\pm$0.02    & 88.5\% \\
\hline      
HC$_3$N (10-9) & int-CLEAN    & -0.79$\pm$0.03    & 24.0\% & -0.59$\pm$0.05    & 43.6\% & -0.67$\pm$0.04    & 38.1\% \\
(broad)      & Feather      & -0.27$\pm$0.04    & 72.2\% & -0.09$\pm$0.03    & 86.4\% & -0.18$\pm$0.02    & 81.0\% \\
             & MACF         & -0.20$\pm$0.02    & 75.6\% & -0.04$\pm$0.04    & 92.4\% & -0.11$\pm$0.02    & 90.5\% \\
\hline                  
\end{tabular}
\tablefoot{($\star$) Mean $\pm$ standard deviation A-par values. ($\star\star$) Percentage of the total flux recovered (FR) respect to the one detected in our single-dish data. We note that most of the best results are usually produced  by the MACF method.
}
\end{table*}

Beyond the visual comparisons in maps and spectra (see above), we performed a quantitative analysis of the QA of our molecular datasets implementing the assessment metrics presented in \citet{Plunkett2023}. In particular, we focused our assessments on the analysis of the Accuracy parameter (A-par):
\begin{equation} \label{eq:Anu}
    A_\nu(x,y)=\frac{I_\nu(x,y)-R_\nu(x,y)}{|R_\nu(x,y)|}.
\end{equation}
Intuitively, A-par measures the relative error (in percentage) between a reference ($R_\nu(x,y)$) and a target ($I_\nu(x,y)$) image while its sign indicates whether the flux is overestimated ($A_\nu(x,y)>0$) or underestimated ($A_\nu(x,y)<0$). A-par can be evaluated in space (x,y) and velocity ($\nu$) and applied to maps and cubes alike within a dedicated assessment region (see Plunkett et al for a full discussion). 

In the case of our molecular observations, we adopt our IRAM-30m datasets as reference and evaluate the performance of the target int-CLEAN, Feather, and MACF reductions. For that, we convolved and regridded each combined map ($\theta=$4\farcs5 resolution) into the single-dish frame ($\theta=$30") and applied an assessment threshold of approximately five times the noise level in our ALMA maps. From the statistical analysis of $A_\nu(x,y)$ we then aim to quantitatively measure the amount of flux recovered by our data combinations in comparison with those found in the single-dish data, assumed to represent the true sky flux distribution at least at the single-dish resolution. Reduced with the same deconvolution parameters, our QA allows us to systematically compare the goodness of each data reduction method applied in our study (Sect.~\ref{sec:ALMA_calibration}).

We summarize the results of our QA in Table~\ref{table:QA_results} in three representative regions in our sample selected by presenting different line intensities and source structure: OMC-3 (bright and extended), Flame Nebula (compact and bright), and NGC~2023 (weak and extended). We display the values of A-par statistics (mean and standard deviation) obtained from the analysis of all voxels within the assessment area as well as the fraction of total flux recovered (FR) in the final combined datacubes of our three lines of interest (N$_2$H$^+$, HNC, and HC$_3$N). We display the results for both high- (top three rows; narrow) and low- (bottom two rows; broad) spectral resolution.

Our QAs confirm the trends suggested during the visual inspection of our maps (Fig.~\ref{fig:OMC3_DCs}). 
All interferometric-alone reductions (int-CLEAN) are subject of widespread filtering missing $>$70\% of the flux both per pixel (A$<$-0.70) and in absolute (FR$\lesssim$40\%) terms. On the contrary, data combination techniques (Feather and MACF) exhibit an significant improved performance combining an enhanced of the local accuracy (|A|$\lesssim$0.10) and flux recovery (FR$\gtrsim$85\%) in most targets and lines. Data combination is particularly relevant in regions with bright, extended emission (e.g., OMC-3 and Flame Nebula).  Nonetheless, data combination shows some limitations in targets with either sharp features (e.g. HC$_3$N in OMC-3) or weak extended (e.g., HC$_3$N in NGC~2023) emission, probably connected to the quality of our single-dish data.
Regardless of these minor issues, our QAs statistically quantify the great impact of advanced data combination techniques during the reduction of ALMA observations \citep[see also ][]{Plunkett2023}. 

The differences between them Feather and MACF are more subtle.
The results in Table~\ref{table:QA_results} indicate some systematic differences between the Feather and MACF results. Overall, MACF produces the most stable results per pixel with A-par values close to |A|$\lesssim$0.05 and a better total flux recovery (FR$>92\%$). 
In some cases, MACF appears to produce noisier edges, although this effect is easily mitigated by a sensible PB cut after combination (Sect.~\ref{sec:ALMA_calibration}).
Based on these results, and for consistency with the analysis in OMC-1 and OMC-2 \citep{Hacar2018}, we have adopted the MACF method as our standard reduction.
Yet, we remark that in specific cases Feather may produce similar or even slightly better (by few percent) results.


While both Feather and MACF produce satisfactory results, we find systematic improvements on their data products when comparing high- (narrow) and low- (broad) spectral resolution reductions (see results of HNC and HC$_3$N in Table~\ref{table:QA_results}). Reductions including single-dish data at high spectral resolutions consistently increase the accuracy (A-par) and flux recovery (FR) of the combined datasets. Resolving the line structure (and its possible filtering in different channels) seems to help on recovering high fidelity images. Further analysis are needed to confirm this trend.

\subsection{Velocity-dependent filtering effects in molecular line observations}\label{appendix:QA_spectra}

\begin{figure*}
\centering
\includegraphics[width=0.8\textwidth]{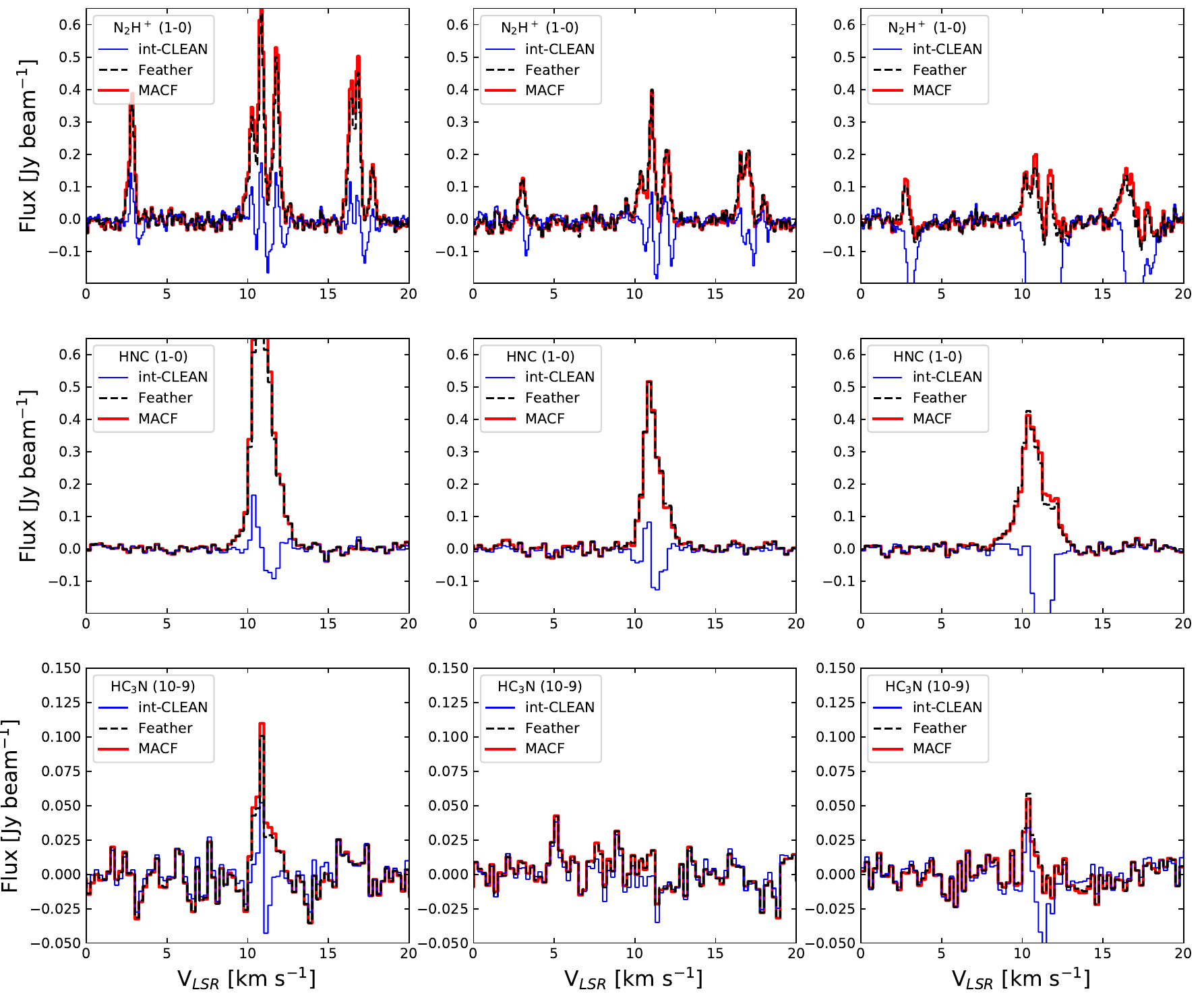}
      \caption{ALMA spectra obtained using int-CLEAN (black), Feather (dashed blue), and MACF (red) data combination methods. We display the (narrow) N$_2$H$^+$~(1-0) {\bf (Top)}, HNC~(1-0) {\bf(Middle)}, and HC$_3$N~(10-9) {\bf(Bottom)} spectra (rows) in three representative positions (columns) within the OMC-3 region (see Fig.~\ref{fig:OMC3_DCs}): {\bf (Left panels)} ($\alpha,\delta$)=(05:35:19.6,-05:00:29.5),
        {\bf (Cental panels)} ($\alpha,\delta$)=(05:35:25.5,-05:04:19.0),
        and {\bf (Right panels)} ($\alpha,\delta$)=(05:35:20.7,-05:01:11.9).
        Note how the filtering effects experienced in the int-CLEAN maps are unpredictable and change depending on position, velocity, and tracer consider in each case.
              }
\label{fig:OMC3_spectra}
\end{figure*}

\citet{Plunkett2023} discussed how the lack of short-spacing information can have a critical impact on the recovered emission both in space and velocity (i.e. spectral cubes). 
We illustrate these effects in Fig.~\ref{fig:OMC3_spectra} by comparing the int-CLEAN (blue), Feather (dashed black), and MACF (red) spectra in three representative positions in our OMC-3 region. As seen in the hyperfine structure of our N$_2$H$^+$ spectra (top panels), filtering can act selectively in velocity changing the emission profile of all our lines independently of their brightness. 

\begin{figure*}
\centering
\includegraphics[width=1.\textwidth]{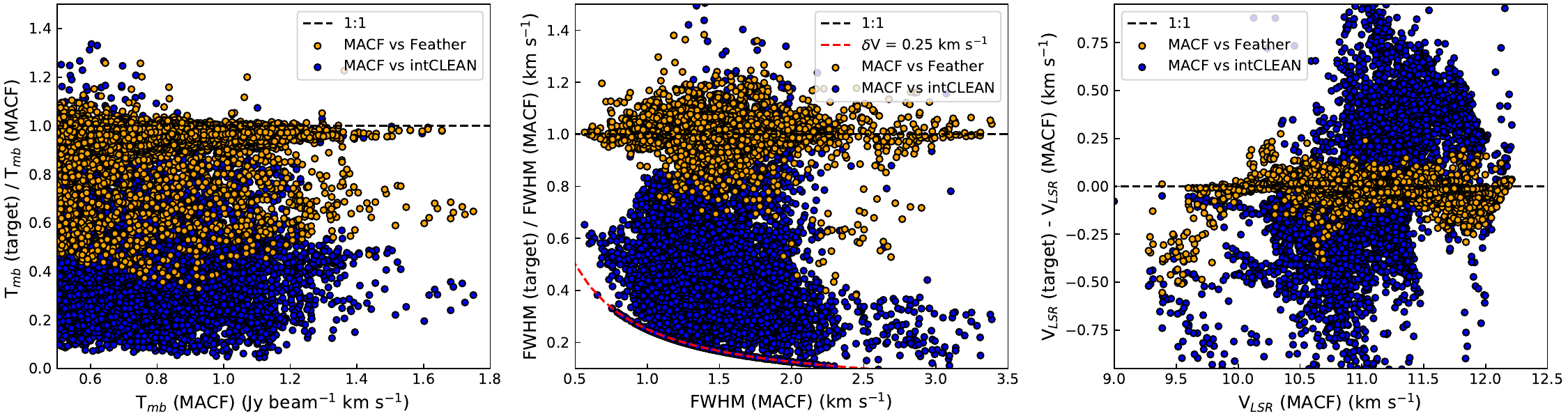}
      \caption{HNC (1-0) line properties derived from the automatic fit of one gaussian component in OMC-3 obtained from our int-CLEAN (blue points) and Feather (yellow points) reductions (targets) in comparisons with MACF. For simplicity, only fits with I(HNC, MACF)$\geq 0.5$~Jy~beam$^{-1}$ and S/N$\geq$5 are shown.
      From left to right: 
      {\bf (Left panel)} Line peak ratio; 
      {\bf (Central panel)} Full-Width-Half-Maximum ratio;
      {\bf (Right panel)} Velocity difference.
      The lower envelope of the central panel is determined by the spectral resolution of our (narrow) reductions with $\delta$V=0.25~km~s$^{-1}$.       }
\label{fig:OMC3_fitresults}
\end{figure*}

Filtering can critically alter all line properties (line flux and peak, centroid and FWHM, hyperfine distribution...) and distort any physical property derived from them. This is clear in many int-CLEAN spectra exhibiting physically unrealistic hyperfine ratios, artificially narrow FWHM, and/or show negative artefacts. 
To explored the effects of filtering on the line parameters we automatically fitted one gaussian component to all HNC (1-0) spectra in OMC-3 obtained by our different data combination methods. In Fig.\ref{fig:OMC3_fitresults} 
we show a pixel-by-pixel comparison of the peak temperature (T$_{peak}$; left panel), Full-Width-Half-Maximum (FWHM; central panel), and line velocity (V$_{LSR}$; right panel) between the int-CLEAN and Feather results with respect to MACF, the latter assumed as reference. As expected from the results in Sect.~\ref{appendix:QA_results}, we find an excellent agreement between the line properties derived in Feather and MACF methods with most variations within the spectral resolution (e.g., see FWHM). The results of the int-CLEAN reduction are however dramatic. Interferometric filtering produces significant changes not only in the line peak (similar to the flux), but also completely corrupts the recovered FWHM and V$_{LSR}$ parameters. As shown from the comparison in velocity (right panel), pure filtering artefacts can alter the entire velocity structure of the cloud producing artificially narrow profiles and systematic drifts across the field. 

Selective filtering effects can be also tracer dependent. The different brightness, velocity distribution, and emission structure of lines and continuum produces differential effects on the recovered emission at the same position. This is seen in Fig.~\ref{fig:OMC3_spectra} when comparing N$_2$H$^+$ (top panels) and HNC (middle panels) spectra in similar positions. The bright and extended HNC distribution across the entire OMC-3 cloud completely suppresses the emission of this molecule in different parts of the cloud more efficiently than in the case of N$_2$H$^+$ (e.g., left middle panel). Other regions can not compensate the negative sidelobes coming from nearby bright spots and produce more prominent negative features (e.g., right middle panel).  We remark here that these effects are also seen in the weaker HC$_3$N spectra, despite this molecule showing a more compact distribution (see bottom panels).

We also notice that large-scale filtering produced by sources with extended and complex emission structure can be highly deceptive. Negative sidelobes can be compensated by bright emission features at similar velocities producing almost flat spectra (e.g. Fig.~\ref{fig:OMC3_spectra}, left middle panel).  Analysis in such spectra could lead into measurements showing artificially low noise values giving the false impression of high quality data reductions. On the contrary, Feather and MACF reductions actually show bright emission features in many of these positions. These effects are apparent in the case of our continuum int-CLEAN maps. There it is clear that most of the continuum emission is filtered out while our maps show a surprisingly low-noise level. 
 
The above maps (Sect.~\ref{appendix:QA_maps}) and spectra (this section) illustrate the perils of using interferometric-alone reductions for the analysis of spatially resolved sources. 
Our results discourage the use of interferometer-alone reductions (int-CLEAN) when characterizing the emission properties of (resolved) physical structures larger than few times the interferometric beamsize. Projects exploring similar targets are strongly encouraged to carefully evaluate the impact of the missing short-spacing information in both maps and spectra. 
In an accompanying paper \citep[][]{BonanomiPaperII} we quantify how the lack of the single-dish information severely corrupts the characterization of the physical properties (masses, sizes, widths, etc...) of cores and filaments in our ALMA observations. Data combination becomes therefore essential for our line emission and kinematic studies in our ALMA sample.

\section{Data products}\label{appendix:DP}

\begin{figure*}
\centering
\includegraphics[width=1.0\textwidth]{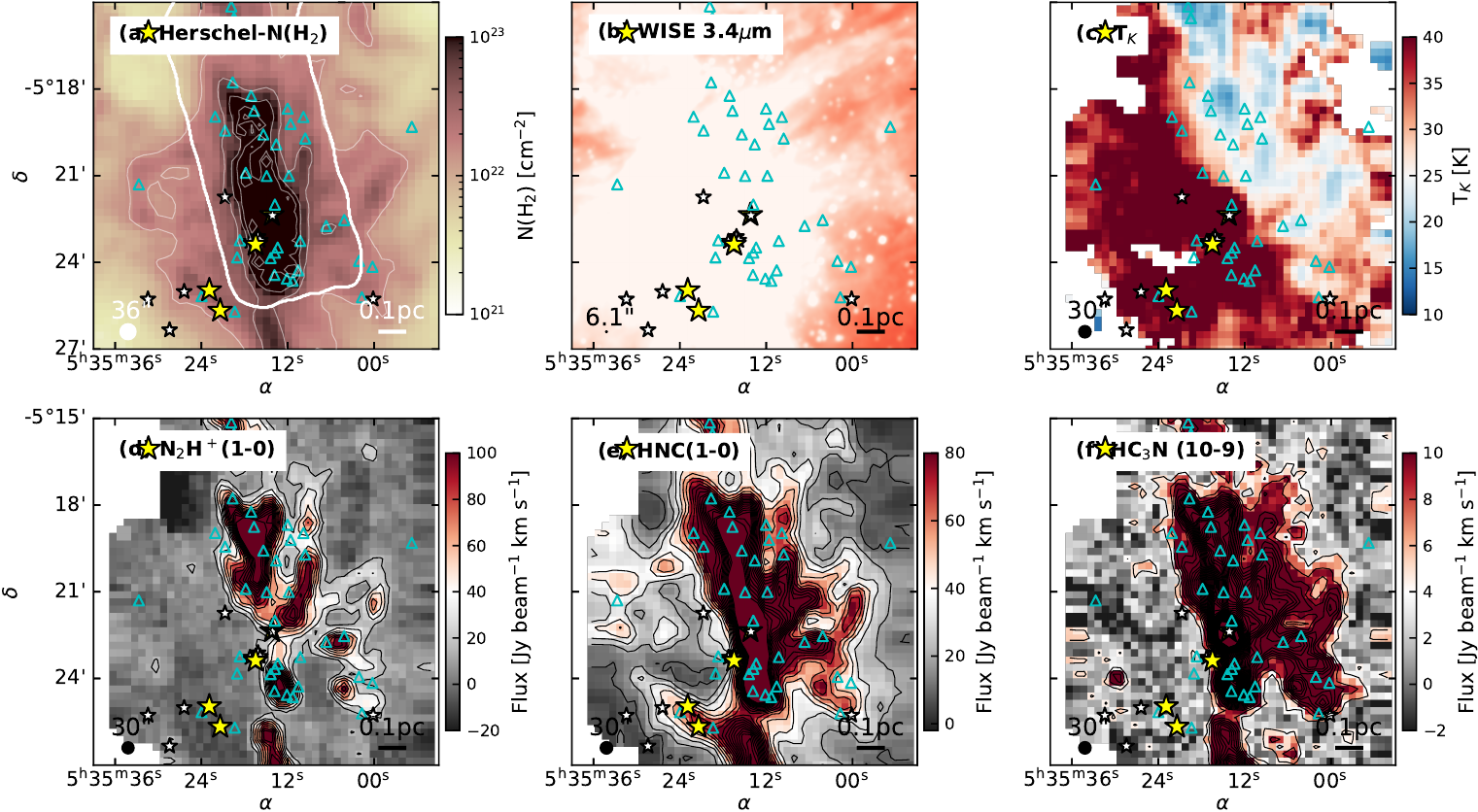}
      \caption{Low-resolution observations in OMC-1 similar to Fig.~\ref{fig:OMC3_IRAM30m}
              }
\label{fig:DP_OMC1_IRAM}
\end{figure*}

\begin{figure*}
\centering
\includegraphics[width=0.8\textwidth]{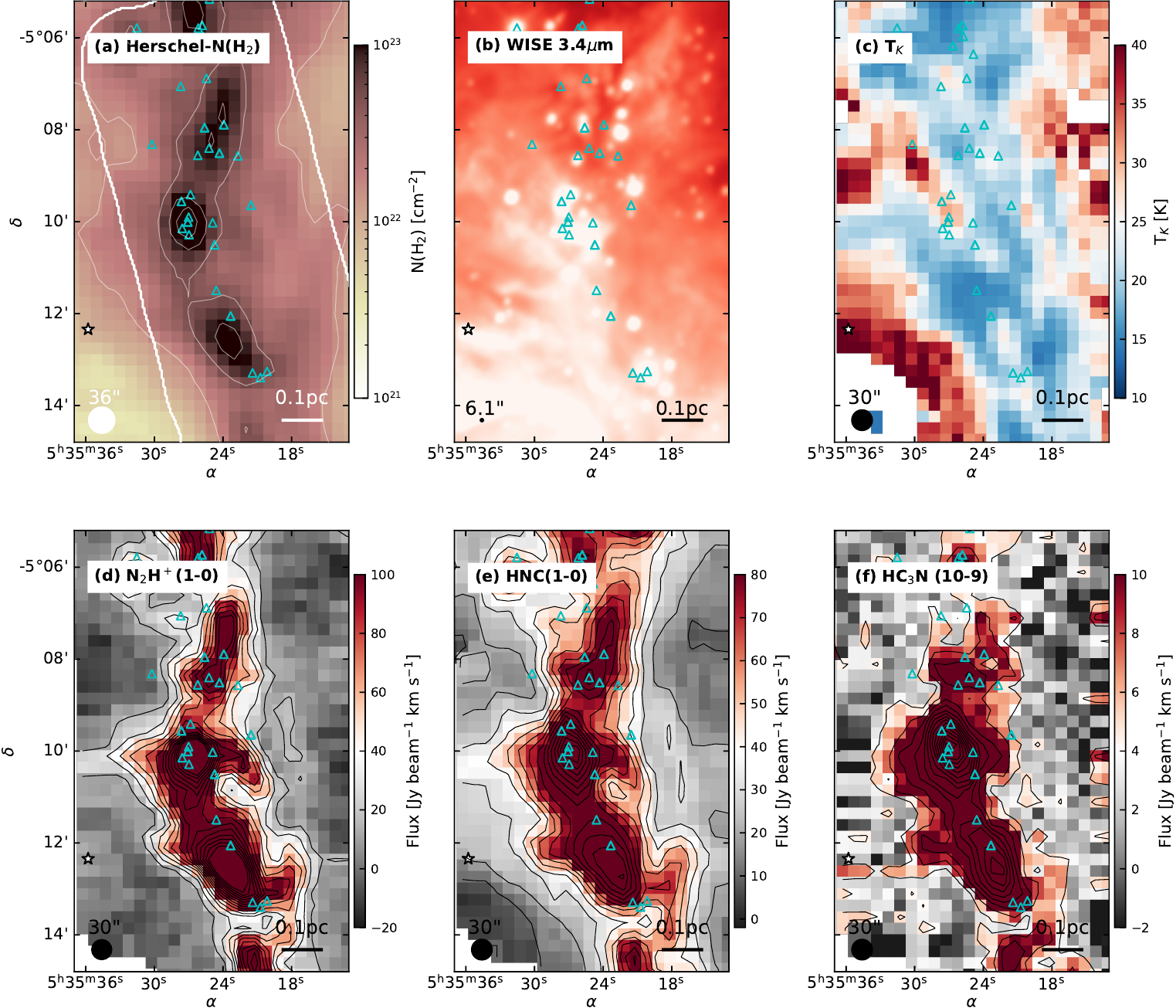}
      \caption{Low-resolution observations in OMC-2 similar to Fig.~\ref{fig:OMC3_IRAM30m}
              }
\label{fig:DP_OMC2_IRAM}
\end{figure*}

\begin{figure*}
\centering
\includegraphics[width=1.0\textwidth]{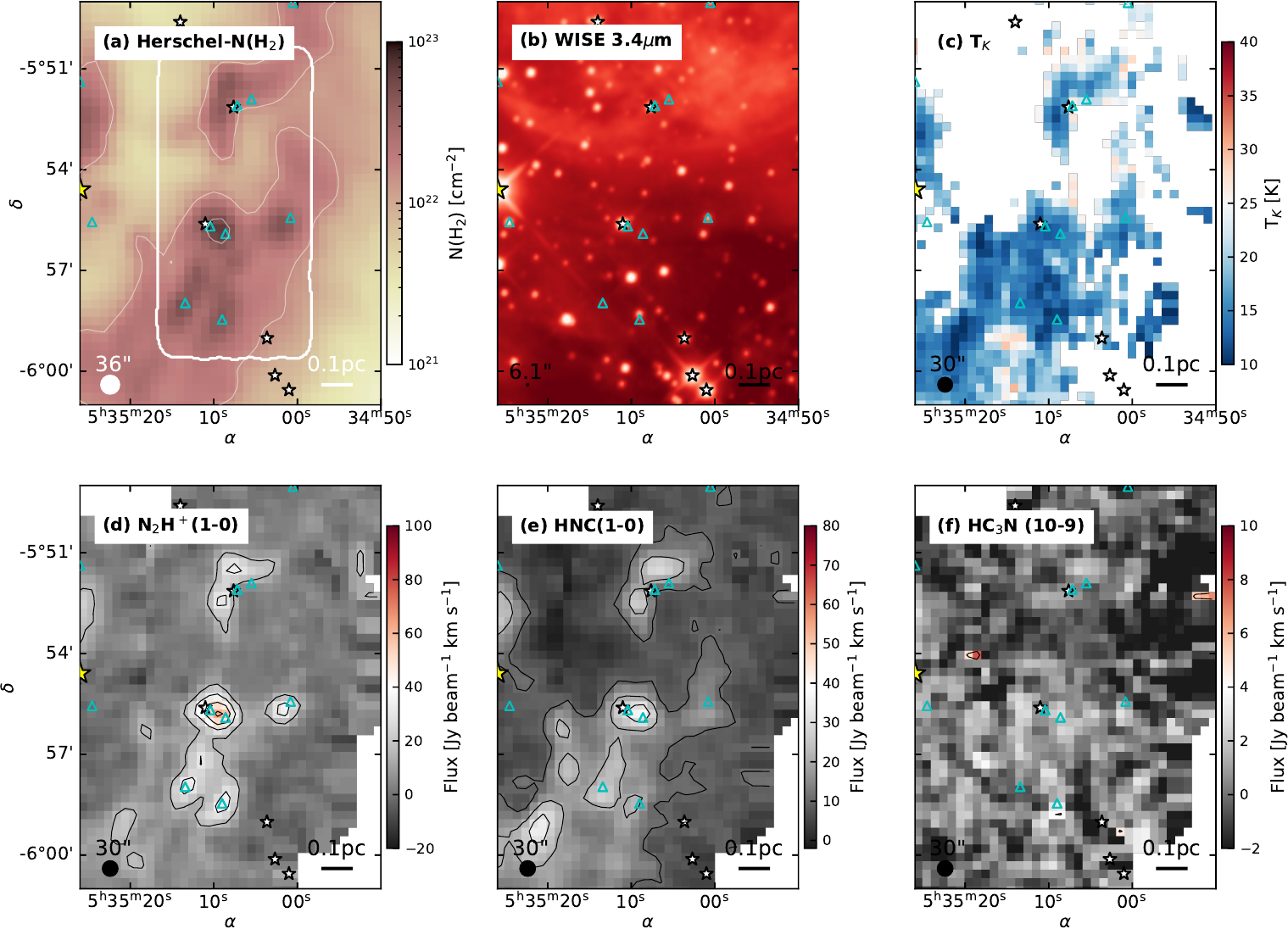}
      \caption{Low-resolution observations in OMC-4~South similar to Fig.~\ref{fig:OMC3_IRAM30m}
              }
\label{fig:DP_OMC4_IRAM}
\end{figure*}

\begin{figure*}
\centering
\includegraphics[width=1.0\textwidth]{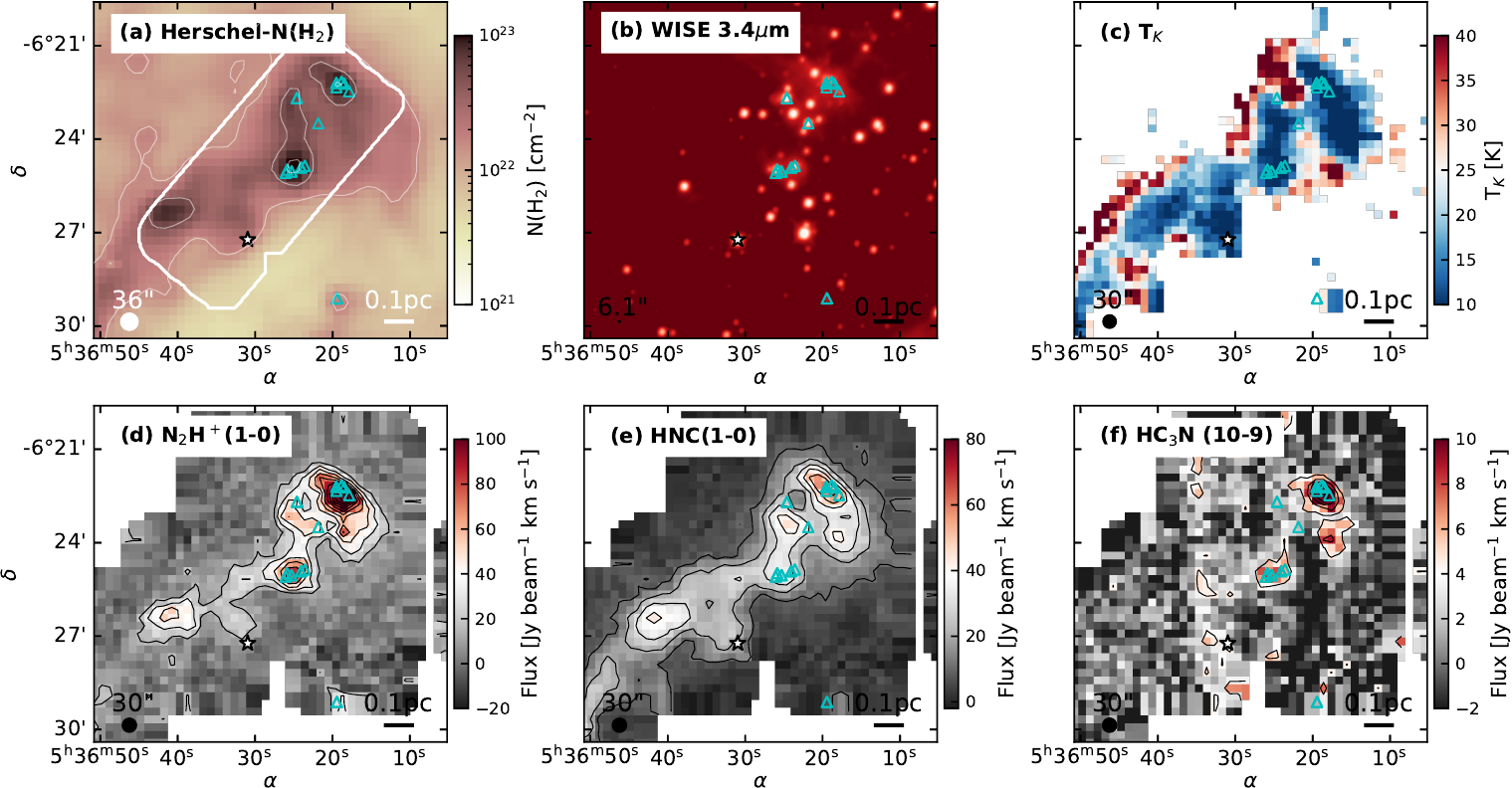}
      \caption{Low-resolution observations in LDN~1641N similar to Fig.~\ref{fig:OMC3_IRAM30m}
              }
\label{fig:DP_LDN1641N_IRAM}
\end{figure*}

\begin{figure*}
\centering
\includegraphics[width=1.0\textwidth]{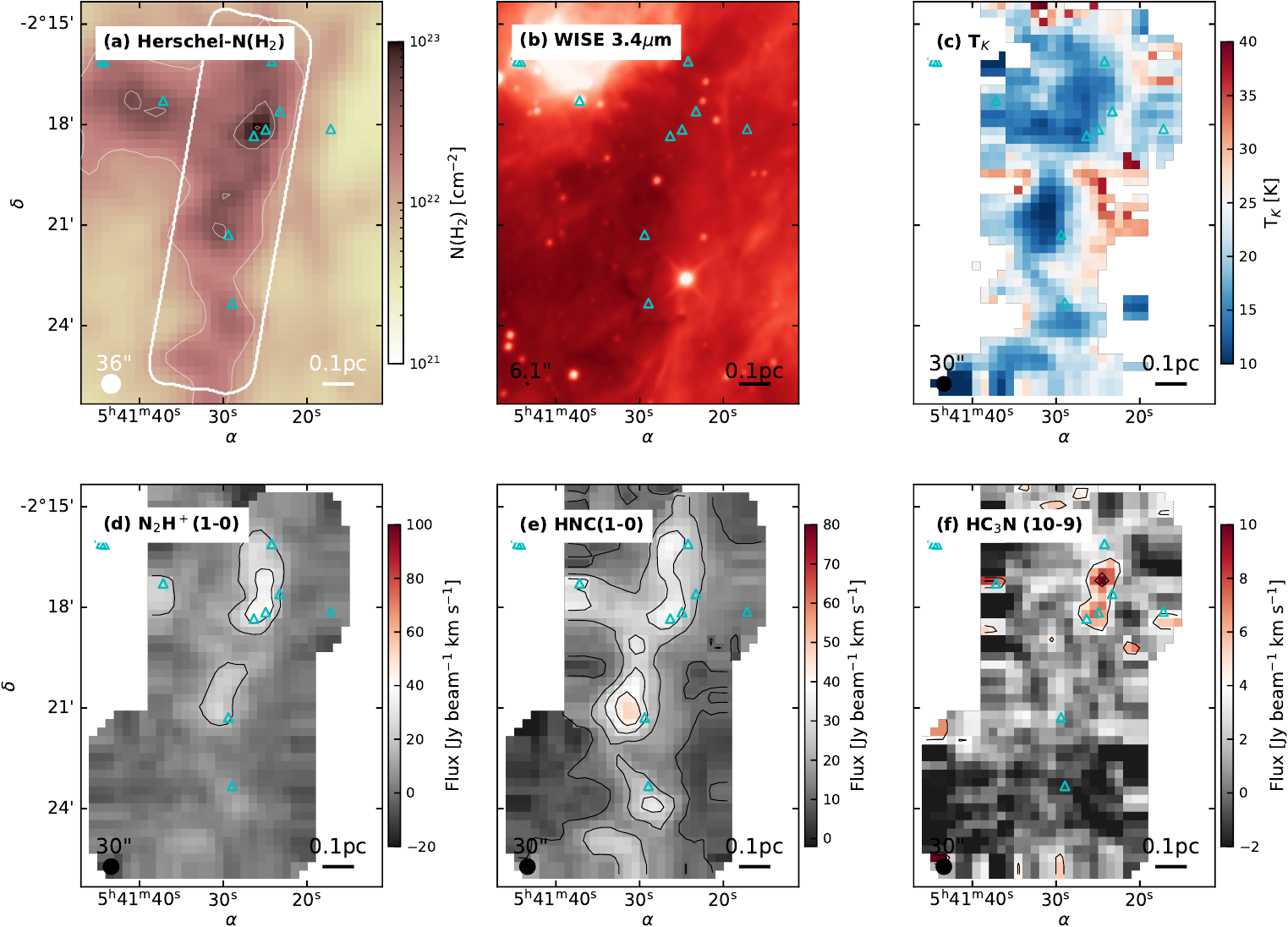}
      \caption{Low-resolution observations in NGC~2023 similar to Fig.~\ref{fig:OMC3_IRAM30m}
              }
\label{fig:DP_NGC2023_IRAM}
\end{figure*}

\begin{figure*}
\centering
\includegraphics[width=0.8\textwidth]{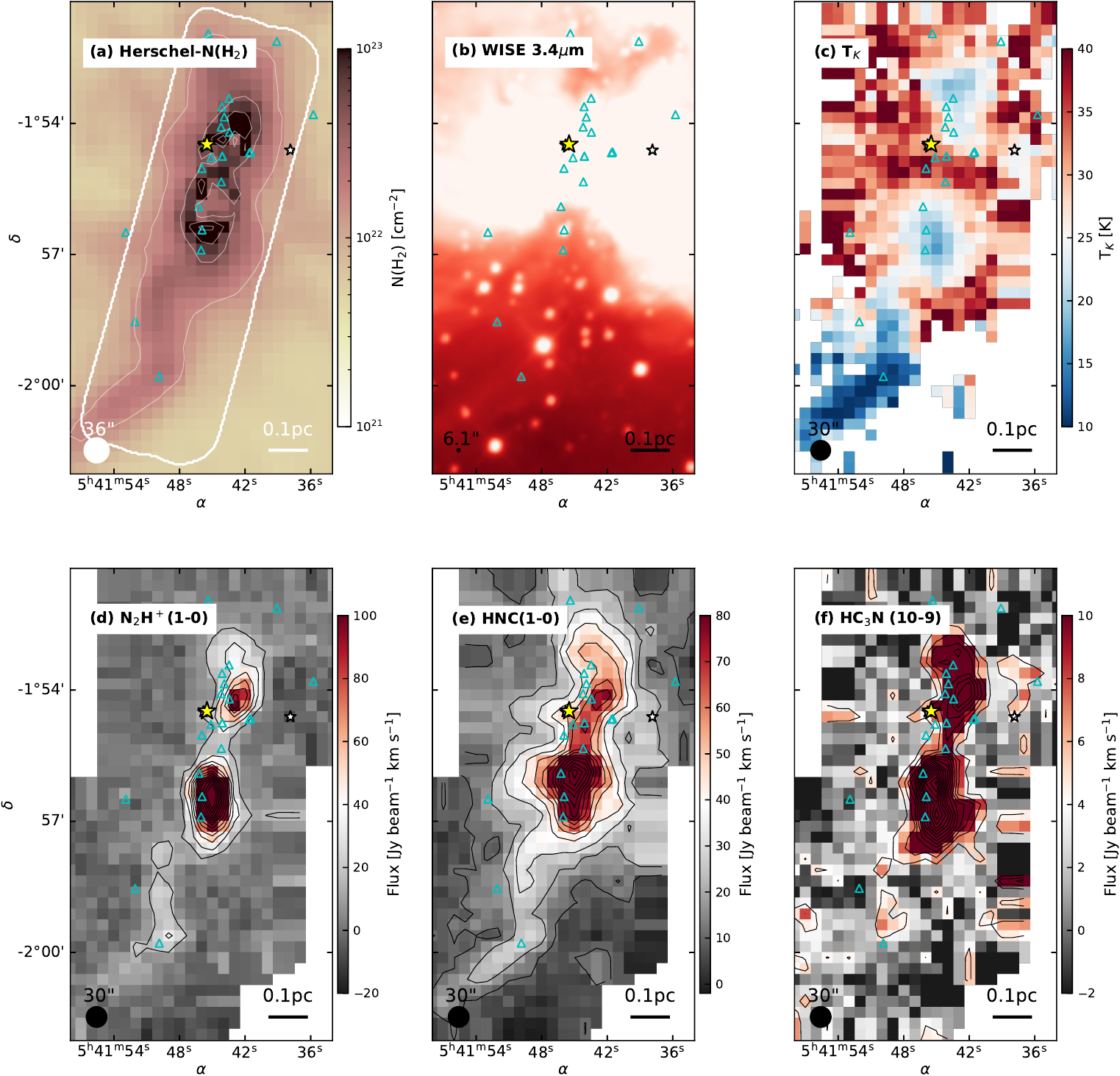}
      \caption{Low-resolution observations in Flame Nebula similar to Fig.~\ref{fig:OMC3_IRAM30m}
              }
\label{fig:DP_FlameNebula_IRAM}
\end{figure*}

\begin{figure*}
\centering
\includegraphics[width=1.0\textwidth]{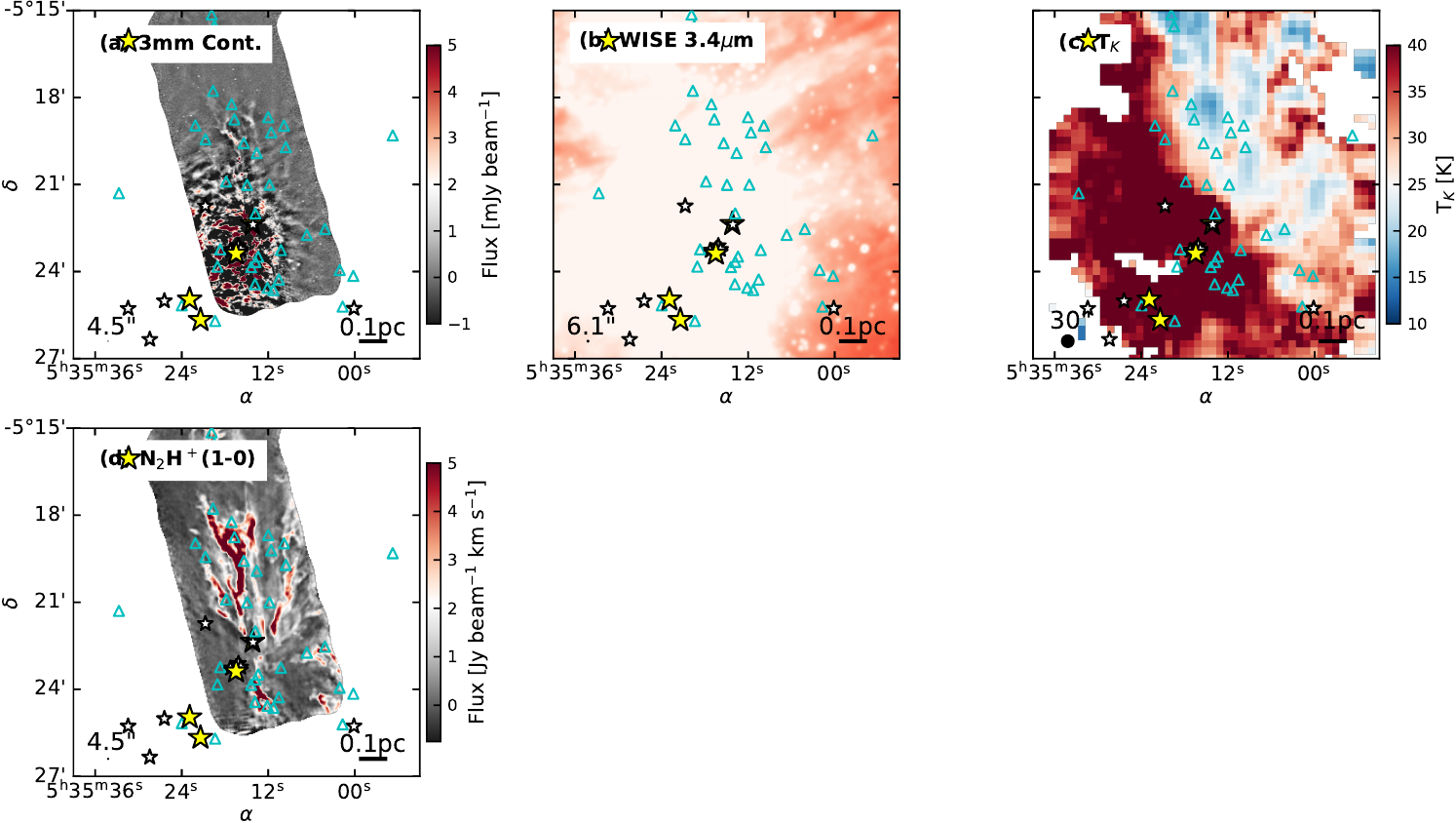}
      \caption{High-resolution observations in OMC-1 similar to Fig.~\ref{fig:OMC3_ALMA+IRAM30m}. Note that no ALMA maps are available in the case of the HNC (1-0) nor HC$_3$N (10-9) lines (see Sect.~\ref{sec:obs_ALMA}).
              }
\label{fig:DP_OMC1_ALMA+IRAM}
\end{figure*}

\begin{figure*}
\centering
\includegraphics[width=1.0\textwidth]{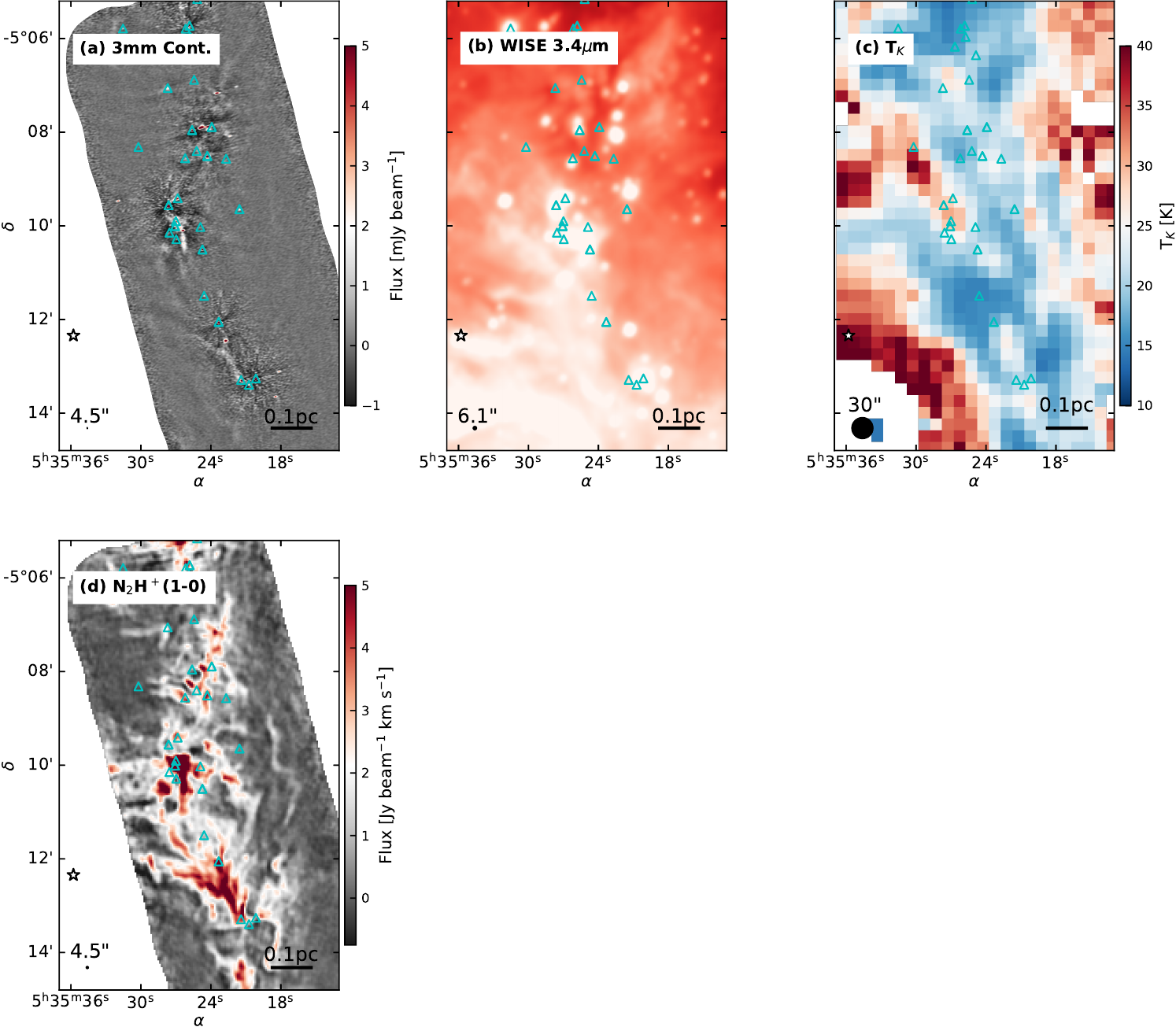}
      \caption{High-resolution observations in OMC-2 similar to Fig.~\ref{fig:OMC3_ALMA+IRAM30m}. Note that no ALMA maps are available in the case of the HNC (1-0) nor HC$_3$N (10-9) lines (see Sect.~\ref{sec:obs_ALMA}).
              }
\label{fig:DP_OMC2_ALMA+IRAM}
\end{figure*}

\begin{figure*}
\centering
\includegraphics[width=1.0\textwidth]{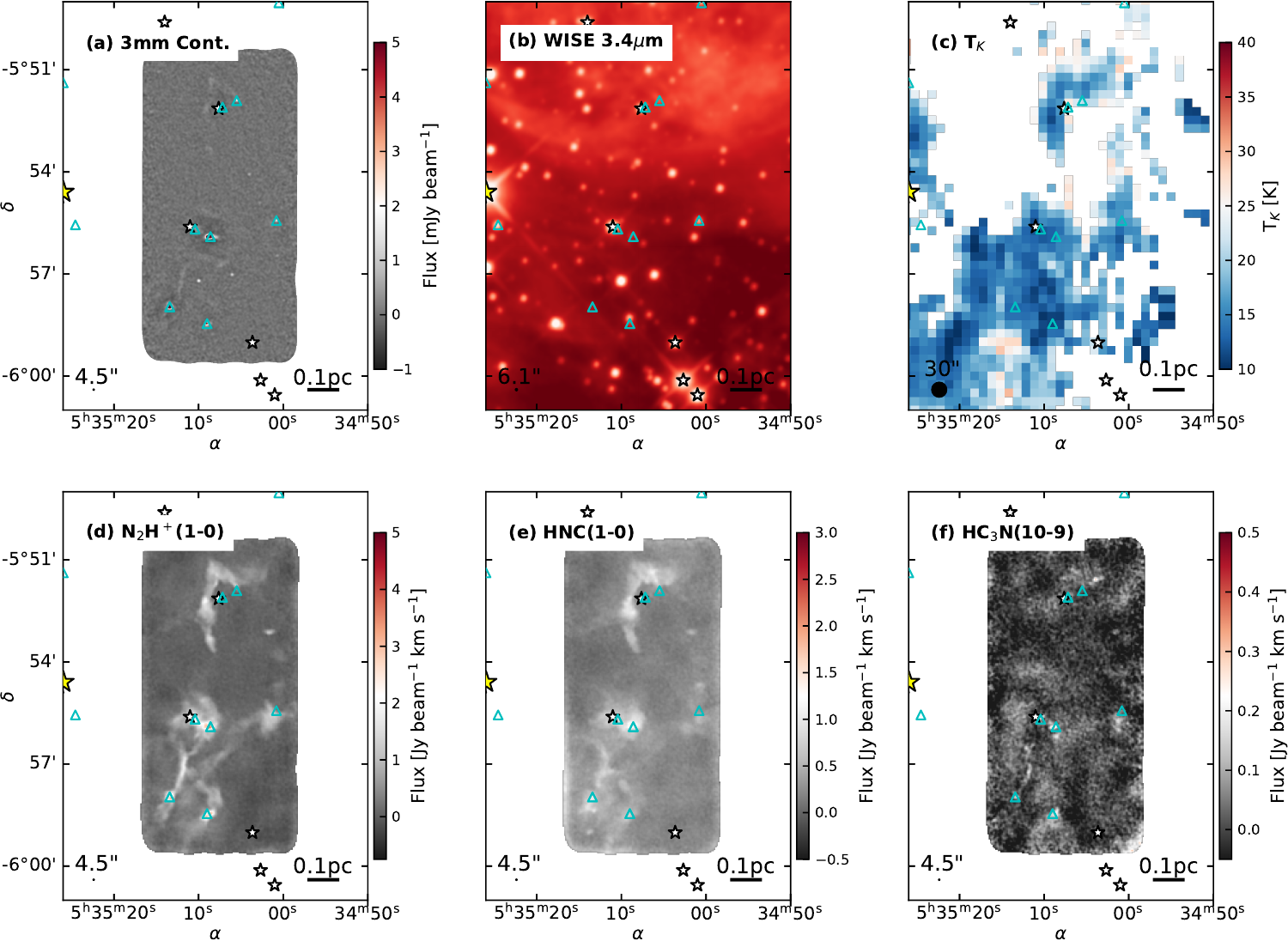}
      \caption{High-resolution observations in OMC-4~South similar to Fig.~\ref{fig:OMC3_ALMA+IRAM30m}.
              }
\label{fig:DP_OMC4_ALMA+IRAM}
\end{figure*}

\begin{figure*}
\centering
\includegraphics[width=1.0\textwidth]{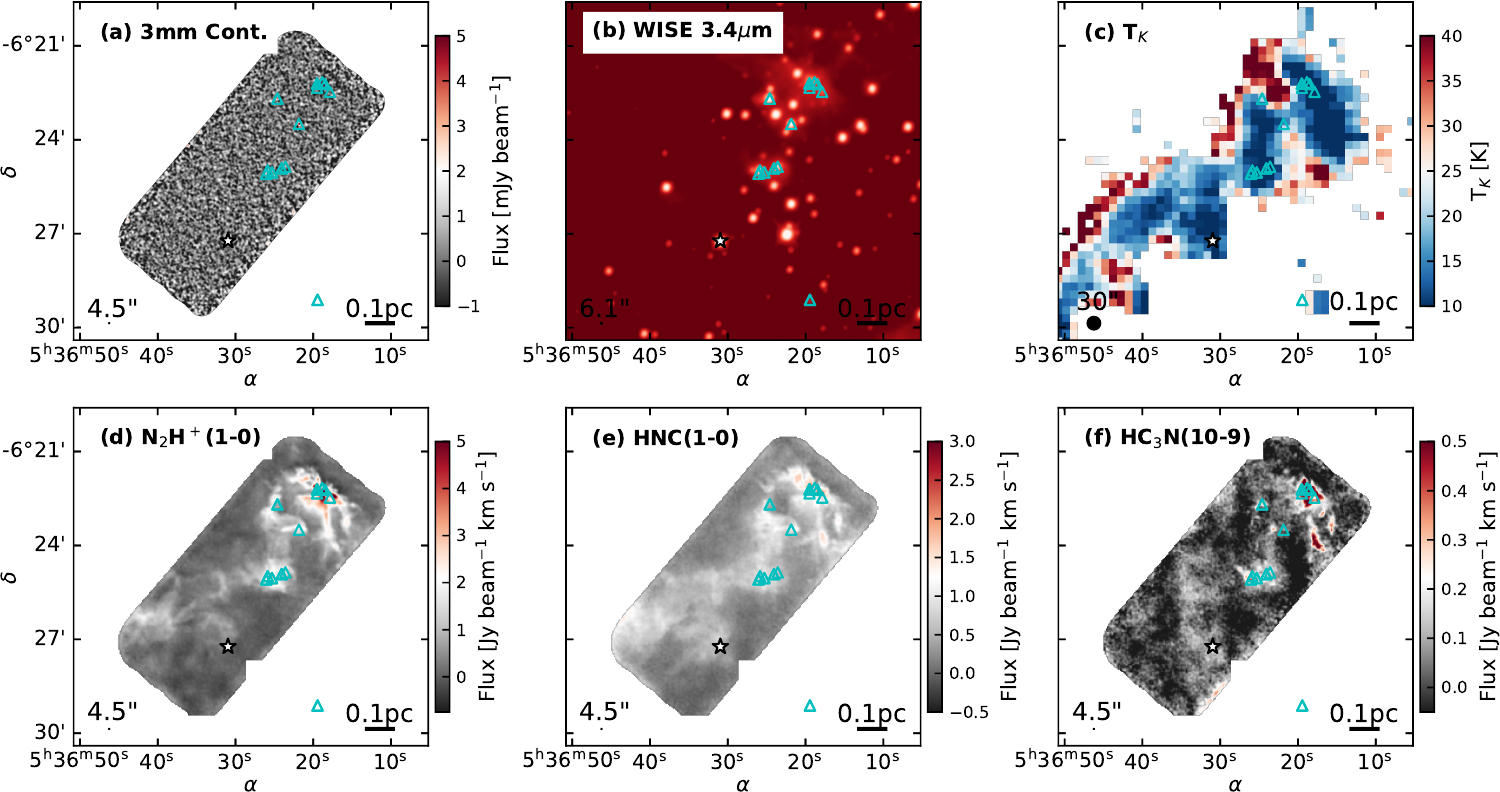}
      \caption{High-resolution observations in LDN~1641N similar to Fig.~\ref{fig:OMC3_ALMA+IRAM30m}.
              }
\label{fig:DP_LDN1641N_ALMA+IRAM}
\end{figure*}

\begin{figure*}
\centering
\includegraphics[width=1.0\textwidth]{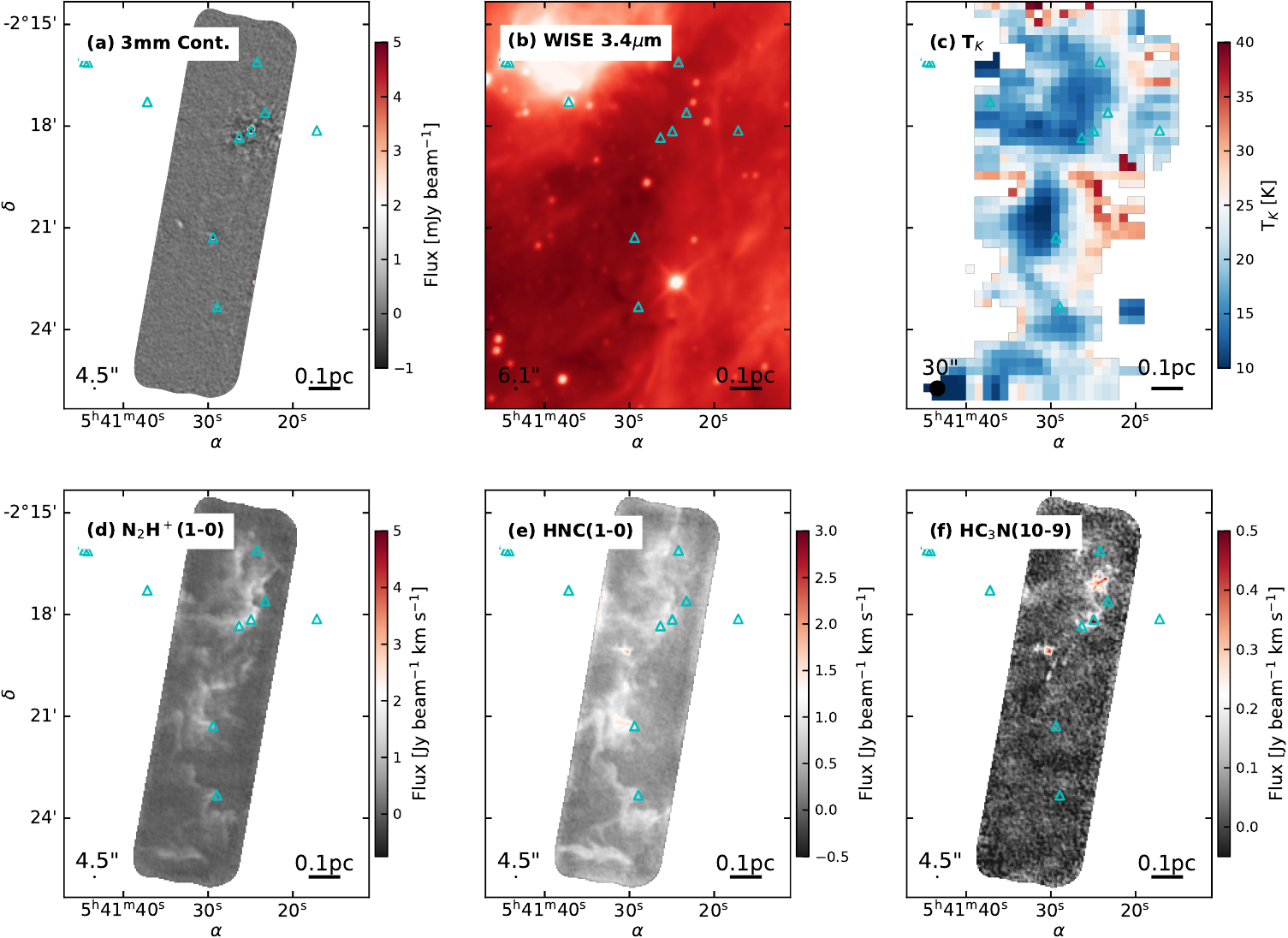}
      \caption{High-resolution observations in NGC~2023 similar to Fig.~\ref{fig:OMC3_ALMA+IRAM30m}.
              }
\label{fig:DP_NGC2023_ALMA+IRAM}
\end{figure*}

\begin{figure*}
\centering
\includegraphics[width=1.\textwidth]{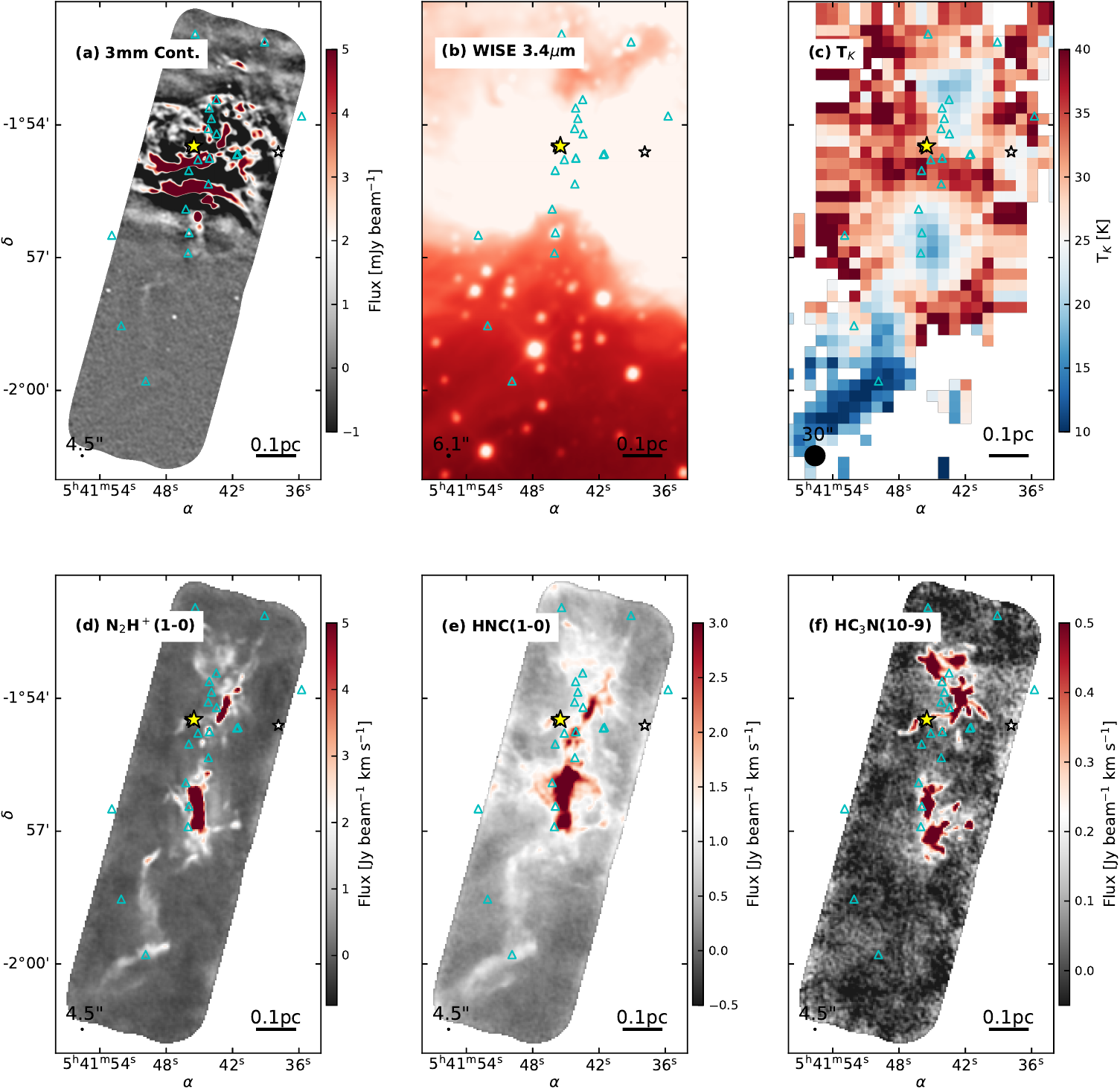}
      \caption{High-resolution observations in Flame Nebula similar to Fig.~\ref{fig:OMC3_ALMA+IRAM30m}.
              }
\label{fig:DP_FlameNebula_ALMA+IRAM}
\end{figure*}

\begin{figure*}
\centering
\includegraphics[width=1\textwidth]{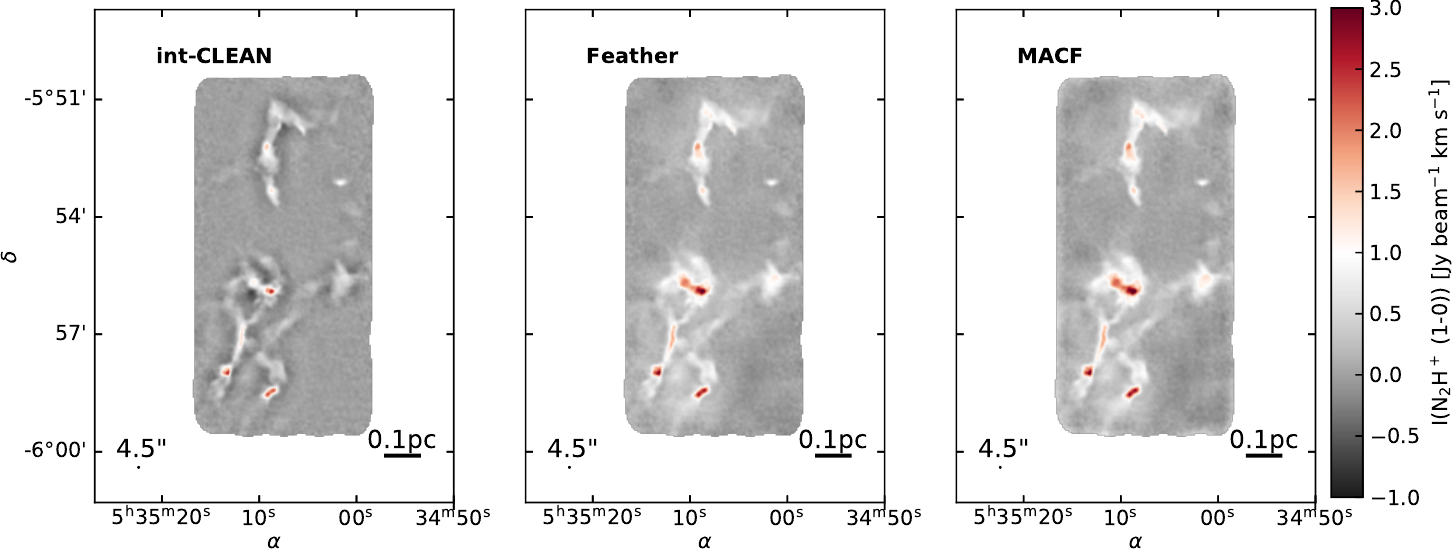}\\
\includegraphics[width=1\textwidth]{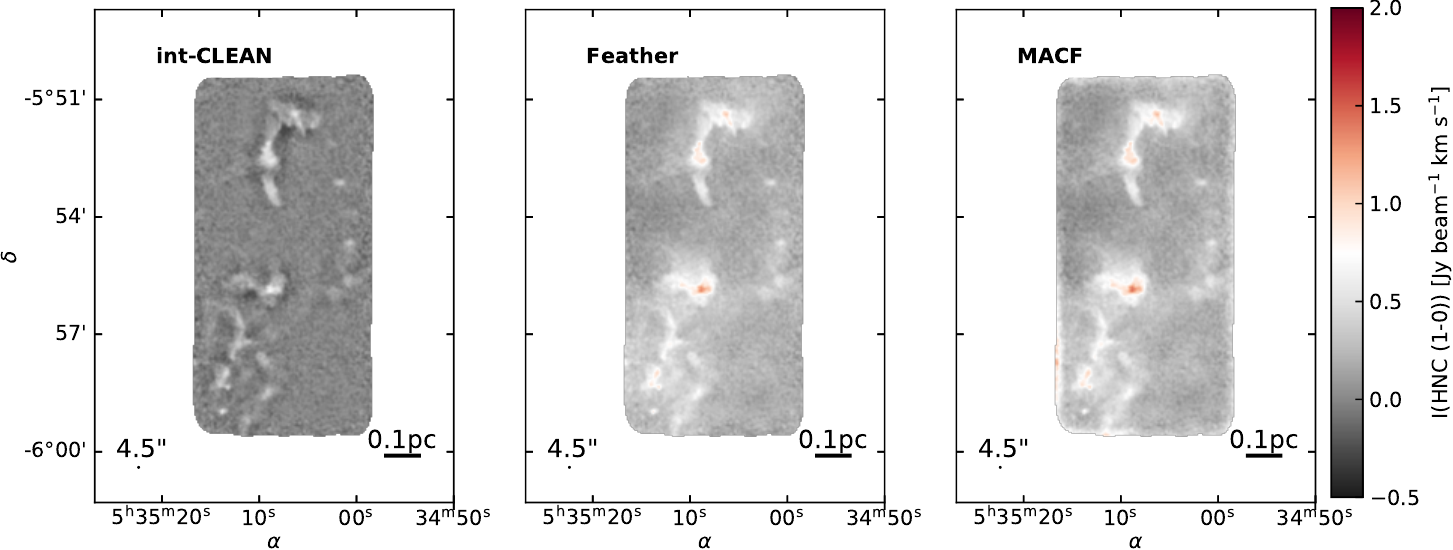}\\
\includegraphics[width=1\textwidth]{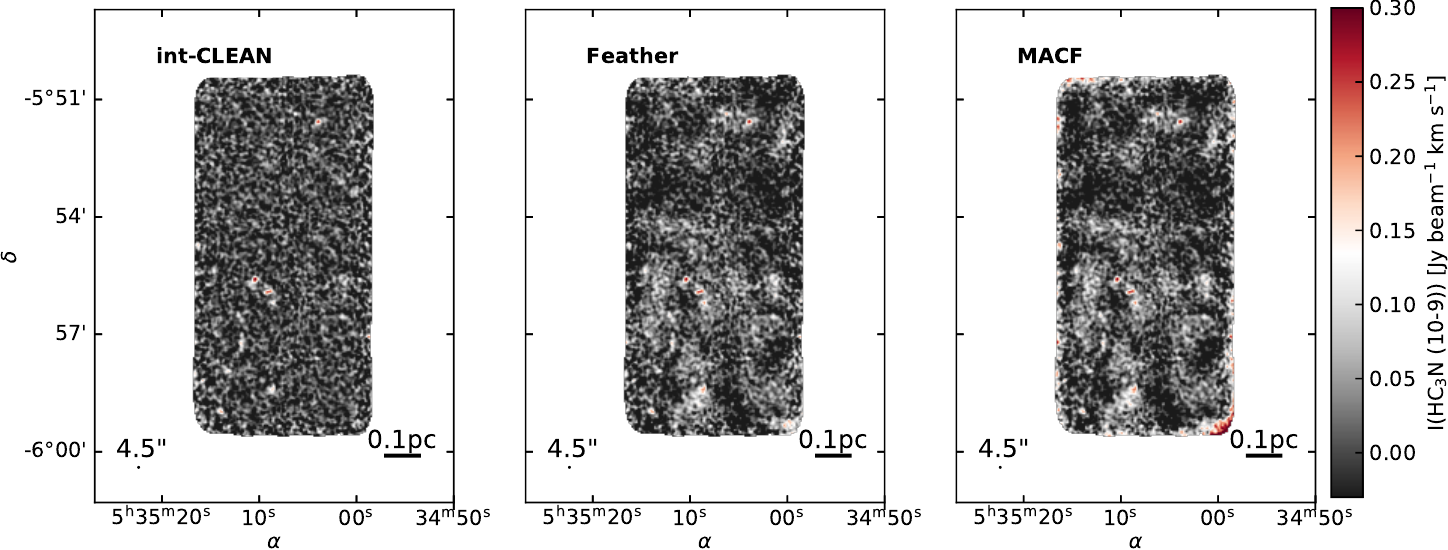}
      \caption{Comparison of the different data combination methods in OMC-4 South, similar to Fig.~\ref{fig:OMC3_DCs}}
\label{fig:OMC4_DCs}
\end{figure*}

\begin{figure*}
\centering
\includegraphics[width=1\textwidth]{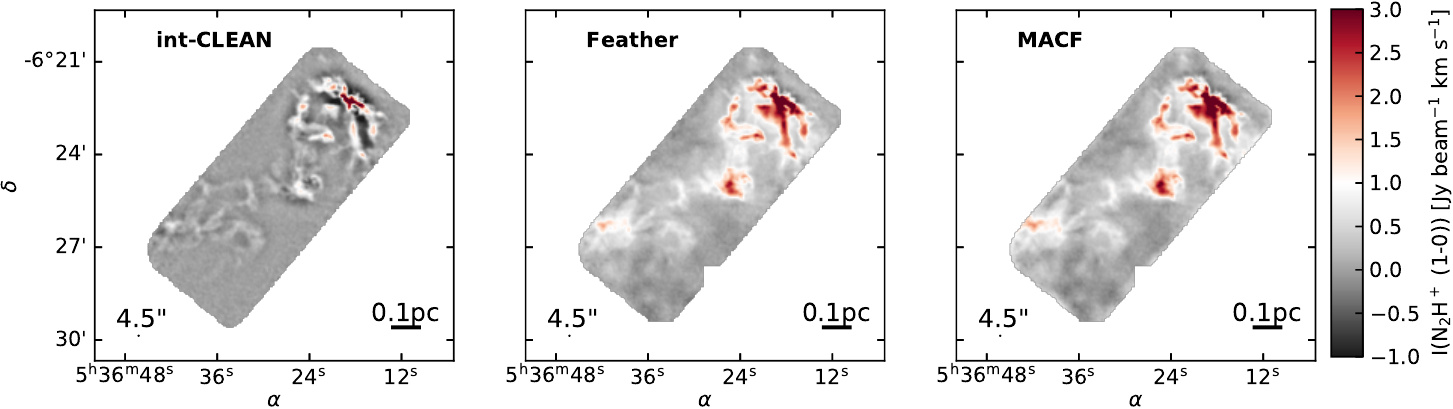}\\
\includegraphics[width=1\textwidth]{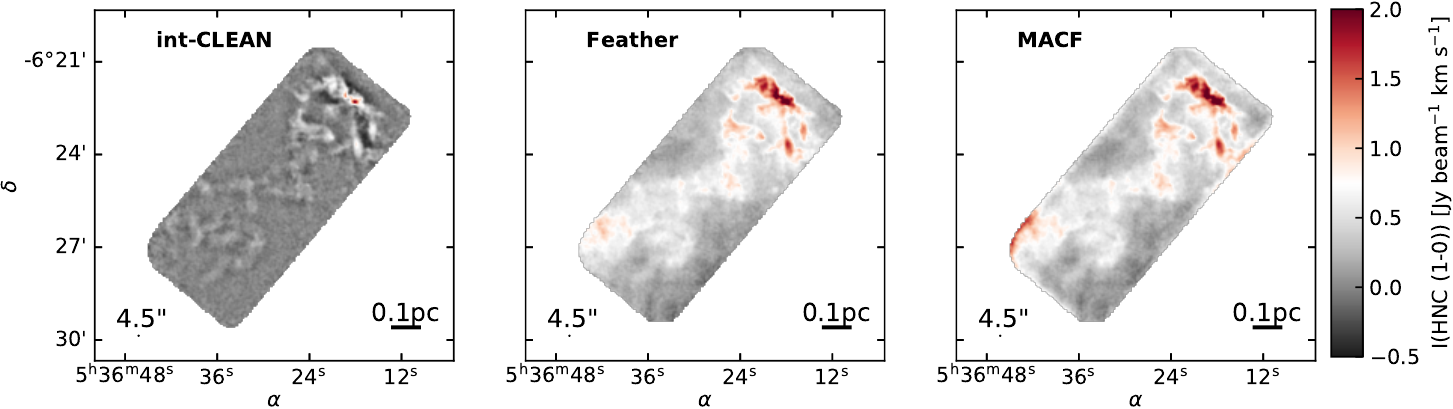}\\
\includegraphics[width=1\textwidth]{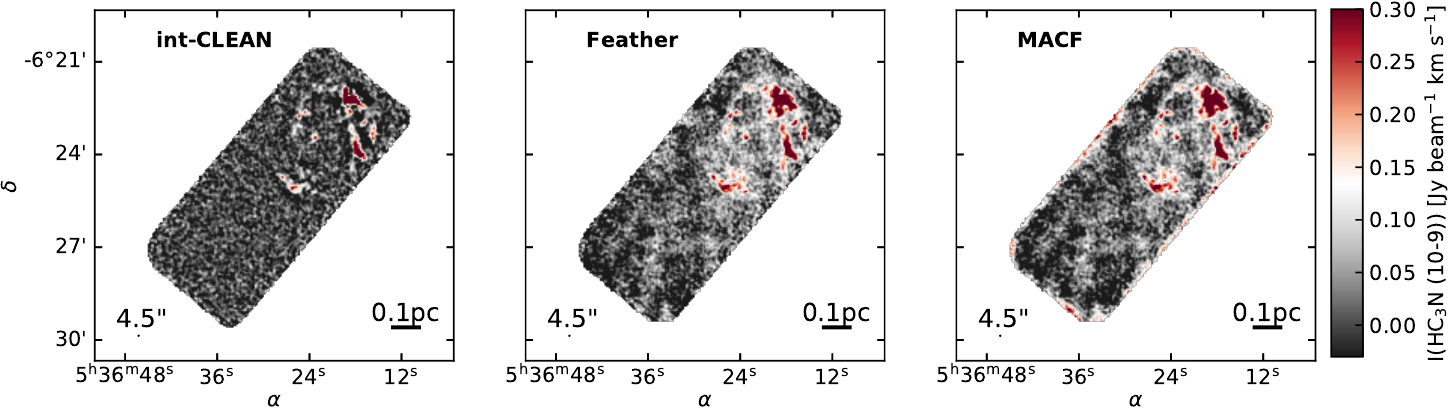}
      \caption{Comparison of the different data combination methods in LDN~1641N, similar to Fig.~\ref{fig:OMC3_DCs}}
\label{fig:LDN1641N_DCs}
\end{figure*}

\begin{figure*}
\centering
\includegraphics[width=0.95\textwidth]{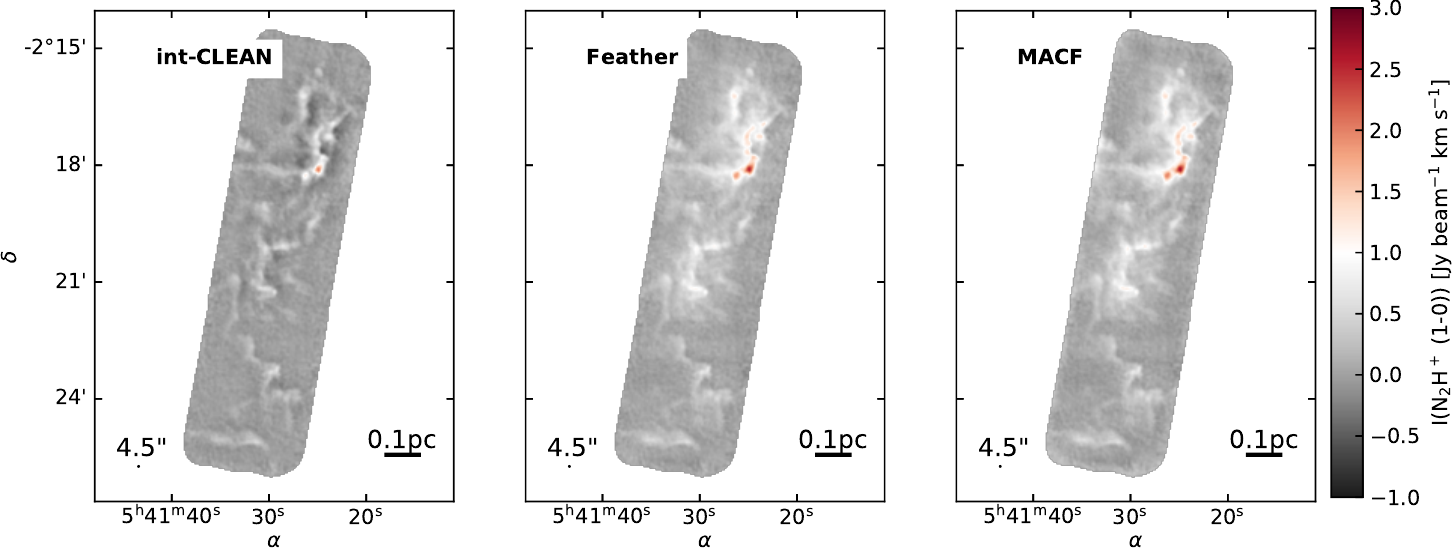}\\
\includegraphics[width=0.95\textwidth]{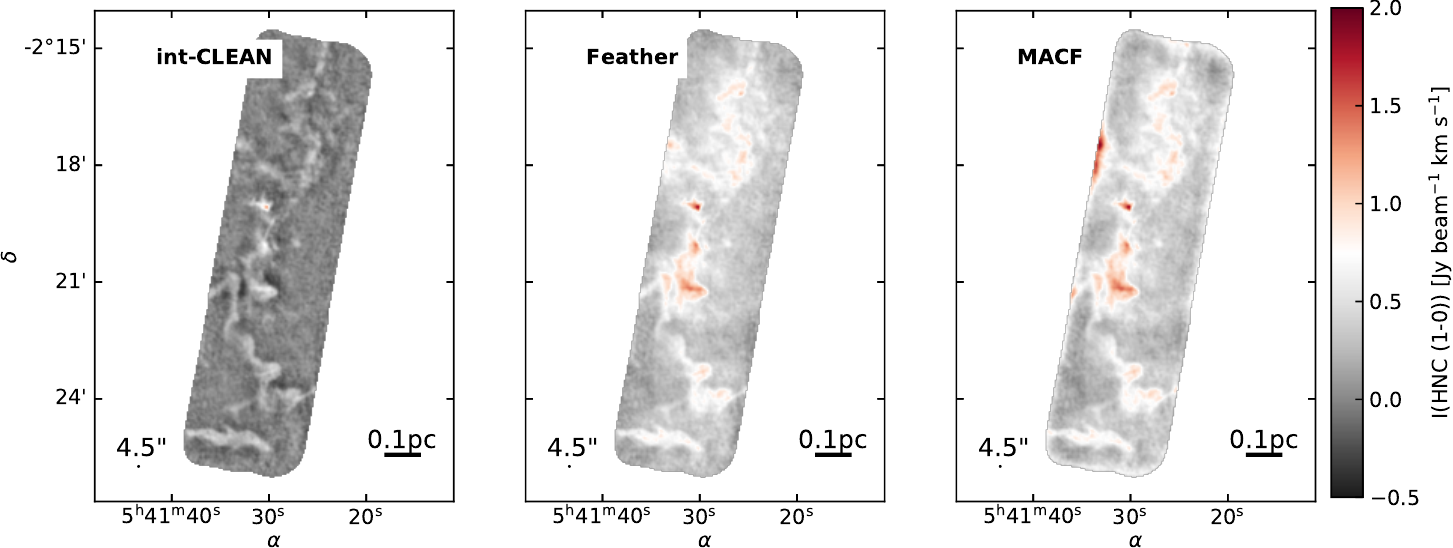}\\
\includegraphics[width=0.95\textwidth]{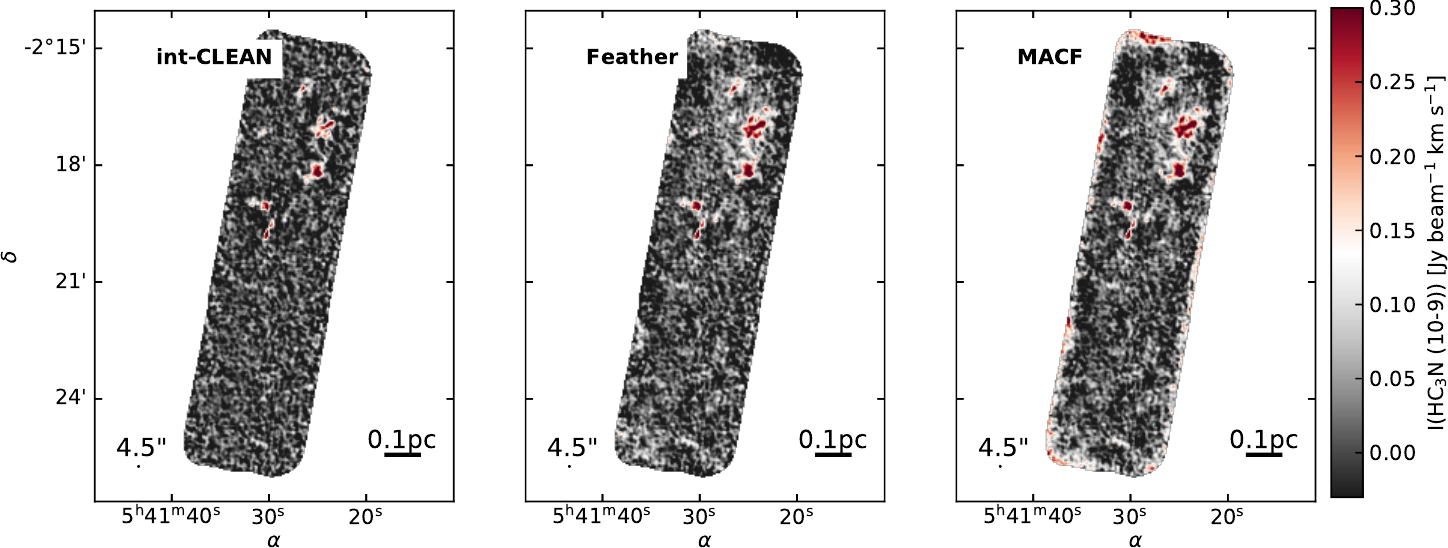}
      \caption{Comparison of the different data combination methods in NGC~2023, similar to Fig.~\ref{fig:OMC3_DCs}}
\label{fig:NGC2023_DCs}
\end{figure*}

\begin{figure*}
\centering
\includegraphics[width=0.89\textwidth]{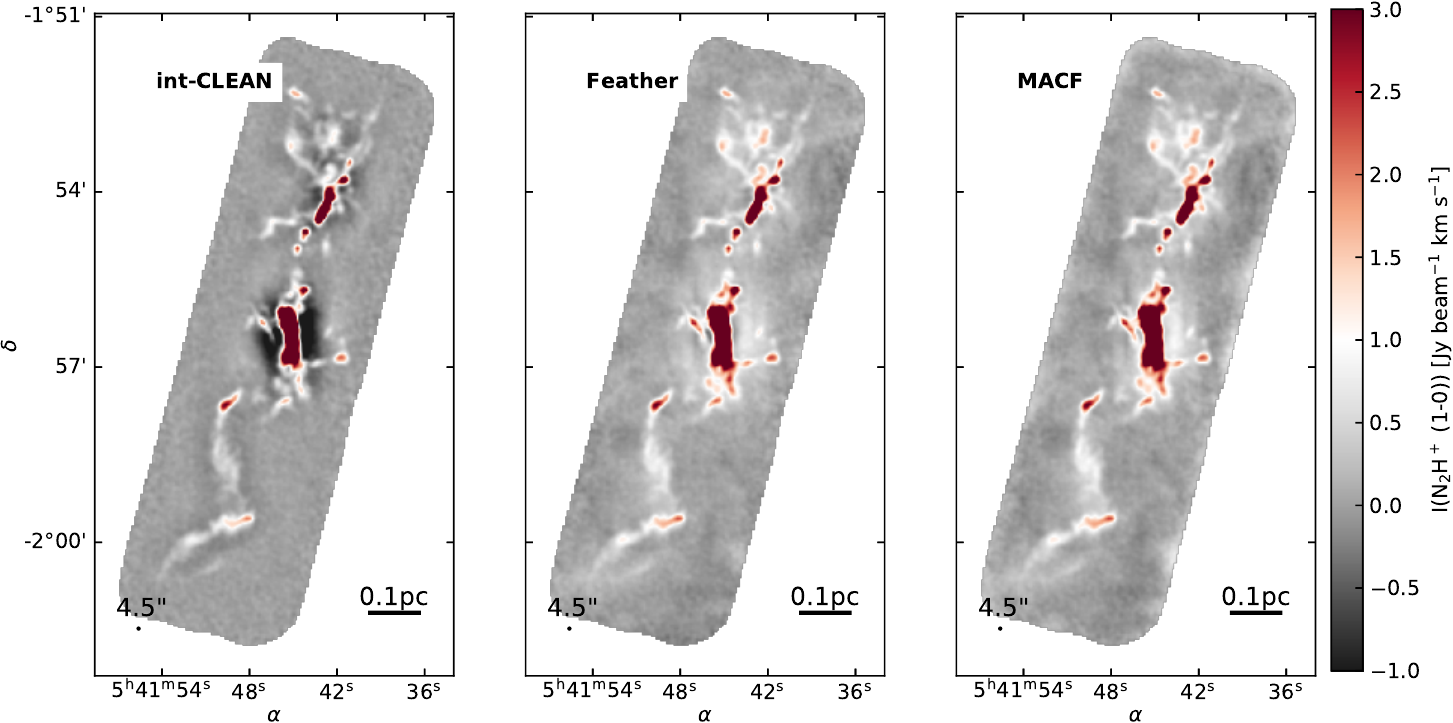}\\
\includegraphics[width=0.89\textwidth]{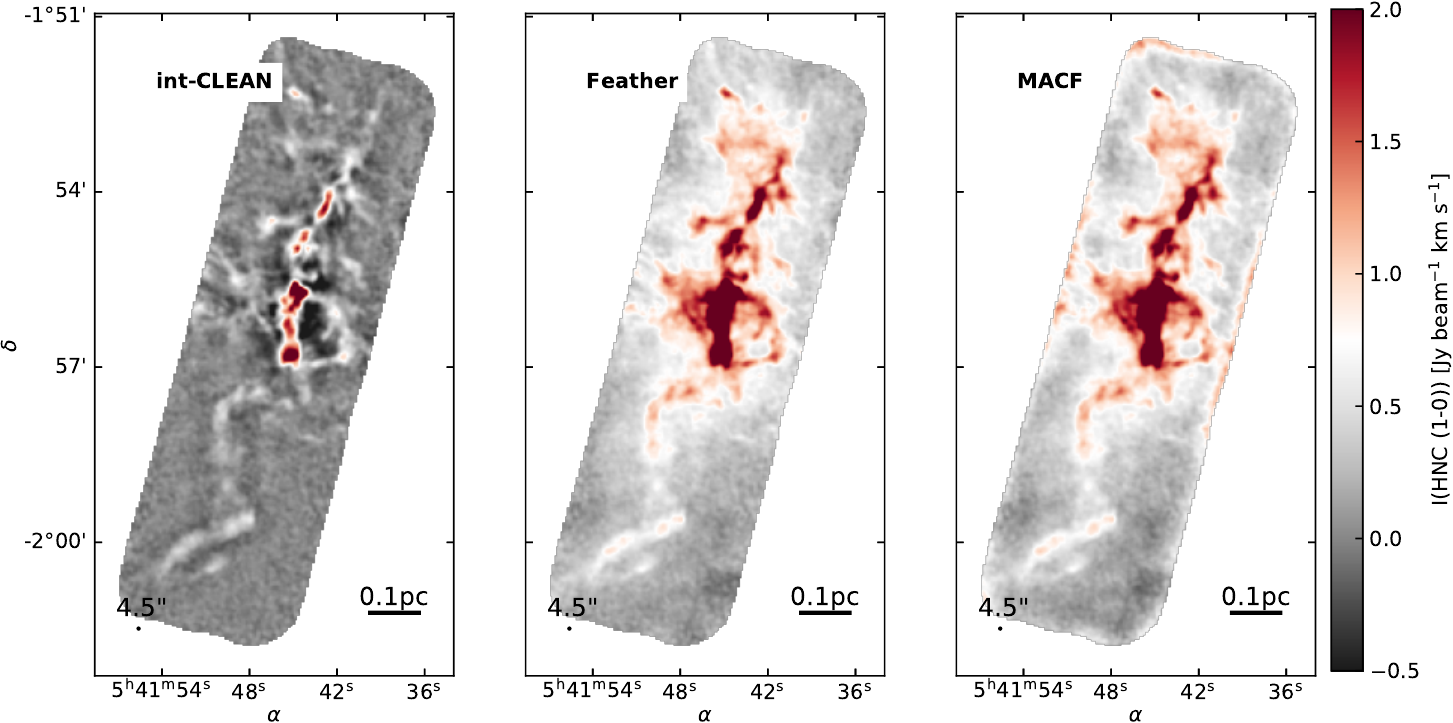}\\
\includegraphics[width=0.89\textwidth]{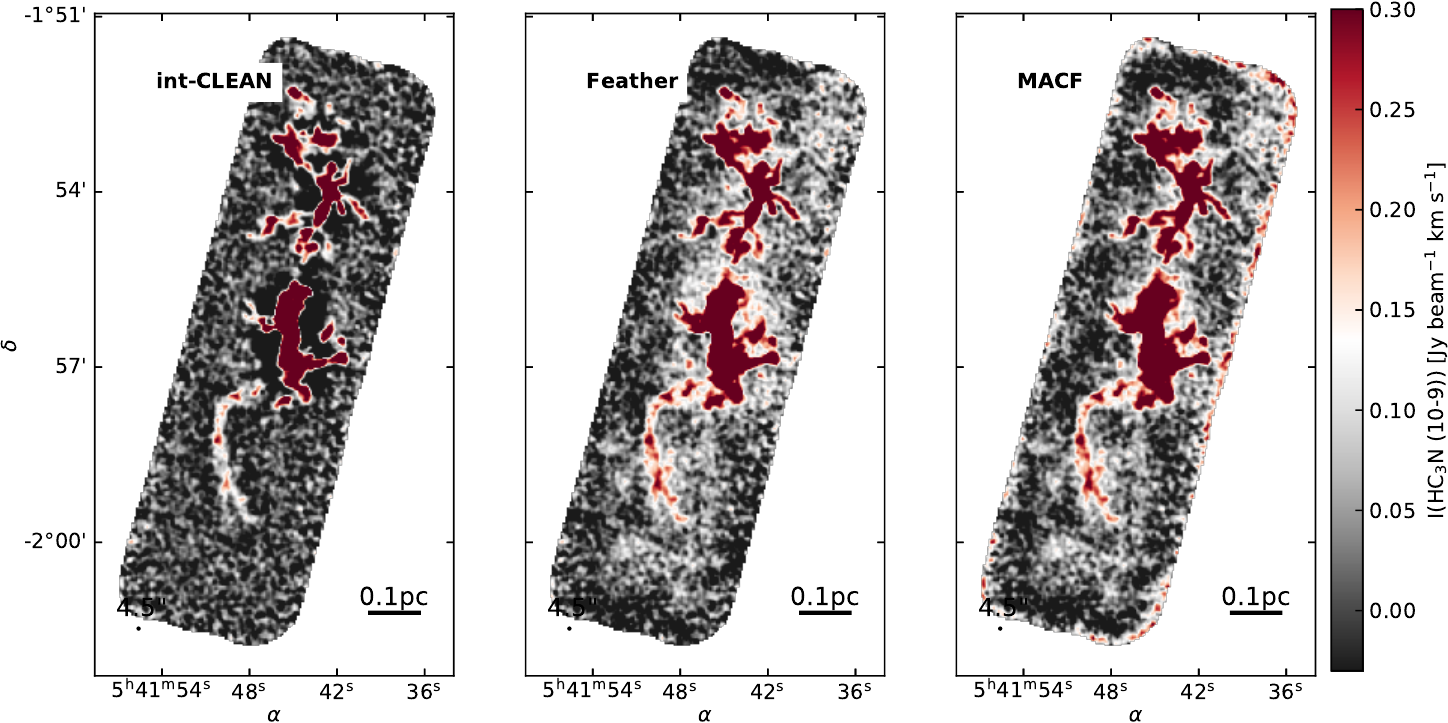}
      \caption{Comparison of the different data combination methods in Flame Nabula, similar to Fig.~\ref{fig:OMC3_DCs}}
\label{fig:Flame_Nebula_DCs}
\end{figure*}

In this appendix we present all the data products of our EMERGE Early ALMA survey in OMC-1, OMC-2, OMC-4 South, LDN~1641N, NGC~2023, and Flame Nebula (see also Figs.~\ref{fig:OMC3_IRAM30m} and \ref{fig:OMC3_ALMA+IRAM30m} for those on OMC-3). We include all the integrated intensity and continuum maps of our IRAM-30m and ALMA observations together with the additional {\it Herschel} and {\it WISE} data.
We display the individual molecular maps of each of these regions using low spatial resolution (single-dish) observations in Figs.~\ref{fig:DP_OMC1_IRAM}-\ref{fig:DP_FlameNebula_IRAM} (see Sect.\ref{sec:EMERGE_globalprop}).
Similar maps for the high spatial resolution (ALMA+IRAM-30m) datasets can be found in Figs.~\ref{fig:DP_OMC1_ALMA+IRAM}-\ref{fig:DP_FlameNebula_ALMA+IRAM} (see Sect.\ref{sec:EMERGE_ALMAprop}).  
Complementing Fig.~\ref{fig:OMC3_DCs}, we display the comparisons between the different data combination methods for all our ALMA fields observed in N$_2$H$^+$ (top panels), HNC (middle panels), and HC$_3$N (lower panels) in Figs.~\ref{fig:OMC4_DCs}-\ref{fig:Flame_Nebula_DCs}.
To facilitate the comparison between targets we display all regions within the same plotting ranges in each of the panels.

\end{appendix}

\end{document}